\documentclass[12pt]{article}
\usepackage{amsmath,amssymb,theorem,cite,epsfig,url,psfrag,eepic,amsmath,mathtools,mathrsfs,amsbsy,dsfont,esint,braket,cancel}
\usepackage{indentfirst}
\usepackage{graphicx}
\usepackage{tikz,pgfplots}
\usepackage[bookmarks=true,hyperfigures=true,colorlinks=true,linkcolor=blue,citecolor=blue,urlcolor=blue,bookmarksnumbered]{hyperref}
\numberwithin{equation}{section}
\advance \topmargin by -\headheight
\advance \topmargin by -\headsep     
\evensidemargin \oddsidemargin
\usepackage{bm,upgreek}
\makeatletter
\def\Appendix{\appendix
  \def\@seccntformat##1{Appendix~\csname the##1\endcsname.~~}}
\makeatother
\def\XXint#1#2#3{{\setbox0=\hbox{$#1{#2#3}{\int}$}
\vcenter{\hbox{$#2#3$}}\kern-.5\wd0}}

\begin{document}
\title{\textbf{QCD$_2$ 't Hooft model: 2-flavour mesons spectrum}\vspace*{0.3cm}}%.3cm
\date{}
\author{Aleksandr Artemev$^{1,2}$\thanks{artemev.aa@phystech.edu}, Alexey Litvinov$^{1}$\thanks{A.Litvinov@skoltech.ru}\; and Pavel Meshcheriakov$^{1,2,3}$\thanks{meshcheriakov.pa@phystech.edu}
%\vspace*{1.5cm}
\\[\medskipamount]
\parbox[t]{0.85\textwidth}{\normalsize\it\centerline{1. Krichever Center, Skolkovo Institute of Science and Technology, 121205 Moscow, Russia}}
\\
\parbox[t]{0.85\textwidth}{\normalsize\it\centerline{2. Landau Institute for Theoretical Physics, 142432 Chernogolovka, Russia}}
\\
\parbox[t]{0.85\textwidth}{\normalsize\it\centerline{3. Moscow Institute of Physics and Technology, 141700 Dolgoprudny, Russia}}}
\maketitle
\begin{abstract}
    We continue analytical study of the meson mass spectrum in the large-$N_c$ two-dimensional QCD, known as the 't Hooft model, by addressing the most general case of quarks with unequal masses. Based on our previous work, we develop non-perturbative methods to compute spectral sums and systematically derive large-$n$ WKB expansion of the spectrum.  Furthermore, we examine the behavior of these results in various asymptotic regimes, including the chiral, heavy quark, and heavy-light limits, and establish a precise coincidence with known analytical and numerical results obtained through alternative approaches. 
\end{abstract}

%%%%%%%%%%%%%%%%%%%%%%%%%%%%%%%%%%%%%%%%%%%%%%%%%%%%%%%%%%%%%%%%%%%%%%%%%%%%%%%%%%%%%%%%%%%%%%%%%\newpage
\tableofcontents
\section{Introduction}
One of the most fundamental and long-standing challenges in theoretical physics is the understanding of the mechanism of confinement in gauge theories. In particular, the inability to analytically describe quark confinement in strongly coupled quantum chromodynamics (QCD) in $3+1$ dimensions remains a major obstacle. More generally, the confinement and the dynamics of chromoelectric flux tubes in non-abelian gauge theories continue to evade a complete theoretical explanation. To make progress, it can be helpful to study lower-dimensional toy models that capture key qualitative features of more complex and realistic four-dimensional gauge theories but allow for more controlled analytical approaches. This simplification in two spacetime dimensions arises primarily because the gauge field is non-dynamical, unlike in four dimensions, where the dynamics of propagating gluon degrees of freedom are physically important. 

One of the well-studied examples of such toy models that we focus on in this paper is the 't Hooft model \cite{THOOFT1974461}, which describes the double scaling limit of $1+1$ dimensional version of QCD: Yang-Mills theory coupled to $N_f$ flavours of quarks in the fundamental representation of the gauge group $SU(N_c)$. 2D QCD is described by the following Lagrangian density
\begin{equation}
    \mathcal{L}_{QCD_2} =-\frac{1}{2}\text{tr\;}F_{\mu\nu}F^{\mu\nu} +\sum\limits_{k=1}^{N_f}\bar{\psi}_k(i\gamma^\mu D_\mu-m_k)\psi_k.
\end{equation}

In his original paper \cite{THOOFT1974461} G. 't Hooft considered the double scaling limit \cite{tHooft:1973alw} of this theory, where the number of colors $N_c$ tends to infinity, while keeping the 't Hooft coupling  $g^2 = g^2_{\text{YM}}N_c$ fixed. In this limit, all Feynman diagrams are divided into equivalence classes according to the genus of the surface $h$ on which they can be drawn without self-intersections, and also according to the number of fermionic loops $L$ in the diagram. Each diagram is proportional to $N_c^{2-2h-L}$. Accordingly, at the leading order at $N_c\rightarrow\infty$ only planar diagrams without fermionic loops are relevant, and diagrams of higher genus are suppressed by the powers of $1/N_c$. In the planar limit, some characteristic features of QCD$_4$, such as the absence of deconfined quark states and the Regge-like behavior of meson spectra, become apparent, making the 't Hooft model one of the most intriguing and yet relatively simple low-dimensional toy models for studying the mechanism of confinement.

In the 't Hooft model, mesons are composite particles whose masses and wavefunctions are precisely determined by the Bethe-Salpeter equation. For a meson consisting of a quark and an antiquark with bare (Lagrangian) masses $m_{1,2}$ this equation assumes the following form (known as 't Hooft equation) \cite{THOOFT1974461,Callan:1976PhysRev}
\begin{equation}\label{'tHooft-eq}
    2\pi^2\lambda_n\;\phi^{(n)}_{12}(x)=\mathcal{H}\phi^{(n)}_{12}(x)=\left[\frac{\alpha_1}{x}+\frac{\alpha_2}{1-x}\right]\phi^{(n)}_{12}(x)-\fint_0^1\limits dy\frac{\phi^{(n)}_{12}(y)}{(x-y)^2}.
\end{equation}
This equation is an eigenproblem for the operator $\mathcal{H}$. Here the parameters\footnote{Here, for simplicity, we use the indices ‘‘$1,2$" to denote flavours. In fact, everything we are going to discuss will work for any quark/antiquark pair. We also note that these parameters are dimensionless since, in $d=1+1$, the coupling constant $g$ has the dimension of mass.} $\alpha_{1,2}$ and $\lambda_n$ are related to the 't Hooft coupling constant $g$, the masses of quark and antiquark $m_{1,2}$ and the meson masses $M_n$ as follows
\begin{equation}\label{mass-notation}
    \alpha_i=\frac{\pi m^2_i}{g^2}-1,\quad M_n^2=2\pi g^2\lambda_n.
\end{equation}

Although the 't Hooft equation can be solved numerically with high precision, allowing key features of the theory to be extracted directly \cite{THOOFT1974461,Hanson:1976ey,Brower:1979PhysRevD,Anand:2021qnd,Kochergin:2024quv}, an analytical treatment remains valuable. A deeper analytical understanding of the eigenproblem not only provides insight into the structure of the solution but also offers a more systematic approach to studying its properties. For this reason, we believe that further analytical investigation of the 't Hooft equation \eqref{'tHooft-eq} is necessary and worthwhile. 

Another motivation for developing analytical approaches to the 't Hooft model comes from a different two-dimensional gauge theory relevant to QCD, that has attracted renewed attention in recent years. It is the $SU(N_c)$ Yang-Mills theory coupled to an adjoint Majorana fermion, commonly referred to as adjoint QCD$_2$ 
\cite{Komargodski:2020mxz,Popov:2022vud,Dempsey:2021xpf,Dempsey:2023fvm,Damia:2024kyt}. Adjoint fermions exhibit non-trivial dynamics of the confining flux tubes, modelling their transverse degrees of freedom in higher dimensions. Obtaining analytical results for adjoint QCD$_2$ is challenging even in the large $N_c$ limit. It is typically studied using numerical methods, with exact results being accessible only in the limit of very heavy adjoint fermions \cite{Asrat:2022aov}. Interestingly, there are indications that the dynamics of the confining flux tubes in this model may become integrable in certain regimes \cite{Dubovsky:2018dlk,Donahue:2019adv,Donahue:2019fgn,Donahue:2022jxu}. Exploring this possibility in detail requires a solid understanding of simpler, yet analytically tractable theories. In this regard, the 't Hooft model provides an ideal testing ground, where exact results can be explicitly derived. By studying its integrable structures, we aim to gain insights that could potentially extend to more complex gauge theories, including adjoint QCD$_2$.

In their pioneering work \cite{Fateev:2009jf} Fateev, Lukyanov and Zamolodchikov (FLZ) introduced new non-perturbative analytical approach to the 't Hooft model, demonstrating that in a very special case\footnote{In \cite{Fateev:2009jf} the authors announced without explanation also some preliminary results for general $\alpha$, but unfortunately their next publication did not follow.} $\alpha_1=\alpha_2=0$, the 't Hooft equation can be recast as  Baxter's TQ equation, a structure commonly associated with integrable models in $1+1$ dimensions. This discovery was a significant step towards understanding the analytical properties of the 't Hooft model. Thanks to this consideration, the authors of \cite{Fateev:2009jf} were able to compute first few spectral sums $G^{(s)}_{\pm}$ (for $s\leq13$)
\begin{equation}\label{spectral_sums_def}
    G_{+}^{(s)} = \sum \limits_{n=0}^\infty \left(\frac{1}{\lambda_{2n}^s} - \frac{\delta_{s,1}}{n+1} \right),\quad G_{-}^{(s)} = \sum \limits_{n=0}^\infty \left(\frac{1}{\lambda_{2n+1}^s} - \frac{\delta_{s,1}}{n+1} \right),
\end{equation}
as well as the asymptotic large--$n$ WKB expansion. More recently, in  \cite{Litvinov:2024riz} last two authors of the current paper extended the FLZ method to the case $\alpha_1 = \alpha_2 = \alpha$ and successfully generalized FLZ's results for the spectral sums as well as for the asymptotic expansions for the spectrum\footnote{Note that the generalization of the FLZ method to the case $\alpha_1=\alpha_2=\alpha$ was also considered in the recent paper \cite{Ambrosino:2023dik}. Initially, in the first version of this paper, which appeared earlier than \cite{Litvinov:2024riz}, the authors used solutions of the TQ equation that did not correctly account for all the required analytic properties. However, in the revised version of \cite{Ambrosino:2023dik} the correct analitycity has been perturbatively restored in the mass parameter $\alpha$ (see below). Thus, the difference between the results obtained in \cite{Ambrosino:2023dik} and \cite{Litvinov:2024riz} is that those of \cite{Litvinov:2024riz} are fully non-perturbative in $\alpha$}. This achievement motivated us to further investigate the case of arbitrary quark masses. 

In this work, following \cite{Fateev:2009jf,Litvinov:2024riz}, we extend previous analyses and show that the reformulation in terms of Baxter's TQ equation  holds for arbitrary quark and antiquark mass parameters $\alpha_{1,2}$, supporting the idea that hidden integrable structures may underlie the 't Hooft model beyond specific parameter choices.

This paper is organized as follows. In Section \ref{From-tHooft-to-TQ}, we review the key properties of the 't Hooft equation and show how its solutions can be reformulated as solutions of Baxter's TQ difference equation. We then extend this reformulation to arbitrary values of $\lambda$ away from the spectrum by introducing an inhomogeneous term in the Fredholm integral equation. In Sections \ref{small-lambda-expansion} and \ref{large-lambda-expansion} we perturbatively\footnote{By this we mean that we construct it as a series expansion in $\lambda$ or $\lambda^{-1}$. We emphasize that we do not use perturbation theory in the 't Hooft equation parameters $\alpha_{1,2}$ and the analytic answers we obtain are exact in $\alpha_{i}$.} construct special solutions of the TQ equation in two distinct limiting cases:  $\lambda \to 0$ and  $\lambda \to -\infty$, respectively. In Section \ref{Q-D-relations}, we describe the procedure for extracting spectral data from the TQ equation, extending the discussion in \cite{Fateev:2009jf,Litvinov:2024riz} to the case of different quark masses. The main idea is to identify nontrivial relations that allow to express the spectral determinants in terms of solutions of the TQ equation presented in Sections \ref{small-lambda-expansion} and \ref{large-lambda-expansion}. Section \ref{Analytical-results} focuses on analytical results, including formulas for the spectral sums $G^{(s)}_\pm$ and the systematic large-$n$ expansion (WKB). In Section \ref{limiting-cases}, we analyze our results in several physically significant limiting cases: the chiral limit $\alpha_i \to -1$, the heavy quark limit $\alpha_i \to \infty$, and the heavy-light limit, when one of quark masses is finite and the other goes to infinity. This allows us to verify our results analytically by comparing them with known results and confirming their consistency. Section \ref{numerics} is devoted to the numerical verification of our results. We test them by comparing with the numerical solution of \eqref{'tHooft-eq}, which is based on the decomposition of functions in the basis of Chebyshev polynomials. Finally, in Section \ref{Discussion}, we present our concluding remarks and explore potential directions for future research. Additional technical details are provided in the appendix.   
%%%%%%%%%%%%%%%%%%%%%%%%%%%%%%%%%%%%%%%%%%%%%%%%%%%%%%%%%%%%%%%%%%%%%%%%%%%%%%%%%%%%%%%%%%
\section{'t Hooft equation: properties and relation to integrability} \label{From-tHooft-to-TQ}
The purpose of this section is to introduce and describe the basic properties of the 't Hooft equation. We will explore the analytical structure of the wave functions and define the main object of our study, the Q-functions. Through a general consideration of the analytical properties of the Q-functions, we will derive the TQ equation and relate its solutions to solutions of an inhomogeneous integral equation.
%%%%%%%%%%%%%%%%%%%%%%%%%%%%%%%%%%%%%%%%%%%%%%%%%%%%%%%%%%%%%%%%%%%%%%%%%%%%%%%%%%%%%%%%%
\subsection{Setup}
Since the eigenvalue problem \eqref{'tHooft-eq} serves as the basis of our study, it is important to discuss its key features. Let us note that hermiticity of the Hamiltonian $\mathcal{H}$ requires that the wave functions have the following boundary behavior \cite{THOOFT1974461}
\begin{equation}\label{phi-boundary-conditions}
    \phi_{12}(x)\sim
    \begin{cases}
        x^{\beta_1},\quad &x\to 0;\\
        (1-x)^{\beta_2},\quad &x\to 1,\\
    \end{cases}
\end{equation}
where $\beta_i$ are the roots of the transcendental equation
\begin{equation}\label{boundary-cond-eq}
    \pi\beta_i\cot{\pi\beta_i}+\alpha_i=0,\quad 0\leq\beta_i<1.
\end{equation}
Thus, for $\alpha_{1,2}\geq -1$ the Hamiltonian $\mathcal{H}$ is positive-definite and hermitian on the space of functions with the boundary conditions \eqref{phi-boundary-conditions}. More specifically, one has
\begin{multline}\label{hermiciti-of-H}
    (\varphi,\mathcal{H}\psi)=(\mathcal{H}\varphi,\psi)=\int_0^1 \limits dx\left[\frac{1+\alpha_1}{x}+
    \frac{1+\alpha_2}{1-x}\right]\varphi^*(x)\psi(x)+\\+
    \frac{1}{2}\int_0^1\limits dx\int_0^1\limits dy\frac{(\varphi^*(x)-\varphi^*(y))(\psi(x)-\psi(y))}{(x-y)^2}. 
\end{multline}
It follows directly from \eqref{hermiciti-of-H} that the 't Hooft equation \eqref{'tHooft-eq} possesses only positive eigenvalues when $\alpha_{1,2}>-1$. Clearly, tachyonic bound states can only arise if at least one of the original particles (quark or antiquark) is tachyonic, as indicated by definition \eqref{mass-notation}. Additionally, a single massless ($\mathcal{H}\phi^{(0)}_{12}=0$) bound state appears when both quark and antiquark have zero mass (chiral limit) $\alpha_1=\alpha_2=-1$. The corresponding eigenfunction is $\phi^{(0)}_{12}(x)=1$. We will discuss the chiral limit in more detail in Section \ref{limiting-cases}.

It has been shown in \cite{Federbush:1977PRevD} that there exists a natural self-adjoint extension of 't Hooft Hamiltonian $\mathcal{H}$ and that its spectrum is discrete (but not finite). Therefore,  the meson eigenstates  $\phi^{(n)}_{12}(x)$ with energies $2\pi^2\lambda_n$ are complete and orthonormal
\begin{equation}\label{complete_and_orth}
    \sum_{k=0}^{\infty}(\phi^{(k)}_{12}(x))^*\phi^{(k)}_{12}(x')=\delta(x-x'),\quad \int_0^1\limits dx\;(\phi^{(n)}_{12}(x))^*\phi^{(m)}_{12}(x)=\delta_{nm}.
\end{equation}
Moreover, they can be chosen real $\phi^*_{12}(x)=\phi_{12}(x)$. Therefore, they are also eigenstates of the charge conjugation operator $\hat{\mathcal{C}}$.  

In what follows, we introduce different parameters
\begin{equation}\label{alphas-alpha-beta}
    \alpha = \frac{\alpha_1 + \alpha_2}{2},\quad \beta = \frac{\alpha_2 - \alpha_1}{2},
\end{equation}
so that the 't Hooft equation \eqref{'tHooft-eq} is rewritten as follows
\begin{equation}\label{tHooft-eq-a-b}
    2\pi^2\lambda\;\phi(x|\beta)=\left[\frac{\alpha}{x(1-x)} + \frac{\beta\, (2x-1)}{x(1-x)}\right]\phi(x|\beta)-\fint_0^1\limits dy\;\frac{\phi(y|\beta)}{(x-y)^2}.
\end{equation}
The replacement of $\beta \to -\beta$ corresponds to the exchange of two constituents of the meson and leaves the spectrum $\lambda_n$ invariant: $\lambda_n(\alpha,-\beta)=\lambda_n(\alpha,\beta)$. The eigenfunctions transform simply as
\begin{equation} \label{sym}
    \phi^{(n)}(x|\beta) = (-1)^n \phi^{(n)}(1-x|-\beta).
\end{equation}
The sign is conventional and is chosen to agree for $\beta = 0$ with another less trivial property of the eigenfunctions related to the parity symmetry \cite{Callan:1976PhysRev}. It reads
\begin{equation}\label{phi-nontrivial-sym}
    \sqrt{1+\alpha-\beta} \int \limits_0^1 \frac{\phi^{(n)}(x|\beta)}{x} dx = (-1)^n\, \sqrt{1+\alpha+\beta}  \int \limits_0^1 \frac{\phi^{(n)}(x|\beta)}{1-x} dx.
\end{equation}
The eigenvalues corresponding to ‘‘odd/even" eigenfunctions $\phi^{(2n)}(x)/\phi^{(2n+1)}(x)$ are $\lambda_{2n}/\lambda_{2n+1}$.
%%%%%%%%%%%%%%%%%%%%%%%%%%%%%%%%%%%%%%%%%%%%%%%%%%%%%%%%%%%%%%%%%%%%%%%%%%%%%%%%%%%%%%%%%%%%%
\subsection{Fredholm integral equation and TQ equation}
A convenient way to rewrite the 't Hooft equation \eqref{'tHooft-eq} further is in terms of the Fourier transform with respect to the new variable $\vartheta = \frac{1}{2} \log \frac{x}{1-x}$ (in the case of equal masses, physical meaning of $\vartheta$ is the rapidity in the center of mass frame). Namely, we define
\begin{align}\label{Fourier-def}
    &\Psi(\nu|\beta)\overset{\text{def}}{=}\mathcal{F}[\phi(x|\beta)]=\int_{0}^{1}\limits \frac{dx}{2x(1-x)}\;\left(\frac{x}{1-x}\right)^{-\frac{i\nu}{2}}\phi(x|\beta), \quad \\
    &\phi(x|\beta)\overset{\text{def}}{=}\mathcal{F}^{-1}[\Psi(\nu|\beta)]=\int_{-\infty}^{\infty}\limits \frac{d\nu}{2\pi}\;\left(\frac{x}{1-x}\right)^{\frac{i\nu}{2}}\Psi(\nu|\beta).
\end{align}
Then, after multiplying the 't Hooft equation by $x(1-x)$ and Fourier transform we obtain the homogeneous Fredholm integral equation
\begin{equation}\label{'tHooft-eq-Fourier}
    \left(\frac{2\alpha}{\pi}+\nu\coth{\frac{\pi\nu}{2}}\right)\Psi(\nu|\beta) -\frac{2i\beta}{\pi}\fint^{\infty}_{-\infty}\limits d\nu'
    \frac{1}{2\sinh\frac{\pi(\nu-\nu')}{2}}\Psi(\nu'|\beta)
    =\lambda\int^{\infty}_{-\infty}\limits d\nu' \frac{\pi(\nu-\nu')}{2\sinh{\frac{\pi(\nu-\nu')}{2}}}\Psi(\nu'|\beta).
\end{equation}
For the norm $\|\varphi\|^2=\int_{0}^{1}\limits dx\, |\varphi(x)|^2$ to be finite and for the function $\varphi(x)$ in \eqref{'tHooft-eq} to satisfy the boundary conditions \eqref{phi-boundary-conditions}, the solution $\Psi(\nu|\beta)$ must be a smooth function of the real variable $\nu$ that vanishes at $|\nu|\to\infty$.

The symmetry property \eqref{sym} translates into the following symmetry for the eigenfunctions in the Fourier space
\begin{equation}
    \Psi_n(\nu|\beta) = (-1)^n \Psi_n (-\nu| -\beta).
\end{equation}
From now on, we will usually omit the argument $\beta$ for brevity, unless it is necessary to specify it. Define now the $Q$-function:
\begin{equation}\label{Q-def}
    Q(\nu)\overset{\text{def}}{=}\left(\frac{2\alpha}{\pi}\sinh\left(\frac{\pi\nu}{2}\right)+\nu\cosh\left(\frac{\pi\nu}{2}\right)\right)\Psi(\nu)=\sinh\left(\frac{\pi\nu}{2}\right)\,
    \left(\frac{2\alpha}{\pi}+\nu\coth{\frac{\pi\nu}{2}}\right)\Psi(\nu)
\end{equation}
and consider its meromorphic continuation to the maximal domain of analyticity. In terms of $Q(\nu)$ the Fredholm integral equation \eqref{'tHooft-eq-Fourier} reads
\begin{equation}\label{'tHooft-eq-Fourier-Q}
\begin{aligned}
    Q(\nu)=&\;\frac{2i\beta}{\pi}\sinh{\left(\frac{\pi\nu}{2}\right)}\fint^{\infty}_{-\infty}\limits d\nu'
    \frac{1}{2\sinh\frac{\pi(\nu-\nu')}{2}}\frac{Q(\nu')}{\sinh\left(\frac{\pi\nu'}{2}\right)
    \left(\frac{2\alpha}{\pi}+\nu'\coth{\frac{\pi\nu'}{2}}\right)}+
    \\
    &+\lambda\sinh\left(\frac{\pi\nu}{2}\right)\int^{\infty}_{-\infty}\limits d\nu' \frac{\pi(\nu-\nu')}{2\sinh{\frac{\pi(\nu-\nu')}{2}}}\frac{Q(\nu')}{\sinh\left(\frac{\pi\nu'}{2}\right)
    \left(\frac{2\alpha}{\pi}+\nu'\coth{\frac{\pi\nu'}{2}}\right)}.
\end{aligned}
\end{equation}
Equation \eqref{'tHooft-eq-Fourier} and finiteness of the norm of the original function $\|\phi\|<\infty$ imply the following analytic properties of the function $Q(\nu)$ valid in the strip $\textrm{Im}\;\nu\in[-2,2]$:
\begin{enumerate}
    \item it grows slower than any exponential as $|\textrm{Re}\,\nu| \to \infty$, meaning that it remains bounded as  
    \begin{equation}\label{Q-is-bounded}  
        \forall \epsilon > 0\quad Q(\nu) = \mathcal{O}(e^{\epsilon|\nu|}) , \quad |\textrm{Re}\,\nu| \to \infty,
    \end{equation}
    \item it necessarily has
    \begin{equation}\label{essential-zeores}
        \text{zeroes}:\quad 0,\quad \pm2i,\quad \pm i\nu^*_1(\alpha)
    \end{equation}
    \item and might have
    \begin{equation}\label{essential-poles}
        \begin{aligned}
            &\text{poles:}\quad &&i\nu^*_1(\alpha+\beta),\quad-i\nu^*_1(\alpha-\beta),
        \end{aligned}
    \end{equation}
\end{enumerate}
where $i\nu^*_1(\alpha)$ is the first root (closest to the origin) of the equation
\begin{equation}
    \frac{2\alpha}{\pi}+\nu\coth\left(\frac{\pi\nu}{2}\right)=0.
\end{equation}
The presence of the listed zeros is mandatory. This requirement can be understood as ``quantization conditions'': normalizable solutions of \eqref{'tHooft-eq-Fourier-Q} exist only for a discrete set of $\lambda$ and necessarily satisfy the conditions $Q(0) = Q(\pm 2i) = 0$.  Other zeros are also allowed. On the other hand, poles are only allowed at the listed points, but there might be no poles at all. We explain this statement in Appendix \ref{check-analyt-Q}.

\begin{figure}[h!]
    \centering
    \includegraphics[width=0.8\textwidth]{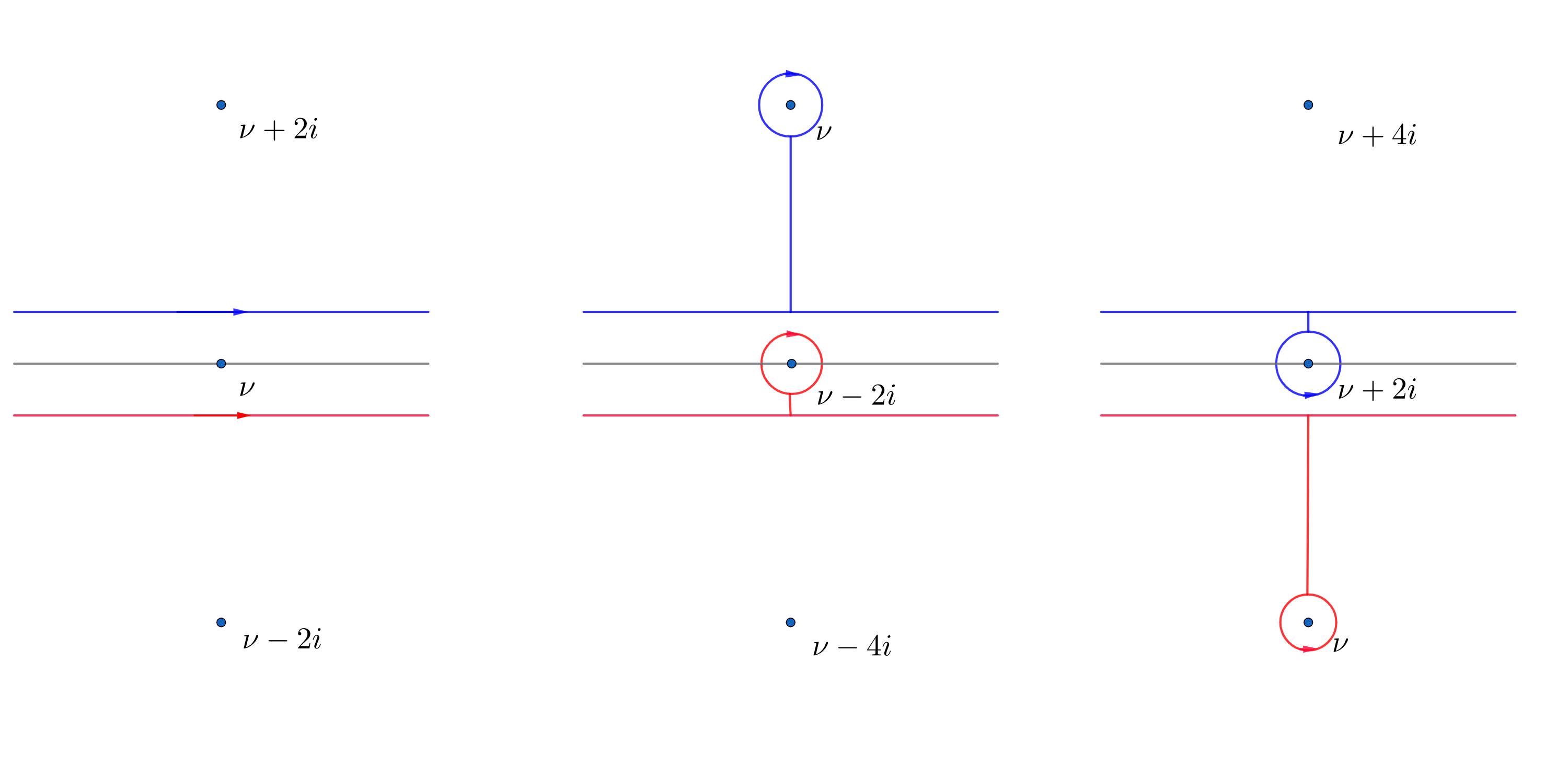}
    \caption{Analytic continuation $\nu \to \nu \pm 2i$ from real values brings additional terms, give by half- residues of the poles crossing the principal value integration
contour}  \label{fig:TQ-complex}
\end{figure}

It turns out that a difference equation for $Q(\nu)$ can be derived. Consider the following linear combination:
\begin{equation}\label{lin-comb}
    Q(\nu + 2i) + Q(\nu-2i) -2 Q(\nu).
\end{equation}
After analytic continuation to $\nu \pm 2i$ in \eqref{'tHooft-eq-Fourier-Q} (see Fig. \ref{fig:TQ-complex}) additional terms appear: half-residues at the poles of $\frac{1}{\sinh \frac{\pi (\nu - \nu')}{2}}$ (at $\nu' = \nu, \nu \pm 2i$) and $\frac{\nu-\nu'}{\sinh \frac{\pi (\nu - \nu')}{2}}$ (at $\nu' = \nu \pm 2i$) that cross the contour. In addition to that, the integral terms cancel each other when we substitute them into \eqref{lin-comb}. The result is the following equation\footnote{We note that the presence of singularities of $Q(\nu)$ in the strip $\textrm{Im}\;\nu\in[-2,2]$ does not affect this derivation.}
\begin{equation}\label{generalized-TQ}
    \left(1+\frac{\beta x}{\nu+2i+\alpha x}\right)Q(\nu+2i)+\left(1-\frac{\beta x}{\nu-2i+\alpha x}\right)Q(\nu-2i)-2Q(\nu)=-\frac{2z}{\nu+\alpha x}Q(\nu),
\end{equation}
where
\begin{equation}\label{x-z-notation}
    x=\frac{2}{\pi}\tanh\left(\frac{\pi\nu}{2}\right),\quad z=2\pi\lambda \tanh\left(\frac{\pi\nu}{2}\right).
\end{equation}

Difference equations of the type \eqref{generalized-TQ}, known as TQ equations, were pioneered by Baxter in his seminal paper \cite{Baxter:1972hz} on the solution of the $8$-vertex model. In exactly solvable lattice models, the Baxter TQ equation is commonly used as a tool to find the eigenvalues of the transfer matrices, and knowing the analytic properties of the $Q$ function is a key ingredient in this approach. In the simplest situation Q-function is a polynomial whose zeroes are to be adjusted in a way that the T-function is a polynomial as well.  In the current setup the situation is similar: one looks for a solution of \eqref{generalized-TQ}  with the desired analytic and asymptotic properties. 

Normalizable solutions of the homogeneous 't Hooft equation necessarily satisfy \eqref{generalized-TQ}. On the other hand,  assume that we have a solution of the TQ equation which admits the required analyticity properties described in \eqref{Q-is-bounded}-\eqref{essential-poles}, but the conditions $Q(0) = Q(\pm 2i) = 0$ are relaxed. Let us define the operator
\begin{equation}
    \hat{K}\cdot f(\nu)\overset{\text{def}}{=}\fint_{-\infty}^{\infty}\limits d\nu'\frac{\pi(\nu-\nu')}{2\sinh{\frac{\pi(\nu-\nu')}{2}}}\frac{1}{\sinh{\frac{\pi\nu'}{2}}}\cdot f(\nu')
\end{equation}
and apply it to the both hand sides of \eqref{generalized-TQ}. The first two terms on the l.h.s., after deforming the principal value integration contour, become
\begin{align}
    &\fint_{-\infty}^{\infty}\limits d\nu'\frac{\pi(\nu-\nu')}{2\sinh{\frac{\pi(\nu-\nu')}{2}}}\frac{Q(\nu'\pm 2i) (\nu' \pm 2i + (\alpha\pm\beta) x')}{(\nu' \pm 2i + \alpha x')\sinh{\frac{\pi\nu'}{2}}} =  \fint_{-\infty}^{\infty}\limits d\nu'\frac{\pi(\nu-\nu'\pm 2i)}{2\sinh{\frac{\pi(\nu-\nu')}{2}}}\frac{Q(\nu') (\nu' + (\alpha\pm \beta) x')}{\sinh{\frac{\pi\nu'}{2}} (\nu' + \alpha x') } \mp \nonumber \\
    & \mp \frac{2\pi i}{2} \left(\underset{\nu' = \mp 2i}{\text{Res}} +\underset{\nu' = 0}{\text{Res}} + \underset{\nu' = \nu\mp2i}{\text{Res}} \right) \frac{\pi(\nu-\nu')}{2\sinh{\frac{\pi(\nu-\nu')}{2}}}\frac{1}{\sinh{\frac{\pi\nu'}{2}}} \frac{Q(\nu'\pm2i) (\nu' \pm 2i + (\alpha\pm\beta) x')}{\nu' \pm 2i + \alpha x'}.
\end{align}
The pole at $\nu' \pm 2i + \alpha x' = 0$ is canceled by the zero of the  $Q$-function and vice versa,  potential poles of $Q(\nu)$ are canceled by zeroes of the expression $(\nu' \pm 2i + (\alpha \pm \beta) x')$. The residue terms above evaluate to
\begin{equation}
    \mp  \frac{i \pi}{\sinh \frac{\pi \nu}{2}} \left(\frac{(\nu \pm 2 i) Q(0) (1+\alpha \pm\beta)}{1+\alpha}+\nu  Q(\pm 2 i)\mp\frac{2 i Q(\nu ) (\nu +x (\alpha \pm \beta ))}{\nu +\alpha  x}\right).
\end{equation}
Collecting everything together, one finds that \eqref{generalized-TQ} becomes
\begin{align} \label{'tHooft-eq-Q-inhomog}
    & \frac{2 i \beta}{\pi} \fint^{\infty}_{-\infty}\limits d\nu'
    \frac{1}{2\sinh\frac{\pi(\nu-\nu')}{2}}\frac{Q(\nu')}{\sinh\left(\frac{\pi\nu'}{2}\right)
    \left(\frac{2\alpha}{\pi}+\nu'\coth{\frac{\pi\nu'}{2}}\right)} -  \frac{Q(\nu)}{\sinh \frac{\pi \nu}{2}} = \nonumber \\
    & = - \lambda \int^{\infty}_{-\infty}\limits d\nu' \frac{\pi(\nu-\nu')}{2\sinh{\frac{\pi(\nu-\nu')}{2}}}\frac{Q(\nu')}{\sinh\left(\frac{\pi\nu'}{2}\right)
    \left(\frac{2\alpha}{\pi}+\nu'\coth{\frac{\pi\nu'}{2}}\right)} - \frac{1}{\sinh \frac{\pi \nu}{2}} \left(Q(0) +  \nu \frac{Q(2i) - Q(-2i) + \frac{2\beta}{1+\alpha} Q(0)}{4i} \right),
\end{align}
which is an inhomogeneous version of \eqref{'tHooft-eq-Fourier-Q}. We see that it becomes homogeneous precisely when $Q$ satisfies the ``quantization conditions'' $Q(0) = Q(\pm 2i) = 0$. 

Our strategy will now be as follows. For the TQ equation \eqref{generalized-TQ} (and, consequently, for the inhomogeneous equation \eqref{'tHooft-eq-Q-inhomog} with the fixed inhomogeneus part  $F(\nu)$) normalizable solutions with proper analyticity exist for any values of $\lambda$. A convenient basis of such solutions $Q_{\pm}(\nu|\alpha,\beta)$ corresponds to the choice 
\begin{equation}
F_{+}(\nu)= \frac{\nu}{\sinh \frac{\pi \nu}{2}},\quad F_{-}(\nu) = \frac{1}{\sinh \frac{\pi \nu}{2}}.
\end{equation}
They can be shown to have the following symmetry properties
\begin{equation}\label{q-symmetry-prop}
    Q_{\pm} (\nu|\alpha, \beta) = \mp Q_{\pm}(-\nu|\alpha,-\beta)
\end{equation}
and satisfy the normalization conditions
\begin{equation}\label{normalization-conditions}
    \begin{aligned}
        &Q_{+}(0|\alpha,\beta) = 0,\quad Q_{+}(2i|\alpha,\beta) - Q_{+}(-2i|\alpha,\beta) = 4i, \\
        &Q_{-}(0|\alpha,\beta) = 1,\quad Q_{-}(2i|\alpha,\beta) - Q_{-}(-2i|\alpha,\beta) = - \frac{2\beta}{1+\alpha}.
    \end{aligned}
\end{equation}
These solutions carry information about the spectrum $\lambda_n$, since they can be expanded in a complete basis of solutions of \eqref{'tHooft-eq-Fourier-Q} as follows:
\begin{equation}\label{Qpm-Kallen–Lehmann}
    Q_{\pm}(\nu,\lambda)=\sum\limits_{n=0}^\infty\frac{c^{\pm}_nQ_n(\nu)}{\lambda-\lambda_n},
\end{equation}
i.e. they exhibit poles in $\lambda$ precisely at the points of the spectrum. We note that compared to $\beta=0$ case, one can not conclude that $Q_{+}(\nu,\lambda)$ has poles only at $\lambda=\lambda_{2n}$ and $Q_-(\nu,\lambda)$ at $\lambda_{2n-1}$. Nevertheless, we were able to find certain linear combinations $\textbf{Q}_{\pm}(\nu,\lambda)$ (see Section \ref{Q-D-relations}) such that the corresponding coefficients $c_n$ are non-zero only for odd or even $n$. In the one-flavour case, when $\beta = 0$, the coefficients $c_n$ for suitable basis functions $Q_{\pm}(\nu,\lambda)$ are proportional to the formfactors of the vector current $J_\mu=\bar{\psi}\gamma_\mu\psi$ and the scalar density $S=\bar{\psi}\psi$, and values of the $Q$-functions themselves at special points are simply related to the polarization operator (see \cite{Fateev:2009jf} and \cite{Callan:1976PhysRev,Einhorn:1976uz}). Similar relations probably exist in the most general two-flavour case. 

In the next two sections, we will describe a systematic procedure of constructing an asymptotic expansion of $Q_\pm(\nu, \lambda)$ in $\lambda$ in two different regimes: $\lambda \to 0$ and $\lambda \to - \infty$.  Then we will describe how these expansions can be used to extract non-perturbative information about the spectrum $\lambda_n$, such as spectral sums and systematic large-$n$ ``WKB'' expansion.
%%%%%%%%%%%%%%%%%%%%%%%%%%%%%%%%%%%%%%%%%%%%%%%%%%%%%%%%%%%%%%%%%%%%%%%%%%%%%%%%%%%%%%%%%%%%%%%%%%%%%%%%%%%%%%%%%%%%%%%%%%%%%%%%%%%%%%%%%%%%%%%%%%%%%%%%%%%%%%%%%%%%%%%%%%%%%%%%%%%%%%%%%%
%%%%%%%%%%%%%%%%%%%%%%%%%%%%%%%%%%%%%%%%%%%%%%%%%%%%%%%%%%%%%%%%%%%%%%%%%%%%%%%%%%%%%%%%%%%%%
\section{Solution of TQ equation in the small $\lambda$ regime}\label{small-lambda-expansion}
We follow the method suggested in \cite{Fateev:2009jf} and further developed in \cite{Litvinov:2024riz}. Namely, we look for a solution of the TQ equation \eqref{generalized-TQ} as a formal power series
\begin{equation}
    Q_{\pm}(\nu,\lambda)=\sum_{k=0}^{\infty}Q_{\pm}^{(k)}(\nu)\lambda^k,
\end{equation}
such that each coefficient in this power series is a meromorphic function in the strip $\textrm{Im}(\nu)\in [-2, 2]$ which satisfies the conditions \eqref{Q-is-bounded}-\eqref{essential-poles}. Moreover, we expect that once the additional conditions \eqref{q-symmetry-prop} and \eqref{normalization-conditions} are specified, the solution of \eqref{generalized-TQ}, obeying the required analytic properties, is actually unique. On the other hand, such a solution should be completely equivalent to the iterative solution of the inhomogeneous integral equation \eqref{'tHooft-eq-Q-inhomog}. However, direct evaluation of the integrals that appear in this solution becomes rather cumbersome. The method described below greatly facilitates calculations. In particular, most of the integrals in an iterative solution appear to be taken from the beginning.

We start by noting that there are two formal solutions of the TQ equation \eqref{generalized-TQ}
\begin{equation}\label{Xi-Sigma}
\begin{aligned}
    &\Xi(\nu|\alpha,\beta)=\left(\nu+\alpha x\right)F\left(\genfrac{}{}{0pt}{1}{1+\frac{i(\nu+(\alpha-\beta)x)}{2}}{2-i\beta x}\biggl|-iz\right),\\
    &\Sigma(\nu|\alpha,\beta)=\frac{\Gamma\left(\frac{i(\nu+(\alpha+\beta)x)}{2}\right)}
    {\Gamma\left(1+\frac{i(\nu+(\alpha-\beta)x)}{2}\right)}\left(\nu+\alpha x\right)F\left(\genfrac{}{}{0pt}{1}{\frac{i(\nu+(\alpha+\beta)x)}{2}}{i\beta x}\biggl|-iz\right),
\end{aligned}
\end{equation}
where the parameters $x$ and $z$ are given by \eqref{x-z-notation}, $\Gamma(y)$ is the gamma function and  $F\left(\genfrac{}{}{0pt}{1}{a}{b}\biggl|y\right)$ is the confluent hypergeometric function (sometimes referred to as $_1F_1(a;b;y)$)
\begin{equation}\label{F-hypergeometric}
    F\left(\genfrac{}{}{0pt}{1}{a}{b}\biggl|y\right)=\sum_{k=0}^{\infty}\frac{(a)_k}{(b)_k}\frac{y^k}{k!}.
\end{equation}
Both of the functions \eqref{Xi-Sigma} do not provide solutions to the TQ equation with required properties. We will gradually modify these formal series by constructing from them the correct solutions of the Baxter's TQ equation \eqref{generalized-TQ}.  

First, we note that the $\Gamma$ factor in the function $\Sigma(\nu|\alpha,\beta)$ exhibits an unpleasant behavior at the points $\nu=\pm i$ where $x(\nu)$ \eqref{x-z-notation} has poles. In order to restore the analyticity we redefine $\Sigma(\nu|\alpha,\beta)$ by replacing
\begin{equation}\label{Gamma-replacement}
    \frac{\Gamma\left(\frac{i(\nu+(\alpha+\beta)x)}{2}\right)}{\Gamma\left(1+\frac{i(\nu+(\alpha-\beta)x)}{2}\right)}\quad\longrightarrow\quad 
    2\pi^2x\frac{ S_0(\nu|\alpha+\beta)}{\nu+(\alpha+\beta)x}\frac{S_0(-\nu|\alpha-\beta)}{\nu+(\alpha-\beta)x},
\end{equation}
where the function $S_0(\nu|\alpha)$, that has been introduced in \cite{Litvinov:2024riz}, is defined by the integral
\begin{equation}\label{S0-integral-representation}
    S_0(\nu+i|\alpha)=
    \exp\left[\frac{i}{4}\int_{-\infty}^{\infty}\limits\log
    \left(\frac{4\pi\tanh\left(\frac{\pi t}{2}\right)}{t+\frac{2\alpha}{\pi}\tanh\left(\frac{\pi t}{2}\right)}\right)
    \left(\tanh\frac{\pi(t-\nu)}{2}-
    \tanh\frac{\pi t}{2}\right)dt\right]
\end{equation}
in the strip $\textrm{Im }\nu\in[0,2]$ and by analytic continuation elsewhere. This function satisfies the shift relation
\begin{equation}\label{S0-shift-eq}
    S_0(\nu+2i|\alpha)=\frac{4\pi \tanh\left(\frac{\pi\nu}{2}\right)}{\nu+\frac{2\alpha}{\pi} \tanh\left(\frac{\pi\nu}{2}\right)}S_0(\nu|\alpha),
\end{equation}
such that the r.h.s. of \eqref{Gamma-replacement} satisfies the same shift relation as the ratio of $\Gamma$-functions in the l.h.s. It can be shown that $S_0(\nu)$ is a unique solution of \eqref{S0-shift-eq} analytic in the strip $\textrm{Im }\nu\in[0,2]$ (in fact, as follows from \eqref{S0-shift-eq}, it is analytic in the larger domain $\textrm{Im }\nu\in[-2,2]$ except for one point $\nu=-2i$).
Also, we note that using the relation
\begin{equation}\label{S0S0-relation}
    S_0(\nu|\alpha)S_0(-\nu|\alpha)=\frac{\nu+\alpha x}{2\pi^2 x},
\end{equation}
that follows immediately from \eqref{S0-integral-representation}, one can rewrite the r.h.s. of \eqref{Gamma-replacement} as
\begin{equation}
    2\pi^2 x\frac{ S_0(\nu|\alpha+\beta)}{\nu+(\alpha+\beta)x}\frac{S_0(-\nu|\alpha-\beta)}{\nu+(\alpha-\beta)x}=
    \frac{G(\nu|\alpha,\beta)}{\nu+(\alpha+\beta)x},
\end{equation}
where 
\begin{equation}\label{G-def}
    G(\nu|\alpha,\beta)\overset{\text{def}}{=}\frac{S_0(\nu|\alpha+\beta)}{S_0(\nu|\alpha-\beta)}.
\end{equation}
It is convenient also to have in mind the relation that follows from \eqref{S0S0-relation}
\begin{equation}\label{G-relation}
    G(\nu|\alpha,\beta)G(-\nu|\alpha,\beta)=\frac{\nu+(\alpha+\beta)x}{\nu+(\alpha-\beta)x}.
\end{equation}

Of course, the replacement \eqref{Gamma-replacement} is not sufficient to construct correct solutions, since the functions $\Xi(\nu|\alpha,\beta)$ and $\Sigma(\nu|\alpha,\beta)$ do not possess the required symmetry properties and also contain multiple poles due to the hypergeometric functions. However, given a solution of TQ equation, even one lacking required analyticity or symmetry properties, one can multiply it by an arbitrary $2i$ periodic function of $\nu$ (quasiconstant) and obtain a new solution. For example, we notice that multiplication by the exponential $e^{\frac{i z}{2}}$ provides new solutions\footnote{These solutions tend in the limit $\beta\rightarrow0$ to $M_{\pm}(\nu|\alpha)$ defined in \cite{Litvinov:2024riz} for $\beta=0$. More precisely, using the expansion
\begin{equation}
    G(\nu|\alpha,\beta)=1+\beta
    \left(\frac{ix}{2}\left(\uppsi_{\alpha}(\nu)+\uppsi_{\alpha}(-\nu)+2\gamma_E+2\log 4\right)+\frac{x}{\nu+\alpha x}\right)+\mathcal{O}(\beta^2),
\end{equation}
where $\gamma_E =0.57721\dots$ is Euler-Mascheroni constant and the function $\uppsi_{\alpha}\big(\nu\big)$ is defined by the integral representation (see \cite{Litvinov:2024riz})
\begin{equation}
    \uppsi_{\alpha}\big(\nu+i\big)=-\gamma_E-\log 4+\frac{1}{2}\int_{-\infty-i\epsilon}^{\infty-i\epsilon}\limits\frac{1}{t+\frac{2\alpha}{\pi}\tanh\frac{\pi t}{2}}
    \left(\tanh\frac{\pi t}{2}-\tanh\frac{\pi(t-\nu)}{2}\right)dt,
\end{equation}
one finds 
\begin{equation}\label{modified-scaling}
    \begin{aligned}
        &M_+(\nu|\alpha,\beta)=M_+(\nu|\alpha)+\mathcal{O}(\beta),\\
        &M_-(\nu|\alpha,\beta)+\frac{i\pi^2\lambda}{2\beta}
        \underbrace{\frac{1}{1-i\beta x}}_{1+i\beta x+\dots}\,
    M_+(\nu|\alpha,\beta)=M_-(\nu|\alpha)+\mathcal{O}(\beta).
    \end{aligned}
\end{equation}}
\begin{equation}
\begin{aligned}
    &M_+(\nu|\alpha,\beta)=e^{\frac{iz}{2}}\Xi(\nu|\alpha,\beta),\\
    &M_-(\nu|\alpha,\beta)=e^{\frac{iz}{2}}\Sigma(\nu|\alpha,\beta),
\end{aligned}
\end{equation}
which  can be shown to satisfy the symmetry properties
\begin{equation}
    M_{\pm}(-\nu|\alpha,-\beta)=\mp M_{\pm}(\nu|\alpha,\beta).
\end{equation}

None of $M_{\pm}(\nu|\alpha,\beta)$ provides the desired solution. First, there are poles of growing order at $\nu=\pm i$ that appear in the $\lambda$ expansion. Moreover, there are new singularities that come from the zeros of the Pochhammer symbols in \eqref{F-hypergeometric}. In order to cure both types of these singularities, we look for solutions of \eqref{generalized-TQ} in a more general form
\begin{equation}\label{Q-ansatz}
    Q_{\pm}(\nu|\alpha,\beta)=A_{\pm}(x|\lambda)M_{\pm}(\nu|\alpha,\beta)+
    B_{\pm}(x|\lambda)M_{\mp}(\nu|\alpha,\beta),
\end{equation}
where $A_{\pm}(x|\lambda)$ and $B_{\pm}(x|\lambda)$ admit the expansion in the parameter $\lambda$. These solutions have to satisfy the generalized symmetry relations
\begin{equation}\label{Qpm-symmetry}
    Q_{\pm}(-\nu|\alpha,-\beta)=\mp Q_{\pm}(\nu|\alpha,\beta),
\end{equation}
as well as the normalization conditions (compare to \eqref{normalization-conditions})
\begin{equation}
\begin{aligned}
    & Q_{+}(2i|\alpha,\beta)=
    %-Q_{+}(-2i|\alpha,\textcolor{red}{-}\beta)
    2i,\qquad &&Q_{+}(0|\alpha,\beta)=0,\\
    &Q_{-}(0|\alpha,\beta)=1,\qquad &&Q_{-}(2i|\alpha,\beta)-Q_{-}(-2i|\alpha,\beta)=-\frac{2\beta}{1+\alpha}.
\end{aligned}
\end{equation}
The role of the functions $A_{\pm}(x|\lambda)$ and $B_{\pm}(x|\lambda)$ is to cancel poles at $\nu=\pm i$ and the Pochhammer poles  
\begin{equation}\label{Q-additional-poles}
    i\beta x_k=\frac{2i\beta}{\pi}\tanh\left(\frac{\pi\nu_k}{2}\right)=k,\quad k\in\mathbb{Z},\quad k\neq 1.
\end{equation}
We note that in the strip $\textrm{Im}\,\nu\in[-2,2]$ for $\beta>0$ there are two solutions to \eqref{Q-additional-poles}
\begin{equation}
    \begin{aligned}
        &\nu_k=-\frac{2i}{\pi}\arctan\frac{\pi k}{2\beta}\quad&\text{and}\quad
        \nu_k=-\frac{2i}{\pi}\arctan\frac{\pi k}{2\beta}+2i\quad&\text{for}\quad k>0,\\
        &\nu_k=-\frac{2i}{\pi}\arctan\frac{\pi k}{2\beta}\quad&\text{and}\quad
        \nu_k=-\frac{2i}{\pi}\arctan\frac{\pi k}{2\beta}-2i\quad&\text{for}\quad k<0.
    \end{aligned}
\end{equation}
In the following we will use the notations
\begin{equation}\label{xi_k-def}
    \arctan\frac{\pi k}{2\beta}=\xi_k\quad k>0,
\end{equation}
and
\begin{equation}\label{g_k-def}
    g_k\overset{\text{def}}{=}G(\nu_k).
\end{equation}
We note that $g_k$ and $g_{-k}$ are not independent, but obey the following relation (see \eqref{G-relation})
\begin{equation}\label{gk*g-k}
    g_kg_{-k}=\frac{2\beta\xi_k+\pi k(\alpha+\beta)}{2\beta\xi_k+\pi k(\alpha-\beta)}\quad\text{for}\quad k>0.
\end{equation}

Taking a closer look at \eqref{Q-ansatz} it becomes clear that the coefficients of $A_{\pm}(x|\lambda)$ and $B_{\pm}(x|\lambda)$ should possess poles at the points \eqref{Q-additional-poles}.  In principle, other singularities are not forbidden. The only requirement is that they cancel in the total $Q$-function \eqref{Q-ansatz}. By trial and error, we have found that one has to include one additional pole at
\begin{equation}
    \frac{2i\beta}{\pi}\tanh\left(\frac{\pi\nu_1}{2}\right)=1,
\end{equation}
in order to cancel all singularities. Thus we  propose the following ansatz for the functions $A_{\pm}(x|\lambda)$, $B_{\pm}(x|\lambda)$:
\begin{equation}
    \begin{aligned}
        &A_+(x|\lambda)=\frac{1}{1-i\beta x}+\frac{a_+^{(1)}x}{1-i\beta x}\lambda+\frac{a_+^{(2)}x+a_+^{(3)}x^2+a_+^{(4)}x^3}{(1-i\beta x)(1+i\beta x)}\lambda^2
        +\dots\\
        &B_+(x|\lambda)=\frac{b_+^{(1)}x}{1-i\beta x}+\frac{b_+^{(2)}x+b_+^{(3)}x^2+b_+^{(4)}x^3}{(1-i\beta x)(2-i\beta x)}\lambda+\frac{b_+^{(5)}x+b_+^{(6)}x^2+b_+^{(7)}x^3+b_+^{(8)}x^4+b_+^{(9)}x^5}{(1-i\beta x)(2-i\beta x)(3-i\beta x)}\lambda^2
        +\dots
    \end{aligned}    
\end{equation}
and 
\begin{equation}
    \begin{aligned}
        &A_-(x|\lambda)=a_-^{(0)}+\frac{a_-^{(1)}x+a_-^{(2)}x^2}{1-i\beta x}\lambda+\frac{a_-^{(3)}x+a_-^{(4)}x^2+a_-^{(5)}x^3+a_-^{(6)}x^4}{(1-i\beta x)(2-i\beta x)}\lambda^2
        +\dots\\
        &B_-(x|\lambda)=\frac{b_-^{(1)}}{1-i\beta x}\lambda+\frac{b_-^{(2)}x+b_-^{(3)}x^2}{(1-i\beta x)(1+i\beta x)}\lambda^2+\frac{b_-^{(4)}x+b_-^{(5)}x^2+b_-^{(6)}x^3+b_-^{(7)}x^4}{(1-i\beta x)(1+i\beta x)(2+i\beta x)}\lambda^2
        +\dots
    \end{aligned}    
\end{equation}
The coefficients $a_{\pm}^{(s)}$ and $b_{\pm}^{(s)}$ are to be adjusted in a way to cancel all the unwanted poles ($\nu=\pm i$ and Pochhammer poles).   It can be shown that the corresponding linear problem in each order in $\lambda$ has a unique solution. Explicit expressions for these coefficients are rather cumbersome. For example
\begin{equation}\label{ap-pm}
\begin{gathered}
    a_+^{(1)}=-\frac{i\pi^2\alpha}{2\beta},\quad
    b_+^{(1)}=-\frac{\pi(\alpha+\beta)+2\beta\xi_1}{\pi g_1},\quad
    a_+^{(2)}=\frac{i\pi^2\left(2\beta\xi_1+\pi(\alpha+\beta)\right)^2}{8\beta^3g_1^2}-\frac{i\pi^4\left(\alpha ^2-\beta ^2\right)}{8\beta ^3},\\
    b_+^{(2)}=\frac{2\pi\alpha\left(2\beta\xi_1+\pi(\alpha+\beta)\right)}{\beta^2g_1}-\frac{2\left(\beta\xi_2+\pi\alpha\right)\left(\beta\xi_2+\pi(\alpha+\beta)\right)}{\beta ^2g_2}
\end{gathered}
\end{equation}
and
\begin{equation}\label{am-pm}
\begin{gathered}
    a_-^{(0)}=\frac{\gamma}{1+\alpha},\quad a_-^{(1)}=\frac{i\pi^2\alpha\gamma}{2\beta(1+\alpha)}-\frac{i\pi\left(2 \beta  \xi _1+\pi(\alpha+\beta)\right)}{2\beta g_1},\quad
    b_-^{(1)}=\frac{i\pi^2}{2\beta},\\
    a_-^{(2)}=\frac{\pi^2\alpha\gamma}{2(1+\alpha)},\quad b_-^{(2)}=\frac{\pi^4\alpha}{4\beta^2}-\frac{\pi^3\gamma\left(
    2 \beta\xi_1+\pi(\alpha+\beta)\right)}{4\beta^2g_1(1+\alpha)},
\end{gathered}
\end{equation}
where we have introduced
\begin{equation}\label{gamma-def}
    \gamma\overset{\text{def}}{=}\sqrt{(1+\alpha+\beta)(1+\alpha-\beta)}.
\end{equation}

Using the method described above, one can construct the solution of the TQ equation \eqref{generalized-TQ} to any desired order in $\lambda$. The search for the coefficients $a_{\pm}^{(s)}$ and $b_{\pm}^{(s)}$ is purely algebraic. The coefficients $a_{\pm}^{(s)}$ and $b_{\pm}^{(s)}$ appear to be rational functions of $\xi_k$, $g_k$ and $\alpha,\beta,\gamma$ with simple denominators similar to \eqref{ap-pm} and \eqref{am-pm}.  We have explicitly calculated these coefficients to the order $s=5$, which is enough to compute the spectral sums $G_{\pm}^{(s)}$ to the order $s\leq4$ (see below).
%%%%%%%%%%%%%%%%%%%%%%%%%%%%%%%%%%%%%%%%%%%%%%%%%%%%%%%%%%%%%%%%%%%%%%%%%%%%%%%%%%%%%%%%%%%%%%%%%%%%%%%%%%%%%%%%%%%%%%%%%%%%%%%%%%%%%%%%%%%%%%%%%%%%%%%%%%%%%%%%%%%%%%%%%%%%%%%%%%%%%%%%%%
\section{Solution of TQ equation in the large $\lambda$ regime}\label{large-lambda-expansion}
In this section we construct the solutions of TQ equation in a different limit $\lambda \to - \infty$, now looking for $Q$-function as a series in $\lambda^{-1}$. During this construction, we pay special attention to the requirement that at each order of the expansion $Q$-functions satisfy the required analyticity properties \eqref{Q-is-bounded}-\eqref{essential-poles} and have the correct $\beta \to 0$ limit, previously found in \cite{Litvinov:2024riz}. As in the case of equal masses of quark and antiquark \cite{Fateev:2009jf,Litvinov:2024riz}, the expansion will be found in several steps.
%%%%%%%%%%%%%%%%%%%%%%%%%%%%%%%%%%%%%%%%%%%%%%%%%%%%%%%%%%%%%%%%%%%%%%%%%%%%%%%%%%%%%%%%%%%%%
\subsection{First step: initial ansatz and its properties}
To derive the large $\lambda$ expansion of the functions $Q_{\pm}(\nu|\alpha,\beta)$, we start with the following ansatz 
\begin{equation}\label{S-ansatz}
    S(\nu|\alpha,\beta)=(-\lambda)^{-\frac{i\nu}{2}}S_1(\nu|\alpha,\beta)\sum_{k=0}^{\infty}\frac{\left(1+\frac{i(\nu+(\alpha-\beta) x)}{2}\right)_k\left(\frac{i(\nu+(\alpha+\beta) x)}{2}\right)_k}{k!}(iz)^{-k}, 
\end{equation}
which satisfies the TQ equation \eqref{generalized-TQ}, given that  $S_1(\nu)$ obeys the functional relation
\begin{equation}\label{S1-shift-eq}
    S_1(\nu+2i|\alpha,\beta)=\frac{\nu+2i+\alpha x}{\nu+2i+(\alpha+\beta)x}\frac{2\pi^2x}{\nu+\alpha x}S_1(\nu|\alpha,\beta). 
\end{equation}

The analyticity requirements for the functions $Q_{\pm}(\nu|\alpha,\beta)$ cannot be satisfied within the ansatz \eqref{S-ansatz} due to the presence of poles of increasing order originating from $(iz)^{-k}$. Moreover, for large positive $\lambda$, the series \eqref{S-ansatz} exhibits exponential growth in the limit $\nu\to\pm\infty$. However, the series \eqref{S-ansatz} serves as a fundamental component in the construction of the asymptotic expansion of $Q_{\pm}(\nu|\alpha,\beta)$ in the case of large negative $\lambda$. 

We are interested in the solution of \eqref{S1-shift-eq} satisfying \eqref{Q-is-bounded}-\eqref{essential-poles} (see there remark about the status of poles) in the strip $\textrm{Im}\,\nu\in[0,2]$. To find such a solution, note that \eqref{S1-shift-eq} can be rewritten as
\begin{equation}
    \frac{\nu+2i+(\alpha+\beta)x}{\nu+2i+\alpha x}S_1(\nu+2i|\alpha,\beta)=\frac{2\pi^2x}{\nu+(\alpha+\beta)x}\cdot \frac{\nu+(\alpha+\beta)x}{\nu+\alpha x}S_1(\nu|\alpha,\beta),
\end{equation}
which is the equation \eqref{S0-shift-eq} solved by $S_0(\nu)$. The function $\frac{\nu+(\alpha+\beta)x}{\nu+\alpha x}S_1(\nu|\alpha,\beta)$ is holomorphic in the strip $\textrm{Im}\,\nu\in[0,2]$ (the factor $\frac{\nu+(\alpha+\beta)x}{\nu+\alpha x}$ cancels the pole of $S_1(\nu|\alpha+\beta)$ at $\nu =i\nu^*_1(\alpha+\beta)$). Therefore, the solution with the properties \eqref{Q-is-bounded}-\eqref{essential-poles} is unique up to a normalization (see appendix B in \cite{Litvinov:2024riz})
\begin{equation}\label{S1-shift-sol}
    S_1(\nu|\alpha,\beta)=\frac{\alpha+\beta}{\alpha}\frac{\nu+\alpha x}{\nu+(\alpha+\beta)x}S_0(\nu|\alpha+\beta),\quad S_1(i|\alpha,\beta)=1,
\end{equation}
where the normalization is chosen in such a way that the Wronskian condition \eqref{Wronsk(0)} is fulfilled. Using \eqref{S1-shift-sol} and \eqref{S0-shift-eq} it can be continued to the bottom strip $\textrm{Im}\,\nu\in(-2,0]$ 
\begin{equation}\label{S1-continuation}
    S_1(\nu-2i|\alpha,\beta)=\frac{\alpha+\beta}{\alpha}\frac{\nu-2i+\alpha x}{2\pi^2x}S_0(\nu|\alpha+\beta).
\end{equation}
The function $S_0(\nu|\alpha+\beta)$ is analytic in the entire strip $\textrm{Im}\,\nu\in(-2,2]$ and contains an \textit{extra} simple pole at $\nu=-2i$. We conclude from \eqref{S1-shift-sol}-\eqref{S1-continuation} that the function $S_1(\nu|\alpha,\beta)$  satisfies the required properties in the strip $\textrm{Im}\,\nu\in(-2,2]$
\begin{itemize}
    \item has zeroes at $\pm i\nu^*_1(\alpha)$;
    \item has  pole at $i\nu^*_1(\alpha+\beta)$ and no extra pole at $-i\nu^*_1(\alpha+\beta)$. 
\end{itemize}
Similarly, we can verify that the function $S_1(-\nu|\alpha,-\beta)$  in the strip $\textrm{Im}\,\nu\in(-2,2]$ 
\begin{itemize}
    \item has zeroes at $\pm i\nu^*_1(\alpha)$;
    \item has pole at $-i\nu^*_1(\alpha-\beta)$ and no extra pole at $i\nu^*_1(\alpha-\beta)$. 
\end{itemize}

In the following analysis, the expansion of the function $S_1(\nu|\alpha,\beta)$ at $\nu=0,\pm i,\pm2i$ will be required. However, due to the relation \eqref{S1-shift-eq}, it suffices to consider the expansion at $\nu=0,i$ only
\begin{equation}\label{S1-expansion}
    \log S_1(\nu|\alpha,\beta)=\sum_{k=0}^{\infty}s_k(\alpha,\beta)\nu^k,\quad \log S_1(\nu|\alpha,\beta)=\sum_{k=0}^{\infty}t_k(\alpha,\beta)(\nu-i)^k.
\end{equation}
From \eqref{S1-shift-sol} we see that the expansion of $S_1(\nu|\alpha,\beta)$ is directly related to the expansion of $S_0(\nu|\alpha+\beta)$
\begin{equation}\label{S0-expansion}
    \log S_0(\nu|\alpha+\beta)=\sum_{k=0}^{\infty}s_k(\alpha+\beta)\nu^k,\quad \log S_0(\nu|\alpha+\beta)=\sum_{k=0}^{\infty}t_k(\alpha+\beta)(\nu-i)^k.
\end{equation}
The method of calculating the expansion of the latter function has been described in \cite{Litvinov:2024riz}, here we give the final results for the first few coefficients. The coefficients $s_{2k}(\alpha,\beta)$ appear to be elementary functions of $\alpha$ and $\beta$
\begin{equation}\label{s-coef-even}
    \begin{aligned}
        &s_0(\alpha,\beta)=s_0(\alpha+\beta)+\log{\left(\frac{\alpha+\beta}{\alpha}\frac{1+\alpha}{1+\alpha+\beta}\right)},\quad &&s_0(\alpha+\beta)=-\frac{1}{2}\log\left(\frac{2\pi^2}{1+\alpha+\beta}\right),
        \\
        &s_2(\alpha,\beta)=s_2(\alpha+\beta)+\frac{\pi^2\beta}{12(1+\alpha)(1+\alpha+\beta)},\quad &&s_2(\alpha+\beta)=\frac{\pi^2}{24(1+\alpha+\beta)},
        \\
        &s_4(\alpha,\beta)=s_4(\alpha+\beta)-\frac{\pi^4\beta(12+7\beta+2\alpha(7+\alpha+\beta))}{1440(1+\alpha)^2(1+\alpha+\beta)^2},\quad &&s_4(\alpha+\beta)=-\frac{\pi^4}{2880}\frac{7+2(\alpha+\beta)}{(1+\alpha+\beta)^2}.
    \end{aligned}
\end{equation}
On the other hand, odd coefficients $s_{2k+1}(\alpha,\beta)$ are more complicated:
\begin{equation}
    s_{2k+1}(\alpha,\beta)=s_{2k+1}(\alpha+\beta),
\end{equation}
where
\begin{equation}\label{s-coef-odd}
\begin{aligned}
  &s_1(\alpha+\beta)=-\frac{i}{2}(1+\log 2\pi+\gamma_E)+\frac{i(\alpha+\beta)}{8}\mathtt{i}_1(\alpha+\beta),\\
  &s_3(\alpha+\beta)=\frac{i}{72}\left(2\pi^2+3\zeta(3)\right)+\frac{i\pi^2(\alpha+\beta)}{96}
  \mathtt{i}_3(\alpha+\beta)
  ,
  \\
  &s_5(\alpha+\beta)=-\frac{i}{7200}\left(14\pi^4+45\zeta(5)\right)+\frac{i\pi^4(\alpha+\beta)}{1920}
  \left(3\mathtt{i}_5(\alpha+\beta)+\mathtt{i}_3(\alpha+\beta)\right)
  ,
  %\\
  %&s_7(\alpha+\beta)=\frac{i}{846720}\left(124\pi^6+945\zeta(7)\right)+\frac{i\pi^6(\alpha+\beta)}{161280}\left(45\mathtt{i}_7(\alpha+\beta)+30\mathtt{i}_5(\alpha+\beta)+2\mathtt{i}_3(\alpha+\beta)\right)
\end{aligned}
\end{equation}
and
\begin{equation}\label{i2k1-def}
    \mathtt{i}_{2k-1}(\alpha)\overset{\text{def}}{=}\fint_{-\infty}^{\infty}\limits\frac{\sinh2t-2t}{t\sinh^{2k-1} t \big(\alpha\sinh t+t\cosh t\big)}dt.
\end{equation}

The coefficients $t_{2k}(\alpha,\beta)$ in \eqref{S1-expansion} arise exclusively from the factor $\frac{\nu+\alpha x}{\nu+(\alpha+\beta)x}$ since $t_{2k}(\alpha+\beta)=0$ and they are elementary functions of $\alpha,\beta$ 
\begin{equation}
\begin{aligned}
    t_0(\alpha,\beta)=&\;0,\\
    t_2(\alpha,\beta)=&\;\frac{\pi^2\beta}{32\alpha^2}\frac{\pi^2(2\alpha+\beta)+8\alpha(\alpha+\beta)}{(\alpha+\beta)^2},\\
    t_4(\alpha,\beta)=&\;\frac{\pi^8}{1024}\left(\frac{1}{(\alpha+\beta)^4}-\frac{1}{\alpha^4}\right)-\frac{\pi^4\beta}{96\alpha^2}\frac{2\alpha(3+\alpha+\beta)+3\beta}{(\alpha+\beta)^2}-\\&-\frac{\pi^6\beta}{192\alpha(\alpha+\beta)^3}\left(9+2\alpha+\frac{\beta(3+\alpha)(3\alpha+\beta)}{\alpha^2}\right).
\end{aligned}
\end{equation}
The odd coefficients $t_{2k+1}(\alpha,\beta)$ are expressed in terms of the same transcendents $\mathtt{i}_{k}(\alpha)$ as $s_{2k+1}(\alpha,\beta)$ 
\begin{equation}
\begin{aligned}
    t_1(\alpha,\beta)=&\;t_1(\alpha+\beta)+\frac{i\pi^2}{4}\frac{\beta}{\alpha(\alpha+\beta)},\\
    t_3(\alpha,\beta)=&\;t_3(\alpha+\beta)-\frac{i\pi^4}{48}\frac{\beta(\alpha(6+\alpha+\beta)+3\beta)}{\alpha^2(\alpha+\beta)^2}-\frac{i\pi^6}{192}\left(\frac{1}{(\alpha+\beta)^3}-\frac{1}{\alpha^3}\right),\\
    t_5(\alpha,\beta)=&\;t_5(\alpha+\beta)+\frac{i\pi^6}{960}\left(\frac{15}{\alpha^3}+\frac{2(5+\alpha)}{\alpha^2}-\frac{2(5+\alpha+\beta)}{(\alpha+\beta)^2}-\frac{15}{(\alpha+\beta)^3}\right)+
    \\&+\frac{i\pi^8}{768}\left(\frac{3}{\alpha^4}+\frac{1}{\alpha^3}-\frac{3+\alpha+\beta}{(\alpha+\beta)^4}\right)+\frac{i\pi^{10}}{5120}\left(\frac{1}{\alpha^5}-\frac{1}{(\alpha+\beta)^5}\right),
\end{aligned}    
\end{equation}
where
\begin{equation}
    \begin{aligned}
        &t_1(\alpha+\beta)=-\frac{i}{2}\left(\gamma_E-1+\log 8\pi\right)+\frac{i(\alpha+\beta)}{8}
        \mathtt{i}_2(\alpha+\beta),\\
        &t_3(\alpha+\beta)=-\frac{i}{72}\left(2\pi^2-21\zeta(3)\right)-\frac{i\pi^2(\alpha+\beta)}{96}
        \mathtt{i}_4(\alpha+\beta),\\
        &t_5(\alpha+\beta)=\frac{i}{7200}\left(14\pi^4-1395\zeta(5)\right)+\frac{i\pi^4(\alpha+\beta)}{1920}
        \left(3\mathtt{i}_6(\alpha+\beta)-\mathtt{i}_4(\alpha+\beta)\right),
    \end{aligned}
\end{equation}
etc and
\begin{equation}\label{i2k-def}
    \mathtt{i}_{2k}(\alpha)\overset{\text{def}}{=}\int_{-\infty}^{\infty}\limits\frac{\sinh t(\sinh2t-2t)}{t\cosh^{2k} t \big(\alpha\sinh t+t\cosh t\big)}dt.
\end{equation}
%%%%%%%%%%%%%%%%%%%%%%%%%%%%%%%%%%%%%%%%%%%%%%%%%%%%%%%%%%%%%%%%%%%%%%%%%%%%%%%%%%%%%%%%%%%%%
\subsection{Second step: symmetry and removing extra singularities}
Our next step is to construct basis solutions $Q^{(\text{b})}_{\pm}(\nu|\alpha,\beta)$ which have proper symmetry under the substitution $\beta\to-\beta$ \eqref{Qpm-symmetry} and have no extra poles at the points $\nu=0,\pm i,\pm2i$.

Following \cite{Fateev:2009jf,Litvinov:2024riz} we represent these basis solutions of TQ equation \eqref{generalized-TQ} in the form
\begin{equation}\label{Q-large-ansatz}
    Q^{(\text{b})}_{\pm}(\nu|\alpha,\beta)=T(c^{-1}|\lambda)R_{\pm}(c|\lambda)S(\nu|\alpha,\beta)\mp T(-c^{-1}|\lambda)R_{\pm}(-c|\lambda)S(-\nu|\alpha,-\beta),
\end{equation}
where $S(\nu|\alpha,\beta)$ is given in \eqref{S-ansatz} and we use the notation
\begin{equation}\label{c-def}
    c=i\pi\coth{\frac{\pi\nu}{2}}.
\end{equation}

Since $c=c(\nu)$ is $2i$-periodic function, the ansatz \eqref{Q-large-ansatz} provides a solution to \eqref{generalized-TQ} for any $T(c^{-1}|\lambda)$ and $R_{\pm}(c|\lambda)$. The purpose of these auxiliary functions is to eliminate poles of increasing order at $\nu=\pm i$ and $\nu=0,\pm 2i$  respectively at each order in $\lambda^{-1}$. We look for the functions $T(c^{-1}|\lambda)$ and $R_{\pm}(c|\lambda)$ in the form of an asymptotic expansion at large $\lambda$ 
\begin{equation}
    T(c^{-1}|\lambda)=1+\sum_{k=1}^{\infty}T^{(k)}(c^{-1})\lambda^{-k}\quad
    R_{\pm}(c|\lambda)=1+\sum_{k=1}^{\infty}
    R_{\pm}^{(k)}\big(c|\log(-\lambda)\big)\lambda^{-k},
\end{equation}
where $T^{(k)}(c^{-1})$ and $R_{\pm}^{(k)}\big(c|\log(-\lambda)\big)$ are polynomials in their respective variables and they can be determined independently from each other. The appearance of $\log{(-\lambda)}$ as an argument of the polynomials $R^{(k)}_{\pm}\big(c|\log(-\lambda)\big)$ is expected. It arises from the necessity of expanding the function $S(\nu)$ about its poles; logarithmic terms emerge due to the factor $(-\lambda)^{-\frac{i\nu}{2}}$. 

Explicit expressions for first several coefficients $T^{(k)}(y)$ (even under $\beta\to-\beta$) have the form
\begin{equation}\label{Ts-coefs}
\begin{aligned}
    &T^{(1)}(y)=\frac{\alpha^2-\beta^2}{2\pi^2}y,
    \\
    &T^{(2)}(y)=\frac{\alpha(\alpha^2-\beta^2)}{4\pi^4}y+\frac{(\alpha^2-\beta^2)^2}{8\pi^4}y^2, 
    \\
    &T^{(3)}(y)=\frac{5\alpha^4-6\alpha^2\beta^2+\beta^4}{24\pi^6}y+\frac{\alpha(\alpha^2-\beta^2)^2}{8\pi^6}y^2+\frac{(\alpha^2-\beta^2)^3}{48\pi^6}y^3,
    \\
    &T^{(4)}(y)=\frac{\alpha(7\alpha^4-10\alpha^2\beta^2+3\beta^4)}{32\pi^8}y+\frac{(13\alpha^2-2\beta^2)(\alpha^2-\beta^2)^2}{96\pi^8}y^2+\frac{\alpha(\alpha^2-\beta^2)^3}{32\pi^8}y^3+\frac{(\alpha^2-\beta^2)^4}{384\pi^8}y^4.
\end{aligned}
\end{equation}
By studying this expansion further, we have found the closed form expression for the function $T(y|\lambda)$ (here $\alpha_k$ are the original parameters related to $\alpha$ and $\beta$ by \eqref{alphas-alpha-beta})
\begin{equation}\label{T-function-explicit}
    T(y|\lambda)=\exp\left[\alpha_1\;f\left(\frac{\alpha_2}{2\pi^2\lambda},\frac{\alpha_1}{\alpha_2}\right)y\right]=\exp\left[\alpha_2\;f\left(\frac{\alpha_1}{2\pi^2\lambda},\frac{\alpha_2}{\alpha_1}\right)y\right],
\end{equation}
where
\begin{equation}\label{f-funtion-explicit}
    \begin{aligned}
        f(t,r)=\int\frac{dt'}{t'}\mathcal{N}(t',r)=&\;\frac{1-2\log(-2)+r-2r\log(-2r)}{2r}+
        \frac{\sqrt{1-2(1+r)t+(1-r)^2t^2}-1}{2rt}-
        \\&-\frac{(1+r)\log t}{r}+
        \frac{\log\left(\sqrt{1-2(1+r)t+(1-r)^2t^2}-1-(1-r)t\right)}{r}+
        \\&+\log\left((1-r)t-1+\sqrt{1-2(1+r)t+(1-r)^2t^2}\right).
    \end{aligned}
\end{equation}
Here $\mathcal{N}(t,r)$ is the generating function for the Narayana numbers $N(n,k)=\frac{1}{n}\binom{n}{k}\binom{n}{k-1}$
\begin{equation}
    \mathcal{N}(t,r)\overset{\text{def}}{=}\frac{1-(1+r)t-\sqrt{1-2(1+r)t+(1-r)^2t^2}}{2rt}=\sum_{n=1}^{\infty}\sum_{k=1}^{n}N(n,k)\;t^nr^k
\end{equation}
and the constant term in the r.h.s. is chosen such that $f(0,r)=0$. Let us stress that in the limit $\beta\to0$ \eqref{T-function-explicit} reproduces our previous result in \cite{Litvinov:2024riz}.

Unfortunately, we were unable to determine  similar explicit form for the functions $R_{\pm}(c|\lambda)$. However, the requirement that poles at the points $\nu=0,\pm2i$ cancel at each order in $\lambda^{-1}$ uniquely determines $R_{\pm}^{(k)}\big(c|\log(-\lambda)\big)$ once the normalization of $Q^{(b)}_{\pm}(\nu|\alpha,\beta)$ is fixed. A convenient choice of normalization is given by $R_{\pm}(0|\lambda)=1$. Then
\begin{equation}
\begin{aligned}\label{R-function-explicit}
    R_\pm(c|\lambda)&=1\pm\frac{c\cdot\gamma}{4\pi^4}\frac{\alpha-\beta}{\alpha+\beta}\lambda^{-2}\pm\frac{c\cdot\gamma}{24\pi^6}\frac{\alpha-\beta}{\alpha+\beta}\Biggl(6c+6q_1\mp\frac{\alpha+\beta}{\alpha-\beta}\gamma\Biggl)\lambda^{-3}+\dots,\\
    q_1&=\;3(1+\alpha)-2\log(-\lambda)-2i\big(s_1(\alpha+\beta)+s_1(\alpha-\beta)\big),
\end{aligned}
\end{equation}
where $\gamma,c$ are defined in \eqref{gamma-def},\eqref{c-def}.
We have determined the coefficients $R_{\pm}^{(k)}\big(c|\log(-\lambda)\big)$ for $k \leq 5$. Again, in the limit $\beta \to 0$, \eqref{R-function-explicit} reproduces our previous result from \cite{Litvinov:2024riz}.
%%%%%%%%%%%%%%%%%%%%%%%%%%%%%%%%%%%%%%%%%%%%%%%%%%%%%%%%%%%%%%%%%%%%%%%%%%%%%%%%%%%%%%%%%%%%%
\subsection{Last step: normalization}
So far, we have constructed solutions $Q^{\text{(b)}}_{\pm}(\nu|\alpha,\beta)$ of the TQ equation possessing correct symmetry \eqref{Qpm-symmetry} and the required analytic properties \eqref{Q-is-bounded}-\eqref{essential-poles}. However, these solutions do not satisfy the required normalization conditions \eqref{normalization-conditions}
\begin{equation}
\begin{aligned}
    &Q^{(\text{b})}_+(2i|\alpha,\beta)\ne2i,\quad &&Q^{(\text{b})}_+(0|\alpha,\beta)\ne0,
    \\
    &Q^{(\text{b})}_-(0|\alpha,\beta)\ne1,\quad &&Q^{(\text{b})}_-(2i|\alpha,\beta)-Q^{(\text{b})}_-(-2i|\alpha,\beta)\ne-\frac{2\beta}{1+\alpha}Q^{(\text{b})}_-(0|\alpha,\beta).
\end{aligned}
\end{equation}
The realization of these conditions is the last step in construction of the correct solutions $Q_{\pm}(\nu|\alpha,\beta)$ of \eqref{generalized-TQ}. The idea is to consider the following linear combinations 
\begin{equation}\label{Q-large-final}
    Q_{\pm}(\nu|\alpha,\beta)=A_\pm(\lambda|\alpha,\beta) Q^{\text{(b)}}_{\pm}(\nu|\alpha,\beta)+B_{\pm}(\lambda|\alpha,\beta)Q^{\text{(b)}}_{\mp}(\nu|\alpha,\beta),
\end{equation}
where the functions $A_{\pm}(\lambda|\alpha,\beta)$, $B_{\pm}(\lambda|\alpha,\beta)$ admit the expansion
\begin{equation}
    A_{\pm}(\lambda|\alpha,\beta)=\sum_{k=0}^{\infty}a^{(k)}_{\pm}(\log(-\lambda)|\alpha,\beta)\lambda^{-k},\quad B_{\pm}(\lambda|\alpha,\beta)=\sum_{k=0}^{\infty}b^{(k)}_{\pm}(\log(-\lambda)|\alpha,\beta)\lambda^{-k}
\end{equation}
and $a^{(k)}_{\pm}(\log(-\lambda)|\alpha,\beta),\;b^{(k)}_{\pm}(\log(-\lambda)|\alpha,\beta)$ are polynomials in $\log(-\lambda)$. The ansatz \eqref{Q-large-final} has the correct analytic properties and the symmetry under $\beta \to - \beta$ is preserved if the functions $A_{\pm}(\lambda|\alpha,\beta)$, $B_{\pm}(\lambda|\alpha,\beta)$ posses the symmetry
\begin{equation}
    A_{\pm}(\lambda|\alpha,-\beta)=A_{\pm}(\lambda|\alpha,\beta),\quad B_{\pm}(\lambda|\alpha,-\beta)=-B_{\pm}(\lambda|\alpha,\beta).
\end{equation}
Solving the linear system \eqref{normalization-conditions} at each order of $\lambda^{-1}$ expansion, we find the functions $A_{\pm}(\lambda|\alpha,\beta)$, $B_{\pm}(\lambda|\alpha,\beta)$ in terms of the coefficients $s_{k}(\alpha\pm\beta)$ \eqref{s-coef-even},\eqref{s-coef-odd} of the expansion of $S_0(\nu)$ \eqref{S0-expansion}. Here we give the first few terms of the expansion after substitution of $s_{k}(\alpha\pm\beta)$
\begin{equation}
\begin{aligned}
    &a^{(0)}_{+}(\log(-\lambda)|\alpha,\beta)=1,\quad b^{(0)}_{+}(\log(-\lambda)|\alpha,\beta)=-\frac{\beta}{1+\alpha+\gamma},
    \\
    &a^{(1)}_{+}(\log(-\lambda)|\alpha,\beta)=\frac{2+2\alpha-\gamma}{2\pi^2}-\frac{i(1+\alpha)}{\sqrt{2}\pi\gamma}\left(\sqrt{1+\alpha+\beta}+\sqrt{1+\alpha-\beta}\right),
    \\
    &b^{(1)}_{+}(\log(-\lambda)|\alpha,\beta)=\frac{1}{2\pi^2}\frac{(1+\alpha-\beta)^{3/2}-(1+\alpha+\beta)^{3/2}}{\sqrt{1+\alpha+\beta}+\sqrt{1+\alpha-\beta}}+\frac{i(1+\alpha)}{\sqrt{2}\pi\gamma}\left(\sqrt{1+\alpha+\beta}-\sqrt{1+\alpha-\beta}\right),
\end{aligned}
\end{equation}
\begin{equation}
\begin{aligned}
    &a^{(0)}_{-}(\log(-\lambda)|\alpha,\beta)=\frac{\pi}{2 \sqrt{2}\gamma}\left(\sqrt{1+\alpha+\beta}+\sqrt{1+\alpha-\beta}\right),
    \\
    &b^{(0)}_{-}(\log(-\lambda)|\alpha,\beta)=-\frac{\pi}{2\sqrt{2}\gamma}\left(\sqrt{1+\alpha+\beta}-\sqrt{1+\alpha -\beta}\right),
    \\
    &a^{(1)}_{-}(\log(-\lambda)|\alpha,\beta)=-\frac{(1+\alpha+\beta)^{3/2}+(1+\alpha-\beta)^{3/2}}{4\sqrt{2}\pi\gamma},
    \\
    &b^{(1)}_{-}(\log(-\lambda)|\alpha,\beta)=\frac{(1+\alpha+\beta)^{3/2}-(1+\alpha-\beta)^{3/2}}{4\sqrt{2}\pi\gamma}.
\end{aligned}    
\end{equation}
We have calculated the coefficients for $k\leq 5$. 

Since $B_\pm(\lambda|\alpha,\beta)$ vanishes for $\beta = 0$, the constructed functions $Q_{\pm}(\nu)$ have the same limit $\beta \to 0$ as $Q^{(\text{b})}_\pm(\nu)$.  Indeed $Q^{(\text{b})}_\pm(\nu)$ reduce to solutions constructed in \cite{Litvinov:2024riz} for the case of equal quark masses up to incorrect normalization at the points $0,2i$ respectively, which can be easily corrected by simple division
\begin{equation}
    Q^{(\text{b})}_+(\nu)\to Q_+(\lambda|\alpha,\beta=0)=2i\frac{Q^{(\text{b})}_+(\nu)}{Q^{(\text{b})}_+(2i)},\quad Q^{(\text{b})}_-(\nu)\to Q_-(\lambda|\alpha,\beta=0)=\frac{Q^{(\text{b})}_-(\nu)}{Q^{(\text{b})}_-(0)}.
\end{equation}
%%%%%%%%%%%%%%%%%%%%%%%%%%%%%%%%%%%%%%%%%%%%%%%%%%%%%%%%%%%%%%%%%%%%%%%%%%%%%%%%%%%%%%%%%%%%%%%%%%%%%%%%%
\section{Relation between Q-functions and spectral determinants}\label{Q-D-relations}
In this section we will discuss how the expansions obtained in Sections \ref{small-lambda-expansion} and \ref{large-lambda-expansion} can be used to study the meson mass spectrum of the  't Hooft model. Let us define the ``spectral determinants'' (Fredholm determinants) as functions whose zeroes are located at even/odd points of the spectrum of the 't Hooft equation \eqref{'tHooft-eq} (or equivalently of the spectrum of the homogeneous Fredholm integral equation \eqref{'tHooft-eq-Fourier}):
\begin{equation}\label{spectral-determinants-definition-1}
    D_+(\lambda)\overset{\text{def}}{=}\left(\frac{8\pi}{e}\right)^{\lambda}\prod_{n=0}^\infty\left(1-\frac{\lambda}{\lambda_{2n}}\right)e^{\frac{\lambda}{n+1}},\quad D_-(\lambda)\overset{\text{def}}{=}\left(\frac{8\pi}{e}\right)^{\lambda}\prod_{n=0}^\infty\left(1-\frac{\lambda}{\lambda_{2n+1}}\right)e^{\frac{\lambda}{n+1}}.
\end{equation}
This infinite product is convergent at least in the vicinity of $\lambda = 0$; in particular, it can be written as
\begin{equation}\label{spectral-determinants-definition-1-2}
    D_{\pm}(\lambda)=\left(\frac{8\pi}{e}\right)^{\lambda}
    \exp\left[-\sum_{s=1}^{\infty}s^{-1}G_{\pm}^{(s)}\lambda^s\right].
\end{equation}
where 
\begin{equation}
    G_{+}^{(s)} = \sum \limits_{n=0}^\infty \left(\frac{1}{\lambda_{2n}^s} - \frac{\delta_{s,1}}{n+1} \right),\quad G_{-}^{(s)} = \sum \limits_{n=0}^\infty \left(\frac{1}{\lambda_{2n+1}^s} - \frac{\delta_{s,1}}{n+1} \right)
\end{equation}
are the ‘‘spectral sums'' (spectral zeta functions)\footnote{The zeta function of the operator $\mathcal{O}$ is defined as $\zeta_{\mathcal{O}}(s) = \text{tr} \;\mathcal{O}^{-s}$ for values of $s$ where this expression is well-defined, and extended to other values through analytic continuation. In this work, we focus only on integer values of $s$. The spectral zeta function can also be expressed in terms of the eigenvalues $\lambda_i$ as $\zeta_{\mathcal{O}}(s) = \sum_{i} \lambda_i^{-s}.$ Note that the spectral determinants \eqref{spectral-determinants-definition-1} are different objects than what is usually called a spectral determinant of $\mathcal{O}$: $\det \mathcal{O} = e^{-\zeta'_{\mathcal{O}}(0)}$.}. Subtraction for $s=1$ is necessary for convergence due to the asymptotic linear behavior of the spectrum: $\lambda_n \sim n/2,\,n \to \infty$.

We will now give and explain two non-perturbative (or, at least, valid in both expansion regimes) relations between the spectral determinants and the constructed functions $Q_{\pm}(\nu|\alpha,\beta)$.
\subsection{Quantum Wronskian relation}
The ‘‘quantum Wronskian'' is an important concept for any second-order difference equations of the form \eqref{generalized-TQ}. It is built from two linearly independent solutions ($Q_+$ and $Q_-$ in our case) as follows
\begin{equation}
    W(\nu)\overset{\text{def}}{=}Q_+(\nu+i)Q_-(\nu-i)-Q_-(\nu+i)Q_+(\nu-i).
\end{equation}
It is a direct consequence of the TQ equation \eqref{generalized-TQ} that $W(\nu)$ satisfies a first order difference equation
\begin{equation}\label{Wronskian-functional-relation}
    W(\nu+i)=\frac{1-\frac{\beta x}{\nu-2i+\alpha x}}{1+\frac{\beta x}{\nu+2i+\alpha x}}W(\nu-i)=\frac{\nu-2i+(\alpha-\beta)x}{\nu-2i+\alpha x}\frac{\nu+2i+\alpha x}{\nu+2i+(\alpha+\beta)x}W(\nu-i).
\end{equation}
A solution of \eqref{Wronskian-functional-relation}, satisfying the necessary analyticity properties (following from the conditions on the functions $Q_\pm(\nu)$ \eqref{Q-is-bounded}, \eqref{essential-poles}), can be constructed in terms of the function $G(\nu)$ given by \eqref{G-def} (for details see Appendix B in \cite{Litvinov:2024riz})
\begin{equation}\label{Wronskian-explicit}
    W(\nu+i)=2i\frac{\gamma}{1+\alpha}\,
    \frac{\nu+\alpha x}{\nu+(\alpha+\beta)x}\frac{\nu+2i+\alpha x}{\nu+2i+(\alpha+\beta)x}G(\nu).
\end{equation}
Normalization has been chosen such that $W(i)=Q_+(2i)Q_-(0)=2i$. One can explicitly check \eqref{Wronskian-explicit} using the constructed expansions for $Q_{\pm}$ in $\lambda$ or $1/\lambda$. 
 
In particular we see that the  quantum Wronskian does not depend on $\lambda$, although $Q_{\pm}(\nu)$ are expected to be singular when $\lambda \to \lambda_n$. Consider then the value of $W(\nu)$ at $\nu = 0$
\begin{equation}\label{Wronsk(0)}
    W(0)=Q_+(i)Q_-(-i)-Q_+(-i)Q_-(i)=2i\frac{\gamma}{1+\alpha}\frac{\alpha^2}{\alpha^2-\beta^2}.
\end{equation}
First, recall the case $\beta = 0$ studied in \cite{Litvinov:2024riz}. Due to the symmetry property \eqref{q-symmetry-prop}, in this case two terms in the l.h.s. are the same and the equation simplifies to 
\begin{equation}
  Q_+(i,\beta=0) Q_-(i,\beta=0) = i.    
\end{equation}
It follows that at this point both $Q_{+}(i,\beta=0)$ and $Q_-(i,\beta=0)$ cannot be singular as $\lambda \to \lambda_n$. Moreover, if one of them has a pole at some point of the spectrum, the other should have a zero. In this case, it is easy to show that $Q_+(i,\beta=0)$ has poles at $\lambda = \lambda_{2n}$ and $Q_-(i,\beta=0)$  at $\lambda = \lambda_{2n+1}$ ($n=0,1,\dots$). There are no other zeroes or poles for $Q_{\pm}(i,\beta=0)$ in the complex plane of $\lambda$, so up to a function with only possible singularity at $\infty$ they are equal to the ratio of spectral determinants \eqref{spectral-determinants-definition-1}. In fact, in this case one has \cite{Litvinov:2024riz}
\begin{equation}\label{Q/Q=D/D}
    \frac{Q_+(i,\beta=0)}{Q_+(2i,\beta=0)}=\frac{1}{2}\frac{D_-(\lambda)}{D_+(\lambda)}\quad\text{and}\quad
    \frac{Q_-(i,\beta=0)}{Q_-(0,\beta=0)}=\frac{D_+(\lambda)}{D_-(\lambda)}.
\end{equation}

For $\beta \neq 0$ the symmetry relation \eqref{q-symmetry-prop} between $Q_{\pm}(i)$ and $Q_{\pm}(-i)$ is more complicated and $W(0)$ does not immediately factorize. However, there still exist two surprisingly simple linear relations between these four functions
\begin{equation}\label{Q-linear-reltion-1}
    \frac{(\alpha+\beta)Q_-(i)-(\alpha-\beta)Q_-(-i)}{(\alpha+\beta)Q_+(i)-(\alpha-\beta)Q_+(-i)}=\frac{i\pi^2\lambda}{1+\alpha}\frac{1+\alpha-\gamma}{2\beta}=\frac{i\beta\pi^2\lambda}{2(1+\alpha)(1+\alpha+\gamma)}
\end{equation}
and
\begin{equation}\label{Q-linear-reltion-2}
    \left(1-\frac{\pi^2\lambda}{\beta}\right)Q_+(i)-2i\,\frac{\alpha-\beta}{\alpha+\beta}\frac{1+\alpha}{\gamma}Q_-(-i)-\frac{\alpha-\beta}{\alpha+\beta}\left(1+\frac{1+\alpha}{\gamma}\frac{\pi^2 \lambda}{\beta}\right)Q_+(-i)=0.
\end{equation}
Using them to express everything in terms of $Q_\pm(i)$, one finds that $W(0)$ acquires a factorized form
\begin{equation}\label{W-factorized-form}
    W(0)=\frac{\pi^2\lambda}{\beta +\pi^2\lambda}\frac{\alpha+\beta}{\alpha-\beta}\frac{1+\alpha+\gamma}{\gamma}\left(Q_-(i)-\frac{i\beta\pi^2\lambda}{2(1+\alpha)(1+\alpha+\gamma)}Q_+(i)\right)\left(Q_+(i)+\frac{2i\beta}{\pi^2\lambda}\frac{1+\alpha}{1+\alpha+\gamma}Q_-(i)\right).
\end{equation}
We see that the linear combinations that appear as factors in \eqref{W-factorized-form}
\begin{equation}\label{Right-Qpm-combination}
\begin{aligned}
    &\mathbf{Q}_{-}(\nu) = Q_-(\nu)-\frac{i\beta\pi^2\lambda}{2(1+\alpha)(1+\alpha+\gamma)}Q_+(\nu); \\
    &\mathbf{Q}_{+}(\nu) = Q_+(\nu)+\frac{2i\beta}{\pi^2\lambda}\frac{1+\alpha}{1+\alpha+\gamma}Q_-(\nu)
\end{aligned}
\end{equation}
reduce to previous expressions $Q_{\pm}(\nu|\beta=0)$ in the limit $\beta \to 0$. In terms of them, the linear relations \eqref{Q-linear-reltion-1}-\eqref{Q-linear-reltion-2} can also be written as
\begin{equation}
\begin{aligned}
  &\frac{\mathbf{Q}_{-}(i)}{ \mathbf{Q}_{-}(-i)} = \frac{\alpha - \beta}{\alpha+\beta}, \\
  &\frac{\mathbf{Q}_{+}(i)}{ \mathbf{Q}_{+}(-i)}  = \frac{\beta + \pi^2 \lambda}{\beta - \pi^2 \lambda} \frac{\alpha - \beta}{\alpha + \beta}.
\end{aligned}
\end{equation}
Now, similar to $\beta=0$ case, we note that if one of the functions $\mathbf{Q}_{\pm}(i)$ develops a pole at some point of the spectrum $\lambda = \lambda_n$, the other should have a zero. We expect that, as in $\beta =0$ case, $\mathbf{Q}_-(i)$ is singular for $\lambda = \lambda_{2n+1}$ and $\mathbf{Q}_+(i)$ for $\lambda = \lambda_{2n}$. Then, we conjecture the following formulas:
\begin{equation}\label{wronsk-rel}
    \begin{aligned}
        &\mathbf{Q}_-(i)=\frac{\alpha}{\alpha+\beta}\frac{\gamma}{1+\alpha} \frac{D_+(\lambda)}{D_-(\lambda)},\\
        &\mathbf{Q}_+(i)=2i\frac{\beta +\pi^2\lambda}{\pi^2\lambda}\frac{\alpha}{\alpha+\beta}\frac{\gamma}{1+\alpha+\gamma}\frac{D_-(\lambda)}{D_+(\lambda)}.
    \end{aligned}
\end{equation}
%%%%%%%%%%%%%%%%%%%%%%%%%%%%%%%%%%%%%%%%%%%%%%%%%%%%%%%%%%%%%%%%%%%%%%%%%%%%%%%%%%%%%%%%%%%
\subsection{Relation with log-derivatives of $\mathbf{Q}$}
Formulas \eqref{wronsk-rel} are not enough to study $D_+(\lambda)$ and $D_-(\lambda)$ individually and they have to be supplemented with other relations. In the special cases considered in \cite{Fateev:2009jf, Litvinov:2024riz}, identities of a different type were found. They incorporate log-derivative of the solution of the TQ equation at the point $\nu=i$. In the case of general $\alpha$ and $\beta=0$, they look as follows \cite{Litvinov:2024riz}:
\begin{align}
    \label{2.17-0-alpha}
    \partial_{\lambda}\log D_{-}(\lambda)-\frac{\alpha}{4}\mathtt{i}_2(\alpha)&=2i\,\partial_{\nu}\log Q_{+}(\nu)\Bigl|_{\nu=i},
    \\\label{2.17-0-alpha-2}
    \partial_{\lambda}\log D_{+}(\lambda)-\frac{\alpha}{4}\mathtt{i}_2(\alpha)&=2i\left(1-\frac{2\alpha}{\pi^2}\lambda^{-1}\right)\partial_{\nu}\log Q_{-}(\nu)\Bigl|_{\nu=i},
\end{align}
where $\mathtt{i}_2(\alpha)$ is defined by \eqref{i2k-def}. These identities are consistent with the ones following from the quantum Wronskian in a sense that poles in $\lambda$ for the r.h.s. come only from the zeroes of $Q_{\pm}(i)$. No rigorous proof of such identities exists even in the special cases considered before; however, they are very useful and pass nontrivial numerical checks.

In the case $\beta \neq 0$, we have managed to find generalizations of both of these identities.  One way to write the generalization of \eqref{2.17-0-alpha} is as follows:
\begin{multline} \label{D-nonsymmetric}
    \partial_{\lambda}\log D_{-}(\lambda)-\frac{1}{\lambda}\left(\frac{\beta^2}{2\alpha(\alpha+\beta)}-\frac{2i\beta G'(i)}{\pi^2}\right)-\frac{\pi^2\beta}{2\alpha(\alpha+\beta)}-\frac{1}{4}(\alpha+\beta)\mathtt{i}_2(\alpha+\beta)= 
    \\=2i\left(1+\frac{\beta}{\pi^2\lambda}\right)\partial_\nu\log\mathbf{Q}_+(\nu)\Bigl|_{\nu=i}.
\end{multline}
Explicitly one has $G'(i) = \frac{i}{8}((\alpha+\beta)\mathtt{i}_2(\alpha+\beta)-(\alpha-\beta)\mathtt{i}_2(\alpha-\beta))$. In this form, it is nontrivial to see the symmetry properties under $\beta \to - \beta$. There is another way to write it where the symmetry becomes manifest:
\begin{multline} \label{D-symmetric}
    \partial_\lambda\log{D_-(\lambda)}-\frac{1}{\lambda}\left(\frac{\beta ^2}{2 (\alpha^2-\beta^2)}-\frac{2i \beta G'(i)}{\pi^2} \right)+\frac{\pi^2\beta^2}{2\alpha\left(\alpha^2-\beta^2\right)}-\frac{1}{8}\Big((\alpha+\beta)\mathtt{i}_2(\alpha+\beta)+(\alpha-\beta)\mathtt{i}_2(\alpha-\beta)\Big)= \\=2i\,\partial_\nu\log\left((\alpha+\beta)\mathbf{Q}_+(\nu)-(\alpha-\beta)\mathbf{Q}_+(-\nu)\right)\Biggl|_{\nu=i}.
\end{multline}
The equivalence of these two formulas follows from the special linear identity on $\mathbf{Q}_+(\pm i)$ and their derivatives
\begin{equation}
    2i \partial_\nu \log \left((\alpha + \beta) \mathbf{Q}_+(\nu) + (\alpha - \beta) \mathbf{Q}_+(-\nu) \right) \Bigl|_{\nu = i} = \frac{\pi ^2 \beta ^2}{2 \alpha\left(\alpha ^2-\beta ^2\right)}+\frac{\pi ^2 \lambda}{2}   \left(-\frac{\pi ^2 }{\alpha ^2-\beta ^2}+\frac{2 i G'(i)}{\beta}\right).
\end{equation}
The generalization of \eqref{2.17-0-alpha-2} is a bit more complicated
\begin{multline} \label{D+nonsymmetric}
    \partial_\lambda\log{D_+(\lambda)}-\frac{1}{\lambda}\left(\frac{\beta^2}{2\alpha}\frac{3\alpha+\beta}{\alpha^2-\beta^2}-\frac{2i\beta G'(i)}{\pi^2}\right)-\frac{\pi^2\beta}{2\alpha(\alpha+\beta)}-\frac{1}{4} (\alpha+\beta)\mathtt{i}_2(\alpha+\beta)=
    \\
    =2i\frac{(1-\frac{\alpha}{\pi^2\lambda})\mathbf{Q}'_-(i)+\frac{\alpha-\beta}{\pi^2\lambda}\mathbf{Q}'_-(-i)}{\mathbf{Q}_-(i)}=2i\left(1-\frac{\alpha}{\pi^2\lambda}\right)\partial_\nu\log{\mathbf{Q}_-(\nu)}\Bigl|_{\nu=i}-2i\frac{\alpha+\beta}{\pi^2\lambda}\partial_\nu\log{\mathbf{Q}_-(-\nu)}\Bigl|_{\nu=i}.
\end{multline}
It can be written in a symmetric form 
\begin{multline} \label{D+symmetric}
    \partial_\lambda\log{D_+(\lambda)}-\frac{1}{\lambda}\left(\frac{\beta^2}{2\alpha^2}\frac{3\alpha^2-\beta^2}{\alpha^2-\beta^2}-\frac{2i\beta G'(i)}{\pi^2}\right)+\frac{\pi^2\beta^2}{\alpha(\alpha^2-\beta^2)}-\frac{(\alpha^2-\beta^2)\left(\mathtt{i}_2(\alpha+\beta)+\mathtt{i}_2(\alpha-\beta)\right)}{8\alpha}= 
    \\
    =2i\left(1-\frac{2\alpha^2-\beta^2}{\pi^2\alpha\lambda}\right)\partial_\nu\log\left((\pi^2\lambda-(2\alpha+\beta))\mathbf{Q}_-(\nu)+(\pi^2\lambda-(2\alpha-\beta))\mathbf{Q}_-(-\nu)\right)\Bigl|_{\nu=i}.
\end{multline}
Equivalence of \eqref{D+nonsymmetric} and \eqref{D+symmetric}, again, is proven using the identity
\begin{equation}
    2i\frac{(\alpha + \beta) \left(1-\frac{\pi^2 \lambda}{\beta} \right) \mathbf{Q}'_-(i) - (\alpha - \beta) \left(1+\frac{\pi^2 \lambda}{\beta} \right) \mathbf{Q}'_-(-i)  }{\mathbf{Q}_-(i) } = \frac{\pi^2 \beta^2}{\alpha (\alpha - \beta)} + \frac{\lambda(\pi^4 \beta-2i\pi^2 G'(i) (\alpha^2 - \beta^2))}{\beta (\alpha - \beta)}.
\end{equation}

We note that the status of the rational \eqref{wronsk-rel} and logarithmic \eqref{D-nonsymmetric}, \eqref{D+nonsymmetric} identities remains the same: they are of conjectural nature, but meet all the numerical and analytical checks with great accuracy (see Sections \ref{limiting-cases},\ref{numerics}). These relations are crucial for $\beta\neq 0$ case, considered in this paper.  For $\beta=0$, there is an alternative way to extract the spectral data from the solutions of the TQ equation described in \cite{Fateev:2009jf}. Namely, an operator with the kernel
\begin{equation}
    K(\nu,\nu')=\frac{1}{\sqrt{f(\nu)f(\nu')}}\frac{\pi(\nu-\nu')}{2\sinh\frac{\pi(\nu-\nu')}{2}}
\end{equation}
belongs to the class of completely integrable operators \cite{Its:1980,Its:1990MPhysB}. It is well known \cite{Its:1980,Its:1990MPhysB,zbMATH01284258} that the kernel of the resolvent of such operators, that is the kernel of $\frac{\hat{K}}{1-\lambda\hat{K}}$,  admits the form 
\begin{equation}
    R(\nu,\nu'|\lambda)=\frac{\sinh\frac{\pi\nu}{2}\sinh\frac{\pi\nu'}{2}}
    {\sinh\frac{\pi(\nu-\nu')}{2}}\sqrt{f(\nu)f(\nu')}\left(
    \Psi_+(\nu'|\lambda)\Psi_-(\nu|\lambda)-\Psi_-(\nu'|\lambda)\Psi_+(\nu|\lambda)\right),
\end{equation}
where $\Psi_{\pm}(\nu|\lambda)$ satisfy an inhomogeneous equation
\begin{equation}
    f(\nu)\Psi_{\pm}(\nu|\lambda)-\lambda\int_{-\infty}^{\infty}\limits
    \frac{\pi(\nu-\nu')}{2\sinh\frac{\pi(\nu-\nu')}{2}}\Psi_{\pm}(\nu'|\lambda)d\nu'=
    e_{\pm}(\nu),
\end{equation}
with $e_{\pm}(\nu)$ being\footnote{Formally, one can take any $e_{\pm}(\nu)$ such that there exists $M(\nu)$ obeying the property
\begin{equation}
    \frac{\pi(e_+(\nu)e_-(\nu')-e_+(\nu')e_-(\nu))}{2(M(\nu)-M(\nu'))}=\frac{\pi(\nu-\nu')}{2\sinh\frac{\pi(\nu-\nu')}{2}}.
\end{equation}
The choice \eqref{epm-choice} corresponds to $M(\nu)=-\coth\frac{\pi\nu}{2}$.}
\begin{equation}\label{epm-choice}
    e_{+}(\nu)=\frac{\nu}{\sinh\frac{\pi\nu}{2}}\quad\text{and}\quad
    e_{-}(\nu)=\frac{1}{\sinh\frac{\pi\nu}{2}}.
\end{equation}
Since this is precisely the r.h.s. of \eqref{'tHooft-eq-Q-inhomog}, the resolvent $R(\nu,\nu'|\lambda)$ is expressed in terms of the functions $Q_{\pm}(\nu)$. In turn, the ratio $D_+(\lambda)/D_-(\lambda)$ and the product $D_+(\lambda)D_-(\lambda)$ of the spectral determinants can be expressed using the resolvent, providing integral relations between $D_{\pm}(\lambda)$ and $Q_{\pm}$. Unlike other relations, this one can be rigorously proven, although it is not as convenient to use in practice. We were unable to find an analogous relation for the case $\beta \neq 0$. 
%%%%%%%%%%%%%%%%%%%%%%%%%%%%%%%%%%%%%%%%%%%%%%%%%%%%%%%%%%%%%%%%%%%%%%%%%%%%%%%%%%%%%%%%%%%%%%
\section{Analytical results}\label{Analytical-results}
\subsection{Spectral sums}
As mentioned near \eqref{spectral-determinants-definition-1-2}, small $\lambda$ expansion coefficients of $\log D_{\pm}(\lambda)$ are given by the spectral sums. Although they do not have any explicit physical meaning, these are good reference ``observables'' to check our analytic results. Also, together with the WKB expansion that we describe in the next subsection, their values can be used to calculate the spectrum numerically with increasing precision: they account for low-lying levels where WKB has the largest error.

Our conjectured formulas give non-perturbative predictions for their values. Taking the logarithm of the Wronskian relation \eqref{wronsk-rel}, one can find the following expressions for the differences of the spectral sums (see definition \eqref{spectral-determinants-definition-1-2})
\begin{equation}
\begin{aligned}
    G^{(1)}_+-G^{(1)}_-=&\;\frac{\pi}{2\beta^2g_1}\left[\pi\alpha g_1-\pi(\alpha +\beta)-2 \beta\xi_1\right],\\
    G^{(2)}_+-G^{(2)}_-=&\;\frac{\pi^2}{2\beta^4g_1g_2}\left[2\pi\alpha (\pi(\alpha+\beta)+2\beta\xi_1)g_2+\pi^2(2\alpha^2-\beta^2)g_1g_2-4(\pi\alpha+\beta\xi_2)(\pi(\alpha+\beta)+\beta\xi_2)g_1\right],\\
    G^{(3)}_+-G^{(3)}_-=&\;\frac{\pi^3}{32\beta^6g^3_1g_2g_3}\Biggl[16\pi^3\alpha(4\alpha^2-3\beta^2)g_1^3g_2g_3+192\pi\alpha(\pi\alpha+\beta \xi_2)(\pi\alpha+\beta(\xi_2+\pi))g_1^3g_3-
    \\&-12\pi^2(\alpha^2+\beta^2)(\pi(\alpha+\beta)+2\beta\xi_1)g_1^2g_2g_3-(\pi(\alpha+\beta)+2\beta\xi_1)^3g_2g_3-
    \\&-9(3\pi\alpha-\pi\beta+2\beta\xi_3)(3\pi(\alpha+\beta)+2\beta\xi_3)(\pi(3\alpha+\beta)+2\beta\xi_3)g_1^3g_2\Biggl],
\end{aligned}
\end{equation}
where $\xi_k,g_k$ are defined in \eqref{xi_k-def}, \eqref{g_k-def}.

On the other hand, logarithmic derivative relations \eqref{D+symmetric} and \eqref{D-symmetric} allow us to calculate exactly the spectral sums separately. The first few of them are
\begin{equation}\label{G1pm-analytical}
    \begin{aligned}
        &G^{(1)}_-=\;\log8\pi-2+\frac{\pi}{4\beta^2g_1}[\pi(\alpha+\beta)-\pi\alpha g_1+2\beta\xi_1]-\frac{(\alpha+\beta)^2\mathtt{i}_2(\alpha+\beta)-(\alpha-\beta)^2\mathtt{i}_2(\alpha-\beta)}{8\beta},
        \\
        &G^{(1)}_+=\;\log8\pi-2-\frac{\pi}{4\beta^2g_1}[\pi(\alpha+\beta)-\pi\alpha g_1+2\beta\xi_1]-\frac{(\alpha +\beta)^2\mathtt{i}_2(\alpha+\beta)-(\alpha-\beta)^2\mathtt{i}_2(\alpha-\beta)}{8\beta},
   \end{aligned}
\end{equation}
\begin{equation}\label{G2pm-analytical}
    \begin{aligned}
        G^{(2)}_-=&\;\frac{\pi^2}{16\beta^4g_1^2g_2}\Biggl[\left(\pi(\alpha+\beta)+2\beta\xi_1\right)^2g_2-8\pi\alpha(\pi(\alpha+\beta)+2\beta\xi_1)g_1g_2+16(\pi\alpha +\beta\xi_2)(\pi(\alpha+\beta)+\beta\xi_2)g_1^2+
        \\&+\Big(4(\pi^2+2\alpha)\beta^2-9\pi^2\alpha^2\Big)g_1^2g_2\Biggl]-\frac{\pi^2(\alpha^2-\beta^2)}{16\beta^3}\Big((\alpha+\beta)\mathtt{i}_2(\alpha+\beta)-(\alpha-\beta)\mathtt{i}_2(\alpha-\beta)\Big),
        \\
        G^{(2)}_+=&\;\frac{\pi^2}{16\beta^4g_1^2g_2}\Biggl[(\pi(\alpha+\beta)+2\beta\xi_1)^2g_2+8\pi\alpha(\pi(\alpha+\beta)+2\beta\xi_1)g_1g_2-16(\pi\alpha+\beta\xi_2)(\pi(\alpha+\beta)+\beta\xi_2)g_1^2-
        \\&-\left(4(\pi^2-2\alpha)\beta^2-7\pi^2\alpha^2\right)g_1^2g_2\Biggl]-\frac{\pi^2(\alpha^2-\beta^2)}{16\beta^3}\Big((\alpha+\beta)\mathtt{i}_2(\alpha+\beta)-(\alpha-\beta)\mathtt{i}_2(\alpha-\beta)\Big),
    \end{aligned}
\end{equation}
\begin{equation}\label{G3pm-analytical}
    \begin{aligned}
        G^{(3)}_-=&\;\frac{\pi^3}{192\beta^6g_1^3g_2g_3}\Biggl[-36\pi\alpha(\pi(\alpha+\beta)+2\beta\xi_1)^2 g_1g_2g_3+3(\pi(\alpha+\beta)+2\beta\xi_1)^3g_2g_3+
        \\&+4(\pi(\alpha+\beta)+2\beta\xi_1)\Big(9\pi^2(\alpha^2+\beta^2)g_2+16 \left(\pi  \alpha +\beta  \xi _2\right) \left(\pi(\alpha+\beta)+\beta\xi_2\right)\Big)g_1^2g_3-
        \\&-576\pi\alpha\left(\pi\alpha+\beta\xi_2\right)\left(\pi(\alpha+\beta)+\beta\xi_2\right)g^3_1g_3-
        \\&-4\pi\Big(5\pi^2(11\alpha^3-8\alpha\beta^2)-24\alpha^2\beta^2+16\beta^4\Big)g^3_1g_2g_3+
        \\&+27\left(3\pi\alpha+2\beta\xi_3-\pi\beta\right)\left(3\pi\alpha+2\beta\xi_3+\pi\beta\right)\left(3\pi(\alpha+\beta)+2\beta\xi_3\right)g^3_1g_2\Biggl]-
        \\&-\frac{\pi^4\alpha(\alpha^2-\beta^2)}{16\beta^5}((\alpha+\beta)\mathtt{i}_2(\alpha+\beta)-(\alpha-\beta)\mathtt{i}_2(\alpha-\beta)),
        \\
        G^{(3)}_+=&\;\frac{\pi^3}{192\beta^6g_1^3g_2g_3}\Biggl[-36\pi\alpha\left(\pi(\alpha+\beta)+2\beta\xi_1\right)^2g_1g_2g_3-3\left(\pi(\alpha+\beta)+2\beta\xi_1\right)^3g_2g_3-\\
        &-4\left(\pi(\alpha+\beta)+2\beta\xi_1\right)\Big(9\pi^2\left(\alpha^2+\beta^2\right)g_2-16\left(\pi\alpha+\beta\xi_2\right)\left(\pi(\alpha+\beta)+\beta\xi_2\right)\Big)g_1^2g_3+\\
        &+576\pi\alpha\left(\pi\alpha+\beta\xi_2\right)\left(\pi(\alpha+\beta)+\beta\xi_2\right)g_1^3g_3-\\
        &-4\pi\Big(-\pi^2(41\alpha^3-32\alpha\beta^2)-24\alpha^2\beta^2+16\beta^4\Big)g_1^3g_2g_3-\\
        &-27\left(3\pi\alpha+2\beta\xi_3-\pi\beta\right)\left(3\pi\alpha+2\beta\xi_3+\pi\beta\right)\left(3\pi(\alpha+\beta)+2\beta\xi_3\right)g_1^3g_2\Biggl]+\\
        &-\frac{\pi^4\alpha(\alpha^2-\beta^2 )}{16\beta^5} ((\alpha+\beta)\mathtt{i}_2(\alpha+\beta)-(\alpha-\beta)\mathtt{i}_2(\alpha-\beta)) .
    \end{aligned}
\end{equation}
\begin{figure}[h!]
    \centering
    \begin{minipage}{0.49\textwidth}
        \centering
        \includegraphics[width=1.\linewidth]{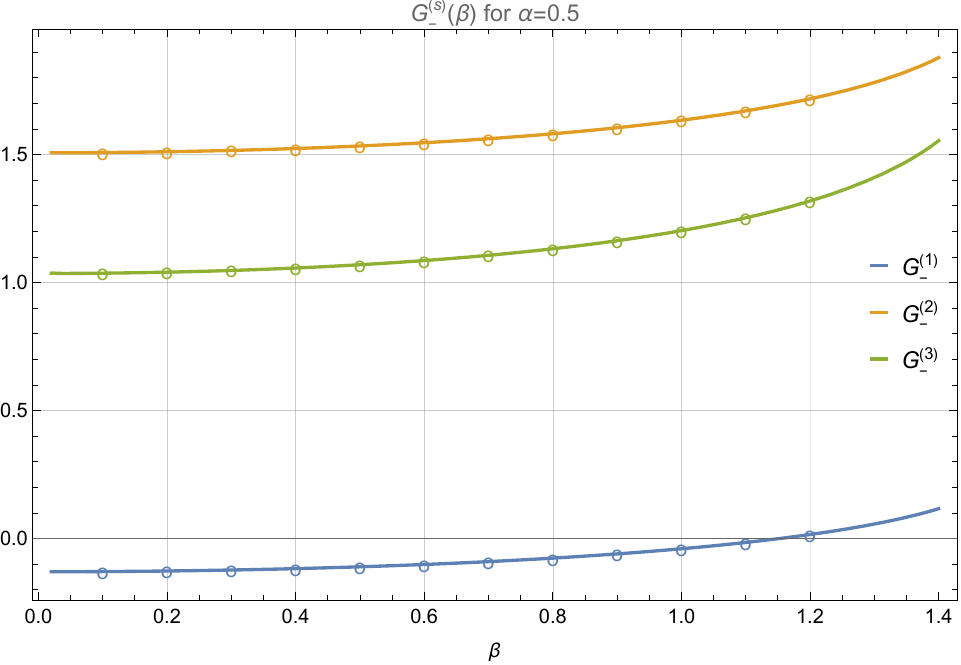}
        %\caption{(a)} % Добавляет подпись к подграфику (опционально)
        \label{Gm_graph}
    \end{minipage}
    \hfill
    \begin{minipage}{0.49\textwidth}
        \centering
        \includegraphics[width=1.\linewidth]{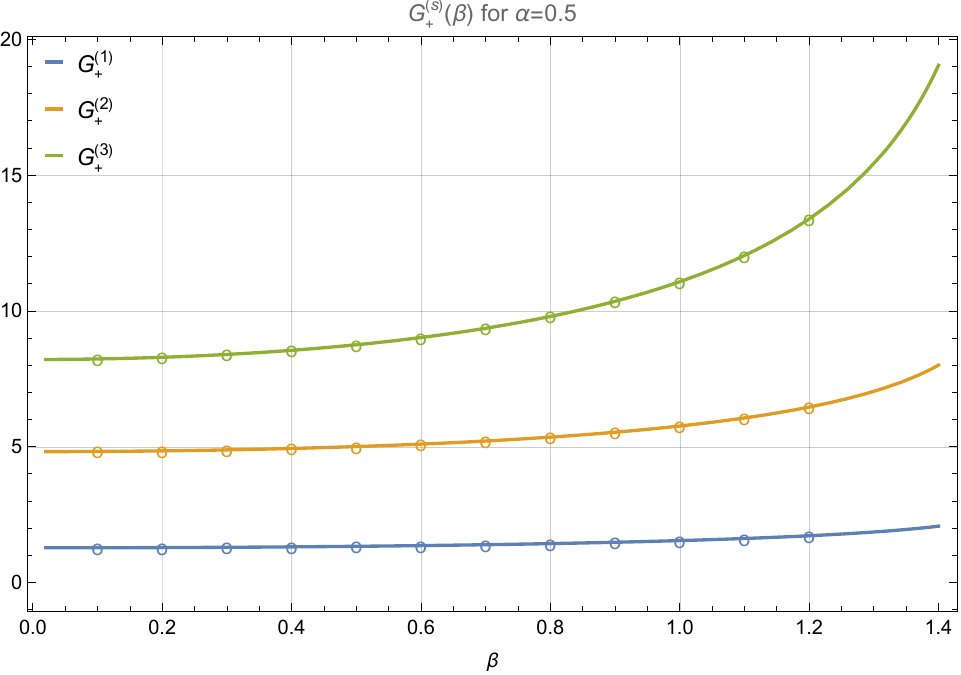}
        %\caption{(b)} % Подпись ко второму графику
        \label{Gp_graph}
    \end{minipage}
    \caption{Spectral sums $G^{(s)}_\pm(\beta)$ for different $s$ and $\alpha = 0.5$. Lines are the analytical predictions for their values, circles are numerical values from Table \ref{Gpm-table}. The region of negative values for $G^{(1)}_-$ is related to the regularization of the first spectral sums in \eqref{spectral_sums_def} and does not contradict the realness of meson masses.}
    \label{fig:G_graphs}
\end{figure}
Note that compared to the case $\alpha_1=\alpha_2=\alpha$, the expressions for spectral sums look more similar to each other (in terms of complexity: for $\beta = 0$ \cite{Litvinov:2024riz}, spectral sums $G^{(k)}_-$ look much simpler than $G^{(k)}_+$). Expressions for higher spectral sums become increasingly bulky. The dependence of the first $s=1,2,3$ spectral sums on $\beta$ for a few values of $\alpha$ is illustrated in Fig. \ref{fig:G_graphs}.

These conjectured formulas pass a few serious consistency checks. First, it is already non-trivial that the difference $G^{(s)}_+ - G^{(s)}_-$ following from \eqref{D+symmetric} and \eqref{D-symmetric} coincides with the one obtained from \eqref{wronsk-rel}. Second, one can show that, although these expressions are very transcendental, in $\beta \to 0$ limit they simplify significantly and agree with the results of \cite{Litvinov:2024riz}; for this, one has to use small $\beta$ expansion of $\xi_k$ and $g_k$ (see Section \ref{limiting-cases}). The last check is the consistency with the numerical results, which will be described in Section \ref{numerics}.

\subsection{WKB expansion}
For sufficiently large $n$ (highly excited states), approximate analytical expressions for both the meson wave functions and the spectrum can be derived using a semi-classical approach \cite{THOOFT1974461,Brower:1979PhysRevD}. Within the region $\frac{1}{n} \lesssim x \lesssim 1 - \frac{1}{n}$ the WKB method provides the following solution for the wave function
\begin{equation}
    \phi^{(n)}_{12}(x)\approx \sqrt{2}\sin{\left(\pi(n+1)x+\delta_{12}^{(n)}(x)\right)}
\end{equation}
and the spectrum reads\footnote{It was previously thought \cite{Brower:1979PhysRevD},\cite{Mondejar:2009td} that the correction to the analytic expression of the meson mass spectrum had an asymptotics of $\mathcal{O}(1/n)$, but in our previous work \cite{Litvinov:2024riz} for the one-flavour case it was found that the correction is actually $\mathcal{O}(\log(n)/n)$ (see \eqref{WKB-alpha}).  As we will establish at the end of the subsection this is also true for the case of two flavours.} 
\begin{equation}\label{WKB-2flavour-old}
    \lambda_n=\frac{1}{2}\left(n+\frac{3}{4}\right)+\frac{\alpha_1+\alpha_2}{\pi^2}\log{n}+\frac{1}{\pi^2}C(\alpha_1)+\frac{1}{\pi^2}C(\alpha_2)+\mathcal{O}\left(\frac{\log{n}}{n}\right).
\end{equation}
The phase shift $\delta^{(n)}$ and the constant\footnote{Note that in \cite{Brower:1979PhysRevD} the expression for $C(\alpha_i)$ was written incorrectly.} $C(\alpha)$ in the spectrum were obtained in \cite{Brower:1979PhysRevD} (see also \cite{Mondejar:2009td}), and they are
\begin{equation}
    \begin{aligned}
        \delta^{(n)}_{12}(x)=\;&\frac{1}{\pi}\Biggl[
        \alpha_2[x\log{n}+\log(1-x)]+x\left(C(\alpha_2)-\frac{\pi^2}{8}\right)-
        \\&-\alpha_1[(1-x)\log{n}+\log{x}]-(1-x)\left(C(\alpha_1)-\frac{\pi^2}{8}\right)\Biggl],
    \end{aligned}
\end{equation}
\begin{equation}
    C(\alpha_i)=\alpha_i+\alpha_i\int_0^\infty\limits dy\left[\frac{\sinh{2y}-2y}{2\cosh{y}(\alpha_i\sinh{y}+y\cosh{y})}-\frac{1}{y+\pi^2}\right].
\end{equation}
In \cite{Fateev:2009jf}, the authors proposed a systematic way to calculate $1/n$ corrections to the WKB formula when $\alpha_1=\alpha_2=0$ 
\begin{equation}\label{WKB_alpha=0}
    \lambda_n=\frac{1}{2}\left(n+\frac{3}{4}\right)-\frac{1}{3\pi^6(n+\frac{3}{4})^3}+\frac{(-1)^{n+1}}{\pi^4(n+\frac{3}{4})^2}\left[1-\frac{4\log{(\pi e^{\gamma_E-\frac{1}{2}}}(n+\frac{3}{4}))}{\pi^2(n+\frac{3}{4})}\right]+\mathcal{O}\left(\frac{\log^2{n}}{n^4}\right)
\end{equation}
and announced preliminary results for the case of nonzero $\alpha$ (see formula 6.1 in \cite{Fateev:2009jf}). 
In our previous work \cite{Litvinov:2024riz} we generalized this method  and presented large-$n$ expansion for any $\alpha$ (more terms can be found in Mathematica notebook \texttt{WKB.nb} attached to \cite{Litvinov:2024riz})
\begin{equation}\label{WKB-alpha}
    \begin{aligned}
        \lambda_n(\alpha)=&\;\frac{1}{2}\mathfrak{n}+\frac{\alpha}{\pi^2}\log{\rho}+\frac{\alpha^2}{\pi^4}\frac{2\log\rho-1}{\mathfrak{n}}-\frac{1}{\pi^6}\frac{1}{\mathfrak{n}^2}\Biggl[2\alpha^3\log^2{\rho}-6\alpha^3\log\rho+3\alpha^3+(-1)^n\pi^2(1+\alpha)\Biggl]+
        \\&+\frac{1}{\pi^8}\frac{1}{\mathfrak{n}^3}\Biggl[\frac{8\alpha^4}{3}\log^3\rho-16\alpha^4\log^2\rho+24\alpha^4\log\rho-\frac{29\alpha^4+\pi^2(1+\alpha)^2}{3}+
        \\&+(-1)^n\pi^2(1+\alpha)\left(4(1+\alpha)\log\rho-(2+8\alpha+8\log{2}+\alpha\mathtt{i}_1(\alpha))\right)\Biggl]+\mathcal{O}\left(\frac{\log^4{\mathfrak{n}}}{\mathfrak{n}^4}\right),
    \end{aligned}
\end{equation}
where
\begin{equation}
    \mathfrak{n}=n+\frac{3}{4}-\frac{\alpha^2}{2\pi^2}\mathtt{i}_2(\alpha),\quad\rho=4\pi e^{\gamma_E}\left(n+\frac{3}{4}-\frac{\alpha^2}{2\pi^2}\mathtt{i}_2(\alpha)\right).
\end{equation}

We now generalize these results to the case of two arbitrary flavours (any values of $\alpha_{1,2}$) and write out the corrections to \eqref{WKB-2flavour-old}. For large negative $\lambda$ we have the following asymptotic expansion for the spectral determinants $D_\pm(\lambda)$
\begin{equation}\label{F_definition}
    D_{\pm}(\lambda)=d_{\pm}\left(8\pi e^{-2+\gamma_E}\right)^{\lambda}(-\lambda)^{\lambda-\frac{1}{8}\pm\frac{1}{4}}
    \exp\Bigl(F^{(0)}_{\pm}(L)+F^{(1)}_{\pm}(L)\lambda^{-1}+F^{(2)}_{\pm}(L)\lambda^{-2}+\dots\Bigr),
\end{equation}
with $F^{(k)}_{\pm}(L)$ being the polynomials in $L = \log (-2\pi \lambda) + \gamma_E$. The factor $(8\pi e^{-2+\gamma_E})^\lambda$ in \eqref{F_definition} follows from log-derivative relations \eqref{D-symmetric}, \eqref{D+symmetric}. It is determined by the difference between $\mathcal{O}(\lambda^0),\;\lambda \to -\infty$ terms in their l.h.s. and r.h.s.. Using the asymptotic expansion for $\mathbf{Q}$-functions and integrating log-derivative relations with respect to $\lambda$, we can find  $F^{(k)}_{\pm}(L)$ explicitly order by order in $1/\lambda$; e.g.
\begin{equation}
\begin{aligned}
    F^{(0)}_{\pm}(L)=&-\frac{\alpha L^2}{2\pi^2}-\frac{\left(16\alpha\log{2}-(\alpha+\beta)^2 \mathtt{i}_2(\alpha+\beta)-(\alpha-\beta)^2\mathtt{i}_2(\alpha-\beta)\right)L}{8\pi^2},\\
    F^{(1)}_{\pm}(L)=&\;\frac{(\alpha^2-\beta^2)L}{2\pi^4}+\frac{\pi^2\alpha+4\alpha^2(4\log{2}-1)-4\beta^2(4\log{2}+1)}{16\pi^4}-
    \\&-\frac{(\alpha^2-\beta^2)((\alpha+\beta)\mathtt{i}_2(\alpha+\beta)+(\alpha-\beta)\mathtt{i}_2(\alpha-\beta))\pm 4\pi^2(\alpha+\gamma)}{16\pi^4}.
\end{aligned}
\end{equation}
Again, \eqref{wronsk-rel} can be used to find the expressions for differences $F^{(k)}_+(L) - F^{(k)}_-(L)$, providing a non-trivial consistency check. It also allows us to find the ratio
\begin{equation}\label{dm/dp}
    \frac{d_-}{d_+}=\frac{\sqrt{2}}{\pi}\frac{\sqrt{1+\alpha+\beta}+\sqrt{1+\alpha-\beta}}{2}
\end{equation}
although we have no way to derive $d_+$ and $d_-$ separately. This is one of the non-trivial predictions of our theory. In the limit $\beta\to 0$, it reduces to the result from \cite{Litvinov:2024riz}. For numerical purposes, the ratio \eqref{dm/dp} can be expressed as a rapidly convergent infinite product \cite{Fateev:2009jf}
\begin{equation}\label{dp/dm-prod}
    \frac{d_-}{d_+}=\frac{\Gamma({\frac{7}{8}})}{\Gamma({\frac{3}{8}})}\left(\prod_{m=0}^{\infty}\limits\,\frac{m+\frac{7}{8}}{\lambda_{2m+1}}\right)\cdot\left(\prod_{m=0}^{\infty}\limits\,\frac{m+\frac{3}{8}}{\lambda_{2m}}\right)^{-1}.
\end{equation}
In Section \ref{numerics}, we numerically verify that this product is equal to \eqref{dm/dp} with high precision (see Table \ref{dm/dp-table}).

The expressions \eqref{F_definition} are valid only for large negative $\lambda$ and can not be used in the vicinity of the spectrum points $\lambda_n>0$, corresponding to meson masses. However, it turns out that at large positive $\lambda$ the spectral determinants $\mathcal{D}_{\pm}(\lambda)$ can be obtained as follows
\begin{equation}
    \mathcal{D}_{\pm}(\lambda)=\frac{1}{2}
    \left(D_{\pm}(-e^{-i\pi}\lambda)+D_{\pm}(-e^{+i\pi}\lambda)\right).
\end{equation}
Here, the two terms are analytic continuations of \eqref{F_definition} through the upper or lower half-plane, respectively. This prescription was introduced in \cite{Fateev:2009jf}, motivated by the requirement for the corresponding $Q$-function to decay at $\nu \to \infty$. It proved to be correct and useful in the previous study \cite{Litvinov:2024riz} as well. In particular, it leads to the following formula
\begin{equation}\label{D-analytical-continuation}
    \mathcal{D}_{\pm}(\lambda)=2d_{\pm}
    \left(8\pi e^{-2+\gamma_E}\right)^{\lambda}
    \lambda^{\lambda-\frac{1}{8}\pm\frac{1}{4}}\exp{\left(\sum_{k=0}^{\infty}\limits\Xi^{(k)}_{\pm}(l)\lambda^{-k}\right)}\cos\left(
    \frac{\pi}{2}\left[2\lambda-\frac{1}{4}\pm\frac{1}{2}+\sum_{k=0}^{\infty}\Phi^{(k)}_{\pm}(l)\lambda^{-k}\right]\right),
\end{equation}
where $\Xi^{(k)}_{\pm}(l)$ and $\Phi^{(k)}_{\pm}(l)$ are again polynomials in
\begin{equation}
    l=\log{(2\pi\lambda)}+\gamma_E,
\end{equation}
which are symmetrized and anti-symmetrized versions of $F^{(k)}_{\pm}(L)$
\begin{equation}
    \Xi^{(k)}_{\pm}(l)\overset{\text{def}}{=}\frac{1}{2}\left(F^{(k)}_{\pm}(l+i\pi)+F^{(k)}_{\pm}(l-i\pi)\right),\quad \Phi^{(k)}_{\pm}(l)\overset{\text{def}}{=}\frac{i}{\pi}\left(F^{(k)}_{\pm}(l-i\pi)-F^{(k)}_{\pm}(l+i\pi)\right).
\end{equation}
Zeroes of $\mathcal{D}_{\pm}(\lambda)$ are determined by the last factor $\cos(\dots)$ and provide the ``quantization conditions'' on $\lambda$
\begin{equation}\label{quantisation-condition-lambda}
    2\lambda-\frac{1}{4}\pm\frac{1}{2}+\Phi^{(0)}_{\pm}(l)+\Phi^{(1)}_{\pm}(l)\lambda^{-1}+\dots=2m+1,\quad m=0,1,2,\dots
\end{equation}
The first few phases $\Phi_{\pm}^{(k)}(l)$ are given by
\begin{equation} \label{phases}
\resizebox{\textwidth}{!}{$
\begin{aligned}
    &\Phi^{(0)}_\pm(l)=-\frac{2\alpha(l+\log{4})}{\pi^2}+\frac{(\alpha+\beta)^2\mathtt{i}_2(\alpha+\beta)+(\alpha-\beta)^2\mathtt{i}_2(\alpha-\beta)}{4\pi^2},\\ &\Phi^{(1)}_\pm(l)=\frac{\alpha^2-\beta^2}{\pi^4},\quad \Phi^{(2)}_\pm(l)=\frac{\alpha(\alpha^2-\beta^2)\pm\pi^2\gamma}{2\pi^6},\\
    &\Phi^{(3)}_{\pm}(l)=\frac{1}{12\pi^8}\Biggl[5\alpha^4-6\alpha^2\beta^2+\beta^4+\pi^2\gamma^2\mp12\pi^2\gamma\left(l-\frac{1}{2}-\frac{3\alpha}{2}-\frac{(\alpha+\beta)\mathtt{i}_1(\alpha+\beta)+(\alpha-\beta)\mathtt{i}_1(\alpha-\beta)}{8}\right)\Biggl],
\end{aligned}
$}
\end{equation}
where $\gamma$ is defined in \eqref{gamma-def} and the integrals $\mathtt{i}_k(\alpha)$ are  defined in \eqref{i2k1-def}, \eqref{i2k-def}.

Then, for large $\lambda$ (or $m$), equation \eqref{quantisation-condition-lambda} can be inverted, giving an asymptotic expansion of $\lambda_n$ in the following form
\begin{equation}
    \begin{aligned}\label{wkb}
        \lambda_{n} =&\; \frac{1}{2} \mathfrak{n} + \frac{\alpha}{\pi^2} \log\rho + \frac{1}{\pi^4}\frac{1}{\mathfrak{n}} \left[2\alpha^2 \log\rho - (\alpha^2 - \beta^2)\right] - \frac{1}{\pi^6}\frac{1}{\mathfrak{n}^2} \Biggl[2\alpha^3\log^2\rho-2\alpha(3\alpha^2 - \beta^2) \log\rho+
        \\&+3\alpha(\alpha^2-\beta^2)+(-1)^n\pi^2\gamma\Biggl]+\frac{1}{\pi^8}\frac{1}{\mathfrak{n}^3}\Biggl[\frac{8\alpha^4}{3}\log^3\rho-4\alpha^2(4\alpha^2-\beta^2)\log^2\rho+
        \\&+\left(8\alpha^3(3\alpha^2-\beta^2)+(-1)^n4\pi^2(1+\alpha)\gamma\right)\log\rho-\frac{29\alpha^4-36\alpha^2\beta^2+7\beta^4+\pi^2\gamma^2}{3}-
        \\
        & -(-1)^n\pi^2\gamma\left(2+8\alpha+8\log{2}+\frac{(\alpha+\beta)\mathtt{i}_1(\alpha+\beta)+(\alpha-\beta)\mathtt{i}_1(\alpha-\beta)}{2}\right)\Biggl]+\mathcal{O}\left(\frac{\log^4{\mathfrak{n}}}{\mathfrak{n}^4}\right),
    \end{aligned}
\end{equation}
where
\begin{equation}
    \mathfrak{n} = n+ \frac{3}{4} -\frac{(\alpha -\beta )^2 \mathtt{i}_2(\alpha -\beta )+(\alpha +\beta )^2 \mathtt{i}_2(\alpha +\beta )}{4 \pi ^2},\quad \rho=4\pi e^{\gamma_E}\mathfrak{n}.
\end{equation}
We present only a few leading terms; however, our technique enables the derivation of an arbitrary number of terms. 

The result \eqref{wkb} naturally leads to two distinct Regge trajectories for odd and even $n$. In $4$D QCD, particles with even and odd momentum $J$ should reside on separate Regge trajectories due to exchange forces in systems of identical constituents. It would be interesting to understand what mechanism leads to a similar phenomenon in the 't Hooft model. Moreover, we observe a difference in the trajectories starting from $1/n^2$, which should indicate the weakness of this mechanism (see also the discussion in Section \ref{Discussion}). 

\begin{figure}[h!]
    \centering
    \includegraphics[width=0.6\linewidth]{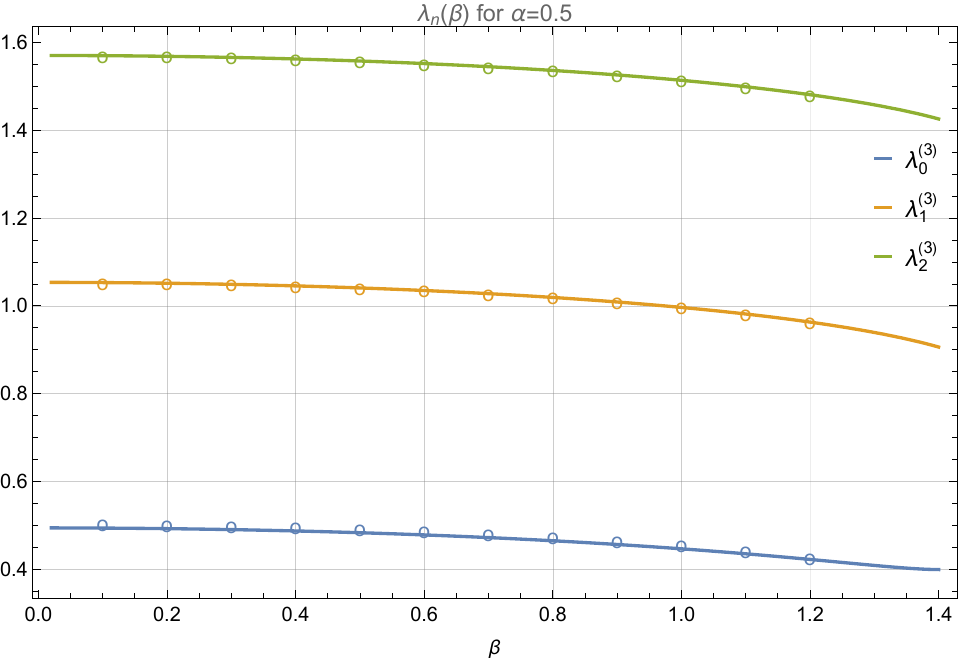}
    \caption{Eigenvalues $\lambda_n(\beta)$ for $\alpha = 0.5$. Lines are the analytical predictions \eqref{wkb} for their values (index $3$ indicates that we truncate \eqref{wkb} beyond the term proportional to $\lambda^{-3}$), circles are numerical values from Table \ref{Table-WKB}.}
    \label{fig:lambda-beta}
\end{figure}

For $\alpha, \beta$ of order $1$, formula \eqref{wkb} works very well even for small $n$; $\beta$--dependence is illustrated in Fig. \ref{fig:lambda-beta}. Numerical results confirming the validity of this formula are presented in Section \ref{numerics}. Note the general trend that $\lambda_n$ decreases with increasing $\beta$. Consequently, the spectral sums increase, as  was demonstrated in the previous subsection. At least for the ground state, which should give the leading contribution to the spectral sums, this can be easily argued from the usual quantum mechanical perturbation theory in $\beta$: the first-order correction vanishes and the second-order one is always negative. 

The question of computing systematic $1/n$ corrections to the wave function of mesons $\phi^{(n)}(x)$ consisting of two different quarks remains open. In \cite{Fateev:2009jf}, the authors proposed an effective approximate formula for the eigenfunctions $\Psi_n(\nu)$ in the special case $\alpha_1=\alpha_2=0$, which was later generalized to the case of arbitrary $\alpha_1=\alpha_2=\alpha$ in \cite{Ambrosino:2023dik}. We are confident that we now have sufficient understanding of the functions $\Psi_\pm(\nu|\beta)=\sum_{n=0}^{\infty}\limits\frac{c^\pm_n\Psi_n(\nu|\beta)}{\lambda-\lambda_n}$ to find an analogous approximation for $\Psi_n(\nu|\beta)$ in the case $\beta \neq 0$; however, we leave this question for future work.  
%%%%%%%%%%%%%%%%%%%%%%%%%%%%%%%%%%%%%%%%%%%%%%%%%%%%%%%%%%%%%%%%%%%%%%%%%%%%%%%%%%%%%%%%%%%%%
%%%%%%%%%%%%%%%%%%%%%%%%%%%%%%%%%%%%%%%%%%%%%%%%%%%%%%%%%%%%%%%%%%%%%%%%%%%%%%%%%%%%%%%%%%%%%
\section{Analysis of results in limiting cases}\label{limiting-cases}
In this section we describe how our exact formulas can be used and simplified in several physically interesting limiting cases. These cases allow us to relate our solution to previously known results in the literature, as well as provide some ways to check our formulas that are complementary to the numerics described in Section \ref{numerics} (our numerical methods are most reliable for quark masses $m_1, m_2$ of order $1$).
\subsection{Warm up: $\alpha_1\approx\alpha_2$}
As a warm-up, let us consider the limit in which the quark masses are close to each other $\alpha_1\approx\alpha_2$ (small $\beta$). The main reason why it is interesting to consider this case is that, compared to the $1$ flavour case, the expressions for the spectral sums $G^{(s)}_\pm$ are given in terms of new objects $\xi_k,g_k$ \eqref{xi_k-def},\eqref{g_k-def} and it is at first glance non-trivial that the answers reproduce the known ones from \cite{Litvinov:2024riz} (formulas (5.3)-(5.5)).

In the limit $\beta\rightarrow0$ we have
\begin{equation}\label{xi-beta-limit}
    \xi_k(\beta)=\frac{\pi}{2}-\frac{2}{\pi k}\beta+\frac{8}{(\pi k)^3}\beta^3-\frac{32}{(\pi k)^5}\beta^5+\mathcal{O}(\beta^7)
\end{equation}
and the coefficients $g_k$ are determined by the expansions of the functions $S_0(\nu|\alpha\pm\beta)$ near $\nu=i$ \eqref{S0-expansion} and are expressed using  \eqref{gk*g-k} and \eqref{xi-beta-limit}. The first few terms of the expansion of $g_k$ in $\beta$ are as follows
\begin{equation}\label{gk-beta-limit}
    g_k(\alpha,\beta)=1+\frac{2\beta}{\alpha}+\frac{\beta^2\left(2\pi^2(k-1)-\alpha^2\mathtt{i}_2(\alpha)-\alpha^3\mathtt{i}_2'(\alpha)\right)}{\pi^2\alpha^2k}+\mathcal{O}(\beta^3).
\end{equation}
Taking into account the expansions \eqref{xi-beta-limit}, \eqref{gk-beta-limit}, the first spectral sums behave as follows 
\begin{equation}
    \begin{aligned}
        G^{(1)}_-=&\;G^{(1)}_-\Big|_{\beta=0}+\frac{\beta ^2}{24 \pi ^2} \Biggl(32-8(3+\alpha)\mathtt{i}_2(\alpha)+3\alpha\mathtt{i}^2_2(\alpha)+8\alpha\mathtt{i}_4(\alpha)-2(3\pi^2+12\alpha+4\alpha^2)\mathtt{i}_2'(\alpha)+
        \\&+6\alpha^2\mathtt{i}_2(\alpha)\mathtt{i}_2'(\alpha )+3\alpha^3(\mathtt{i}'_2(\alpha))^2+8\alpha^2\mathtt{i}_4'(\alpha)-3\pi^2\alpha\mathtt{i}_2''(\alpha)\Biggl)+\mathcal{O}(\beta^4),
        \\
        G^{(1)}_+=&\;G^{(1)}_+\Big|_{\beta=0}-\frac{\beta^2}{24\pi^2}\Biggl(32-8(3+\alpha)\mathtt{i}_2(\alpha)+3\alpha \mathtt{i}^2_2(\alpha)+8\alpha\mathtt{i}_4(\alpha)+2(3\pi^2-12\alpha-4\alpha^2)\mathtt{i}_2'(\alpha)+
        \\&+6\alpha^2\mathtt{i}_2(\alpha)\mathtt{i}_2'(\alpha)+3\alpha^3(\mathtt{i}'_2(\alpha))^2+8\alpha^2\mathtt{i}_4'(\alpha)+9\pi^2\alpha\mathtt{i}_2''(\alpha)+2\pi^2\alpha^2\mathtt{i}'''_2(\alpha)\Biggl)+\mathcal{O}(\beta^4),
    \end{aligned}
\end{equation}
where $'$ means the derivative by $\alpha$. The higher spectral sums \eqref{G2pm-analytical},\eqref{G3pm-analytical} also give the correct limit $\beta\to0$. We do not give here the expressions for the corrections because they are too cumbersome.
%%%%%%%%%%%%%%%%%%%%%%%%%%%%%%%%%%%%%%%%%%%%%%%%%%%%%%%%%%%%%%%%%%%%%%%%%%%%%%%%%%%%%%%%%%%%%%%%%%%%%%%%%%%%%%%%%%%%%%%%%%%%%%%%%%%%%%%%%%%%%%%%%%%%%%%%%%%%%%%%%%%%%%%%%%%%%%%%%%%%%%%%%%%%%%%%
\subsection{Chiral limit: spectral sums and GMOR relation}
From a physical perspective, the chiral limit $m_i\ll g$ is of significant interest. When quark masses are zero, bosonization approach \cite{GEPNER1985481, Frishman:2010zz} to two-dimensional gauge theories predicts that IR effective theory describing the massless degrees of freedom is given by coset/gauged WZW CFT, which in the case of fundamental matter is equivalently $U(N_F)_{N_c}$ WZW model \cite{Delmastro:2021otj}. Introducing the small quark mass can be done by adding a perturbation of the form $\sim m\, \text{tr\,}\left(h + h^\dagger \right)$, where $h$ is group-valued fundamental field in WZW model. 

In \cite{Zhitnitsky:1985um}, an exact result for the condensate in the chiral limit $m_1=m_2=m \to 0$ within the 't Hooft model was derived using the QCD sum rule technique. Later, in \cite{Burkardt:1995eb}, a more general formula applicable to arbitrary, but equal quark masses $m_1=m_2=m$ was obtained\footnote{Note that we have written the expression in notations different from the original work \cite{Burkardt:1995eb}.}
\begin{equation}
    \braket{\bar\psi\psi}=\frac{mN_c}{2\pi}\left\{\log{\frac{\pi}{1+\alpha}}-1-\gamma_E+\alpha\log4-\alpha^2 I(\alpha)\right\},\quad I(\alpha)=\frac{1}{4}\int^{\infty}_{-\infty}\limits\frac{dt}{t^2}\frac{\sinh{2t}-2t}{\cosh{t}(\alpha\sinh t+t\cosh t)}.
\end{equation}
In the chiral limit, $\alpha = -1 + a$, where $a = \frac{\pi m^2}{g^2} \to 0$, it approaches
\begin{equation}\label{condensate}
    \braket{\bar\psi\psi}\big|_{a\to0}=-\frac{mN_c}{2\pi}\left\{\frac{\pi}{\sqrt{3a}}-\log{\left(\frac{\pi}{a}\right)}+\mathcal{O}(1)\right\}\approx-N_c\sqrt{\frac{g^2}{12\pi}}.
\end{equation}

We expect that in this limit a massless meson (we will call it ``pion"\footnote{Pions are commonly referred to as Goldstone bosons; however, since there is no spontaneous symmetry breaking in the 't Hooft model, this term is not entirely accurate.}) will emerge, meaning that the ground-state energy $M_0 \to 0$. Consequently, the even spectral sums $G^{(s)}_+$ will diverge, while the odd sums $G^{(s)}_-$ will remain finite. Additionally, the Gell-Mann–Oaks–Renner (GMOR) relation \cite{Gell-Mann:1968hlm}, which connects the pion mass to the quark condensate, is expected to hold
\begin{equation}\label{GMOR}
    \frac{N_c}{\pi}M^2_0\approx-2(m_1+m_2)\braket{\bar{\psi}\psi}.
\end{equation}

The leading term of the ground state energy for $m_i \to 0$ can be found directly from the 't Hooft equation: it is determined by the boundary behavior of the wave function $\phi^{(0)}_{12}(x)$. We use an approximate ansatz\footnote{In the chiral limit the normalized wave function of the ground state is approximately $\phi^{(0)}_{12}(x)\big|_{m_i\to0}\approx1$.} for the wave function normalized by $1$ and satisfying the boundary conditions \eqref{phi-boundary-conditions}, \eqref{boundary-cond-eq}
\begin{equation}\label{chiral-ansatz}
    \phi^{(0)}_{12}(x)\approx\sqrt{\frac{\Gamma(2+2\beta_1+2\beta_2)}{\Gamma(1+2\beta_1)\Gamma(1+2\beta_2)}}x^{\beta_1}(1-x)^{\beta_2},\quad \beta_i\approx\sqrt{\frac{3}{\pi}}\frac{m_i}{g}.
\end{equation}
Substituting this ansatz \eqref{chiral-ansatz} into the 't Hooft equation\footnote{
Written in equivalent form
\begin{equation}
    2\pi^2\lambda_n\;\phi^{(n)}_{12}(x)=\frac{\pi}{g^2}\left[\frac{m^2_1}{x}+\frac{m^2_2}{1-x}\right]\phi^{(n)}_{12}(x)-\fint_0^1\limits dy\frac{\phi^{(n)}_{12}(y)-\phi^{(n)}_{12}(x)}{(x-y)^2}.
\end{equation}
} \eqref{'tHooft-eq} and integrating over $x$, one finds
\begin{equation}
    \lambda_0=\frac{1}{2\pi g^2}\int_0^1\limits dx\left[\frac{m^2_1}{x}+\frac{m^2_2}{1-x}\right]\phi^{(0)}_{12}(x)\approx \frac{\sqrt{a_1}+\sqrt{a_2}}{2\sqrt{3}\pi},\quad a_i=\frac{\pi m^2_i}{g^2}.
\end{equation}
This asymptotic agrees with GMOR formula \eqref{GMOR}; since $M^2_0=2\pi g^2\lambda_0$, we have
\begin{equation}
    M^2_0\approx\frac{g\sqrt{\pi}}{\sqrt{3}}(m_1+m_2)\quad\Rightarrow\quad \frac{N_c}{\pi}M^2_0\approx-2(m_1+m_2)\braket{\bar{\psi}\psi}.
\end{equation}
The naive ansatz \eqref{chiral-ansatz} cannot be used for computing subleading corrections to the ground state energy.  However, the spectral sums that we have obtained \eqref{G1pm-analytical}-\eqref{G3pm-analytical} allow us to calculate it.  

To determine the asymptotic behavior of the spectral sums for $m_i \to 0$ and extract $\lambda_0$ from them, we utilize the asymptotics of the integrals $\mathtt{u}_{2k-1}(\alpha)$ (see (3.17) and Appendix C in \cite{Litvinov:2024riz}). The integrals $\mathtt{i}_{2k}(\alpha)$ arising in the spectral sums can be expressed as
\begin{equation}\label{i-u-relation}
    \mathtt{i}_{2k}(\alpha)=2\mathtt{u}_{2k-1}(\alpha)+2\alpha \mathtt{u}_{2k+1}(\alpha)-2\mathtt{c}_{2k},\quad \mathtt{c}_{2k}=\int_{-\infty}^{\infty}\limits\frac{\sinh t}{t \cosh^{2k+1}t}dt.
\end{equation}
The integrals $\mathtt{c}_{2k}$ can be calculated analytically, for example
\begin{equation}
    \mathtt{c}_2=\frac{14\zeta(3)}{\pi^2},\quad c_4=\frac{2\left(7\pi^2\zeta(3)+93\zeta(5)\right)}{3\pi^4},\quad \mathtt{c}_6=\frac{2\left(56\pi^4\zeta(3)+930\pi^2\zeta(5)+5715\zeta(7)\right)}{45\pi^6},\quad \dots
\end{equation}
In the chiral limit, the asymptotics of $\mathtt{i}_{2k}(\alpha)$ are (we give the first few terms)
\begin{equation}\label{i2k-chiral-lim}
    \mathtt{i}_{2k}(-1+a)\big|_{a\to0}=-2(\mathtt{c}_{2k}-c^{(0)}_{2k-1}+c^{(0)}_{2k+1})-4\pi\sqrt{3a}+2(c^{(0)}_{2k-1}+c^{(0)}_{2k+1}-c^{(1)}_{2k+1})a+\mathcal{O}\left(a^{\frac{3}{2}}\right),
\end{equation}
where the coefficients $c^{(l)}_{2k-1}$ can only be computed numerically
\begin{equation}
    c^{(l)}_{2k-1}\overset{\textrm{def}}{=}(-1)^l\int_{-\infty+i\epsilon}^{\infty+i\epsilon}\limits\frac{\sinh^{l+2}t}{t\cosh^{2k-1} t \big(t\cosh t-\sinh t\big)^{l+1}}dt.
\end{equation}
Note that $\beta = \frac{a_2 - a_1}{2}$ is also a small parameter (at most of order $a$). Then, since $\nu_k$ is close to $-i$, expansion for the coefficients $g_k$ \eqref{g_k-def} in the chiral limit can be obtained from \eqref{i2k-chiral-lim} and expansion of the $G$-function for $\nu \to -i$, found by combining \eqref{G-def}, \eqref{G-relation} and \eqref{S0-expansion}. The first few terms read
\begin{equation}
    g_k\big|_{a_i\to0}=1-2\beta+\frac{2\sqrt{3}}{\pi k}\beta(\sqrt{a_1}-\sqrt{a_2})+\frac{2\left(c_2-c^{(0)}_{1}+c^{(1)}_{1}+2c^{(0)}_{3}-c^{(1)}_{3}+\pi^2(k(1-x)-1)\right)\beta^2}{\pi^2k}+\mathcal{O}(\beta^{\frac{5}{2}}),
\end{equation}
where $\alpha_i=-1+a_i, a_i=\frac{\pi m_i^2}{g^2}\to0.$ 

In this approach, the spectral sums can be expanded\footnote{In fact, we obtain slightly different linear combinations of $c^{(l)}_k$ and $\mathtt{c}_{2k}$, but numerically they are equal to those given in formulas \eqref{G1pm-chiral},\eqref{G2pm-chiral}. We have written down the expressions in such a form that it is more convenient to compare them with the chiral limit of spectral sums for one flavour \cite{Litvinov:2024riz}.} as follows
\begin{equation}\label{G1pm-chiral}
\begin{aligned}
    G^{(1)}_-=&\;\log(8\pi)-3-\frac{7\zeta(3)}{\pi^2}+\frac{1}{2}(c^{(0)}_{1}-c^{(0)}_{3})-\frac{\pi\sqrt{3}}{2}(\sqrt{a_1}+\sqrt{a_2})+\mathcal{O}(a),
    \\
    G^{(1)}_+=&\;\frac{2\sqrt{3}\pi}{\sqrt{a_1}+\sqrt{a_2}}+\log(8\pi)-1+\frac{7\zeta(3)}{\pi^2}+\frac{1}{2}(c^{(0)}_{1}+c^{(0)}_{3})+\mathcal{O}(\sqrt{a}),
\end{aligned}
\end{equation}
\begin{equation}\label{G2pm-chiral}
\begin{aligned}
    G^{(2)}_-=&\;\frac{10}{3}+\frac{56\zeta(3)}{3\pi^2}-\frac{124\zeta(5)}{\pi^4}+2\left(c^{(0)}_{3}-c^{(0)}_{5}\right)-3 \sqrt{3} \pi  \left(\sqrt{a_1}+\sqrt{a_2}\right)+\mathcal{O}(a),
    \\
    G^{(2)}_+=&\;\frac{12\pi^2}{(\sqrt{a_1}+\sqrt{a_2})^2}+\frac{\sqrt{3} \left(56\zeta(3)+\pi^4+4\pi^2(2+c^{(0)}_{3})\right)}{\pi (\sqrt{a_1}+\sqrt{a_2})}+\mathcal{O}(1).
\end{aligned}
\end{equation}
\begin{equation}\label{G3pm-chiral}
\begin{aligned}
    G^{(3)}_-=&\;\frac{4\pi^2}{9}-\frac{104}{15}+\frac{28(15-\frac{46}{\pi^2})\zeta(3)}{45}+\frac{62(16-3\pi^2)\zeta(5)}{3\pi^4}-\frac{1016\zeta(7)}{\pi^6}+(\pi^2-4)c^{(0)}_3-
    \\&-(\pi^2-8)c^{(0)}_5-4c^{(0)}_7 - \frac{3\sqrt{3}}{2}  \pi^3 \left(\sqrt{a_1}+\sqrt{a_2}\right)+\mathcal{O}(a),
    \\
    G^{(3)}_+=&\;\frac{24\sqrt{3}\pi^3}{(\sqrt{a_2}+\sqrt{a_1})^3}+\frac{9(56\zeta(3)+\pi^4+4\pi^2(2+c_3^{(0)}))}{(\sqrt{a}_1 +\sqrt{a}_2)^2}+\mathcal{O}\left(\frac{1}{a^{\frac{1}{2}}}\right).
\end{aligned}
\end{equation}

Thus, from the spectral sum $G^{(2)}_+$ we can find the leading and subleading asymptotics of the ground state (it is also consistent with the asymptotic of $G^{(3)}_+$)
\begin{equation}
    \lambda_0=\frac{\sqrt{a_1}+\sqrt{a_2}}{2\sqrt{3}\pi}\underbrace{-\frac{56\zeta(3)+\pi^4+4\pi^2\left(2+c^{(0)}_{3}\right)}{48\pi^4}}_{0.0453546\dots}(\sqrt{a_1}+\sqrt{a_2})^2+\mathcal{O}(a^{\frac{3}{2}}).
\end{equation}
The subleading term is the new result\footnote{Ansatz \eqref{chiral-ansatz} gives a prediction for the coefficient in subleading term $\frac{1}{2\pi^2}=0.0506606\dots$}. In principle, by utilizing higher-order spectral sums, one can determine additional terms in the expansion of the ground state energy in the chiral limit.
%%%%%%%%%%%%%%%%%%%%%%%%%%%%%%%%%%%%%%%%%%%%%%%%%%%%%%%%%%%%%%%%%%%%%%%%%%%%%%%%%%%%%%%%%%%%%%%%%%%%%%%%%%%%%%%%%%%%%%%%%%%%%%%%%%%%%%%%%%%%%%%%%%%%%%%%%%%%%%%%%%%%%%%%%%%%%%%%%%%%%%%%%%%%%%%%%%%%
\subsection{Heavy quark limit: spectral sums}
Now, assume that both quark masses are very large: $\alpha, \beta \gg 1$. This limit presumably admits a large $\alpha$ expansion for the eigenvalues along the lines of \cite{ZIYATDINOV:2010ModPhysA}. A way to test the agreement between the two methods would be to calculate the spectral sums, as it has been  demonstrated in \cite{Litvinov:2024riz} for the case of zero $\beta$. We put $\beta = \alpha \cdot \sigma$ with $\sigma$ of order $1$, and expand the spectral sums in $1/\alpha$. The spectral sums are expressed in terms of $\xi_k = \arctan \frac{\pi k}{2\beta},\, g_k = G(-2i\xi_k/\pi)$ and $G'(i)$, which have the expansions of the form
\begin{equation}
\begin{aligned}
    &\xi_k=\frac{\pi k}{2\alpha\sigma}-\frac{\pi^3k^3}{24\alpha^3\sigma^3}+\mathcal{O}\left(\frac{1}{\alpha^5}\right),
    \\ 
    &G(\nu)=\left(\frac{1-\sigma}{1+\sigma}\right)^{\frac{-i(\nu-i)}{2}} \left(1+\mathcal{O}\left(\frac{1}{\alpha}\right)\right),
    \\
    &G'(i)=\frac{i}{4}\left(-2\log\frac{1-\sigma}{1+\sigma}+\frac{\pi^2}{2\alpha}\frac{\sigma}{\sigma^2-1}+\mathcal{O}\left(\frac{1}{\alpha^2}\right)\right).
\end{aligned}
\end{equation}
It turns out that for two leading orders it is enough to know only the terms listed explicitly in the expansion for $G(\nu)$ and $G'(i)$. For example, the second and third spectral sums read as
\begin{equation} \label{gs-largealpha-direct}
\begin{aligned}
    &G_{\pm}^{(2)} =  \frac{\pi ^2 \left(2 \sigma -\left(\sigma ^2-1\right) \log \frac{1-\sigma}{1+\sigma}\right)}{4 \alpha  \sigma ^3} \pm \frac{\pi ^4 \left(2-\sigma ^2-2 \sqrt{(1-\sigma)(1+\sigma)}\right)}{4 \alpha ^2 \sigma ^4} + \mathcal{O}\left(\frac{1}{\alpha^3} \right), \\
    &G^{(3)}_{\pm} = \frac{\pi ^4 \left(6 \sigma-4 \sigma ^3+3(1- \sigma ^2) \log \left(\frac{1-\sigma }{\sigma +1}\right) \right)}{12 \alpha ^2 \sigma ^5} \pm \frac{\pi^6 \left(4-3\sigma^2 + (\sigma^2-4) \sqrt{(1-\sigma)(1+\sigma)} \right)}{4\alpha^3 \sigma^6} + \mathcal{O}\left(\frac{1}{\alpha^4} \right).
    \end{aligned}
\end{equation}
One can verify that in the limit $\sigma \to 0$ expressions from \cite{Litvinov:2024riz} reappear. 

Large $\alpha$ solution similar to \cite{ZIYATDINOV:2010ModPhysA} proceeds as follows (we will not perform the derivation in detail). In $\vartheta$-space, 't Hooft equation looks like (we introduce $A = \frac{\pi^2 \alpha}{2\lambda}$)
\begin{equation} \label{thooft-theta}
    \Omega(\vartheta) \phi(\vartheta) = \frac{\pi^2}{\alpha} \fint_{-\infty}^{\infty}\limits  \frac{d\vartheta'}{2\pi} \frac{\phi(\vartheta')}{\sinh^2 (\vartheta-\vartheta')} ,\quad \Omega(\vartheta) = 1 + \sigma \tanh \vartheta - \frac{1/A}{\cosh^2 \vartheta}.
\end{equation}
For large $\alpha$, at leading order ``WKB-type'' ansatz for $\phi$ can be used
\begin{equation}
    \phi(\vartheta) = \int_{-\infty}^{\infty}\limits d\eta\,e^{2i\alpha S(\eta)/\pi^2} \left(\frac{a_1}{\sinh (\vartheta + \eta - i0)} + \frac{a_2}{\sinh (\vartheta - \eta + i0)}\right).
\end{equation}
To cancel the singular part inside the integral in \eqref{thooft-theta}, the ``action'' must be
\begin{equation}
    S(\vartheta) = \int d\vartheta\, \Omega(\vartheta)  = \vartheta + \sigma \log \cosh \vartheta - \frac{1}{A} \tanh \vartheta.
\end{equation}
After that, the ``Bohr-Sommerfeld quantization condition'' can be derived as the leading order equation for the spectrum. It states that the action between the turning points
\begin{equation}
   \vartheta_{\pm}: \Omega(\vartheta_{\pm}) = 0,\quad \tanh \vartheta_{\pm} = \frac{1}{2}\left(-A \sigma \pm \sqrt{A^2 \sigma^2 - 4 A + 4} \right)
\end{equation}
for even and odd levels respectively is equal to
\begin{equation} \label{ziyat-quant-cond}
    \frac{1}{2}\left(S(\vartheta_-(\lambda_n^{\pm})) - S(\vartheta_+(\lambda_n^{\pm}))\right) = \frac{\pi^2}{2\alpha} \left(n - \left[\frac{1}{2} \pm \frac{1}{4} \right]\right),\quad n=1,2,\dots
\end{equation}
This equation receives subleading corrections in $1/\alpha$, which we will not study. Let us now derive the asymptotics for the spectral sums $G_{\pm}^{(s)} = \sum \limits_{n=1}^\infty (\lambda^{\pm}_n)^{-s}$ from this condition. For $1/\alpha$ expansion, we can use the Euler-Maclaurin formula
\begin{equation}\label{Euler-Maclaurin}
    \sum_{n=1}^{\infty}f_n\approx\int_0^{\infty}\limits dn\, f(n)-\frac{1}{2} f(0) = \int_{1/2 \pm 1/4}^{\infty}\limits dn\, f(n)-\frac{1}{2} f(0) + \int \limits_0^{1/2 \pm 1/4}  dn\, f(n).
\end{equation}
These terms should be enough to check two leading orders of the expansion for the spectral sums, as it was in the case $\beta = 0$ (for higher ones we need to take into account corrections to \eqref{ziyat-quant-cond} anyway). We will, however, only check the leading one. To rewrite the integral, we take a differential of \eqref{ziyat-quant-cond}:
\begin{equation}
    dn \frac{\pi^2}{2\alpha} = \frac{dA}{2} \left(\cancel{\Omega(\vartheta_-)} \frac{\partial \vartheta_-}{\partial A} - \cancel{\Omega(\vartheta_+)} \frac{\partial \vartheta_+}{\partial A} + \frac{1}{A^2} \left(\tanh \vartheta_- - \tanh \vartheta_+ \right) \right) = - \frac{dA}{2A^2} \sqrt{A^2 \sigma^2 - 4 A + 4}.
\end{equation}
The first two terms in the brackets are zero by the definition of $\vartheta_{\pm}$. Now, the first term in the r.h.s. of Euler-Maclaurin expansion \eqref{Euler-Maclaurin} applied to the spectral sums can be rewritten as
\begin{equation}
    \sum \limits_{n=1}^\infty (\lambda^{\pm}_n)^{-s} \approx \frac{1}{2} \left(\frac{\pi^2}{2\alpha} \right)^{s-1}  \int \limits_0^{A_0} dA\,A^{s-2}  \sqrt{A^2 \sigma^2 - 4 A + 4}, 
\end{equation}
where $A_0$ is the solution for \eqref{ziyat-quant-cond} with zero in the r.h.s., i.e. 
\begin{equation}
    \vartheta_-(A_0) = \vartheta_+(A_0),\quad A_0 = \frac{2}{\sigma^2} \left(1-\sqrt{1-\sigma^2}\right).
\end{equation}
The integral can be evaluated explicitly. For $s=2,3$ we have
\begin{equation}
\begin{aligned}
    &G^{(2)}_{\pm} \approx \frac{\pi ^2 \left(\sigma-\frac{1}{2}\left(\sigma ^2-1\right) \log \frac{1-\sigma}{1+\sigma} \right)}{2 \alpha  \sigma ^3},
    \\
    &G^{(3)}_{\pm} \approx \frac{\pi ^4 \left(3 \sigma-2 \sigma ^3-\frac{3}{2} \left(\sigma ^2-1\right) \log \frac{1-\sigma}{1+\sigma} \right)}{6 \alpha ^2 \sigma ^5},
\end{aligned}
\end{equation}
which agrees with \eqref{gs-largealpha-direct}.
%%%%%%%%%%%%%%%%%%%%%%%%%%%%%%%%%%%%%%%%%%%%%%%%%%%%%%%%%%%%%%%%%%%%%%%%%%%%%%%%%%%%%%%%%%%%%
\subsection{Heavy-heavy and heavy-light limit: WKB}
In this subsection we put $g=\sqrt{\pi}$ to conform with the reference \cite{Kochergin:2024quv}.

As derived in Section \ref{Analytical-results}, the spectrum of highly excited states in large-$N_c$ QCD$_2$ follows a linear pattern \eqref{wkb}. While this holds true for a fixed value of $m_{1,2}$ as $n\to\infty$, our focus now shifts to a different limit, where one of the masses approach $m_{i}\to\infty$ first, followed by $n\gg1$. We stress that these limits do not commute.

To study the limit $\alpha \to \infty$ in more detail, we would like to understand how to use our WKB expansion for levels with $n \lesssim \alpha$. It is not quite obvious: if $\alpha$ is large, for small enough $n$ the expansion parameter $\mathfrak{n}$ becomes negative and \eqref{wkb} cannot be used. For such $n$, the corresponding eigenvalue is itself of order $\alpha$, as can be argued from physical grounds. Since the expressions for phases $\Phi_{\pm}^{(k)}$ contain powers of $\alpha$ which grow with $k$, contributions from higher phases are no longer suppressed. The solution would be to resum the leading terms in \eqref{quantisation-condition-lambda} somehow. As it can be seen from the explicit expressions, these leading terms are of order $\frac{\alpha^{k+1}}{\lambda^k}$ (excluding the case $k=0$, which has a $\alpha \log \frac{\alpha}{\lambda}$ contribution), with the subleading terms suppressed by the additional factor $1/\alpha^2$ or less.

Recall first the case $\beta = 0$ ($1$ flavour heavy-heavy limit). The phases presented in \cite{Litvinov:2024riz} up to order $7$ together with the asymptotic expansions for the integrals $\mathtt{u}_k(\alpha)$, $\mathtt{i}_k(\alpha)$ allow us to write \eqref{quantisation-condition-lambda} as follows:
\begin{align}
    \frac{2\alpha}{\pi^2} \left(\frac{2}{A} + \log A - 1 - \log 4 + \frac{A}{4} \left[1+\frac{A}{4}+\frac{5 A^2}{48}+\frac{7 A^3}{128}+\frac{21 A^4}{640}+\frac{11 A^5}{512}+\frac{429 A^6}{28672}+\mathcal{O}\left(A^7\right)\right] \right) + \nonumber
    \\
    +\mathcal{O}\left(\frac{1}{\alpha} \right) = n + \frac{1}{2},
\end{align}
where $A = \frac{2\alpha}{\pi^2 \lambda}$. The expression in square brackets can be recognized as the first terms in Maclaurin expansion of the following hypergeometric series: ${} _3F_2\left(1,1,\frac{3}{2};2,3;A\right)$.  It turns out that the first term is precisely zero for $A = 1$; for $n$ of order $1$, this fixes the leading term of $\lambda_n$ in $1/\alpha$ expansion to be $\frac{2\alpha}{\pi^2}$. This corresponds to $M_n^2 = 4m^2 + \dots$, as expected in the heavy quark limit. Moreover, expansion near $A=1$ gives for the l.h.s.
\begin{equation} \label{a-to-1-quantization-cond}
    \frac{2\alpha}{\pi^2} \left(A^{-1}-1 \right)^{3/2} \left(\frac{4}{3}-\frac{2 (A^{-1}-1)}{5}+\frac{3}{14} (A^{-1}-1)^2 + \dots\right)+\mathcal{O}\left(\frac{1}{\alpha} \right) =  n + \frac{1}{2},
\end{equation}
which leads to
\begin{equation} \label{large-alpha-wkb}
    \frac{\lambda}{2\alpha/\pi^2} = 1 + \left(\frac{3\pi^2}{8 \alpha}(n+1/2) \right)^{2/3} + \dots
\end{equation}
This coincides with the first term of the ``relativistic'' expansion in \cite{ZIYATDINOV:2010ModPhysA}, applied for $1 \ll n \ll \alpha$. 

However, in the strict $\alpha \to \infty$ limit the meson masses are better described in terms of non-relativistic approximation, where the numerical coefficient is proportional to zeroes of the Airy function $\textrm{Ai}(z)$ or its derivative $\mathrm{a}_k$ and $\mathrm{a}_k'$ respectively\footnote{We note in Wolfram Mathematica Language they are spelled as $\mathrm{a}_k=\textrm{AiryAiZero}[k]$ and $\mathrm{a}_k'=\textrm{AiryAiPrimeZero}[k]$}. In particular
\begin{equation} \label{wkb-nonrel-asympt}
    \lim \limits_{\alpha \to \infty} \frac{\frac{\lambda_n}{2\alpha/\pi^2}-1}{\alpha^{-2/3}} = -\left(\frac{\pi}{2} \right)^{2/3}\cdot 
    \begin{cases}
        \mathrm{a}_{k+1}',\quad&n =2k; \\
        \mathrm{a}_{k+1},\quad&n=2k+1.
    \end{cases}
\end{equation}
Although \eqref{large-alpha-wkb} approximates this answer well enough as $n$ becomes large, we might expect that it should coincide with it exactly, since all the terms dropped from the quantization condition are subleading in $1/\alpha$. The resolution of this apparent paradox is that they still might contribute to the leading order answer if they are singular for $A \to 1$. For example, if they are of the form $\frac{a/\alpha}{(A^{-1}-1)^{3/2}}$, the leading order solution to \eqref{a-to-1-quantization-cond} is
\begin{equation}
    A^{-1}-1 \approx \left(\frac{3\pi^2}{8 \alpha}\right)^{2/3} \left(n + \frac{1}{2} + \frac{a \cdot(8/3\pi^2)}{n+1/2} \right)^{2/3} \approx \left(\frac{3\pi^2}{8 \alpha}(n+1/2)\right)^{2/3} \left(1 + \frac{16 a}{9 \pi^2 (n+1/2)^2} + \dots \right).
\end{equation}
The $n$-dependence conforms with the first correction term in the asymptotic series for the zeroes of the Airy function. It would be interesting to resum $\mathcal{O}(1/\alpha)$ corrections to the phases $\Phi^{(k)}_{\pm}$ explicitly to check \eqref{wkb-nonrel-asympt} from our approach.

Now let us go back to the case $\beta \neq 0$, but $\alpha$ and $\beta$ are both large. One can note that there is a specific case where the question greatly simplifies: $\alpha = \beta$. The leading terms in \eqref{phases} for $k \geq 1$ are all proportional to $(\alpha^2 - \beta^2)$ (we conjecture that it holds for all $k$) and thus disappear. 

This value of $\beta$ gives one case when the ``heavy-light'' approximation to `t Hooft equation (see e.g. \cite{Burkardt:2000PhysRevD, Kochergin:2024quv}) is applicable. In this limit, meson masses are known to have the following expansion in $1/M$, $M \approx \sqrt{2\alpha}$ being the mass of the heavy quark: 
\begin{equation}\label{heavy-light-mass-def}
    M_n^2 = M^2 + M \epsilon_n + \dots,    
\end{equation}
where $\epsilon_n$ is finite in $M \to \infty$ limit and is given by the spectrum of a different integral equation. The explicit expression for the subleading term was announced without derivation in \cite{Zhitnitsky:1995qa}. In \cite{Bigi:1998kc,Burkardt:2000PhysRevD}, an integral expression for this correction, as well as other $1/M$ corrections, is obtained. However, this expression does not help to obtain an analytic form of $\epsilon_n$, since it requires knowledge of the wave function of the meson state. With our approach, we establish an analytic dependence of the energy on the state number $n$ and find that it agrees with \cite{Zhitnitsky:1995qa}, but our numerical factor is different and we believe it is correct (see below).

Our quantization condition \eqref{quantisation-condition-lambda} at leading order in $\alpha$ for this case is rewritten as follows:
\begin{equation}
    \frac{1}{8} + \frac{2\alpha}{\pi^2} \left(\frac{2}{A} + \log \frac{A}{2} - 1 \right) + \mathcal{O}\left(\frac{1}{\alpha} \right) = n + \frac{3}{4}.
\end{equation}
The term in the brackets in this case is zero for $A=2$, or $\lambda = \frac{\alpha}{\pi^2}$. This means that $M_n^2 = M^2 + \dots$, as expected. For small $n$, expanding the l.h.s. in $A-2$, we obtain the following expression for the first corrections in $\alpha$
\begin{equation} \label{epsilon-m=1}
    \lambda_n \approx \frac{\alpha}{\pi^2} + \sqrt{\frac{\alpha}{\pi^2} \left(n+\frac{5}{8} \right)}\quad\text{ or }\quad M_n^2 \approx  M^2 +\sqrt{2}\pi M \sqrt{n+\frac{5}{8}}.
\end{equation}
This gives an analytic prediction for $\epsilon_n$ in this particular case.

\begin{figure}[h!]
    \centering
    \includegraphics[width=0.49\linewidth]{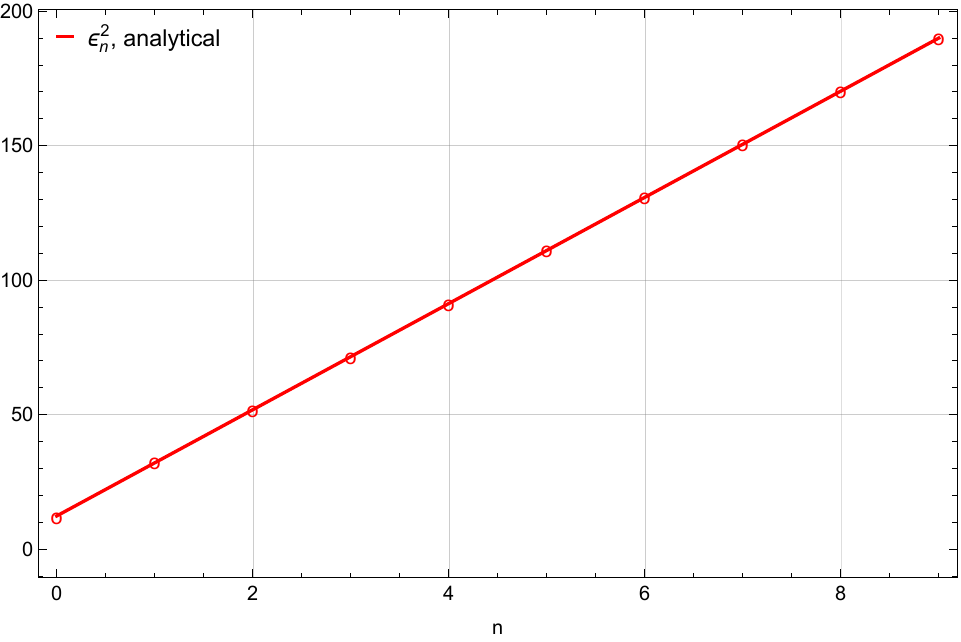}
    \includegraphics[width=0.49\linewidth]{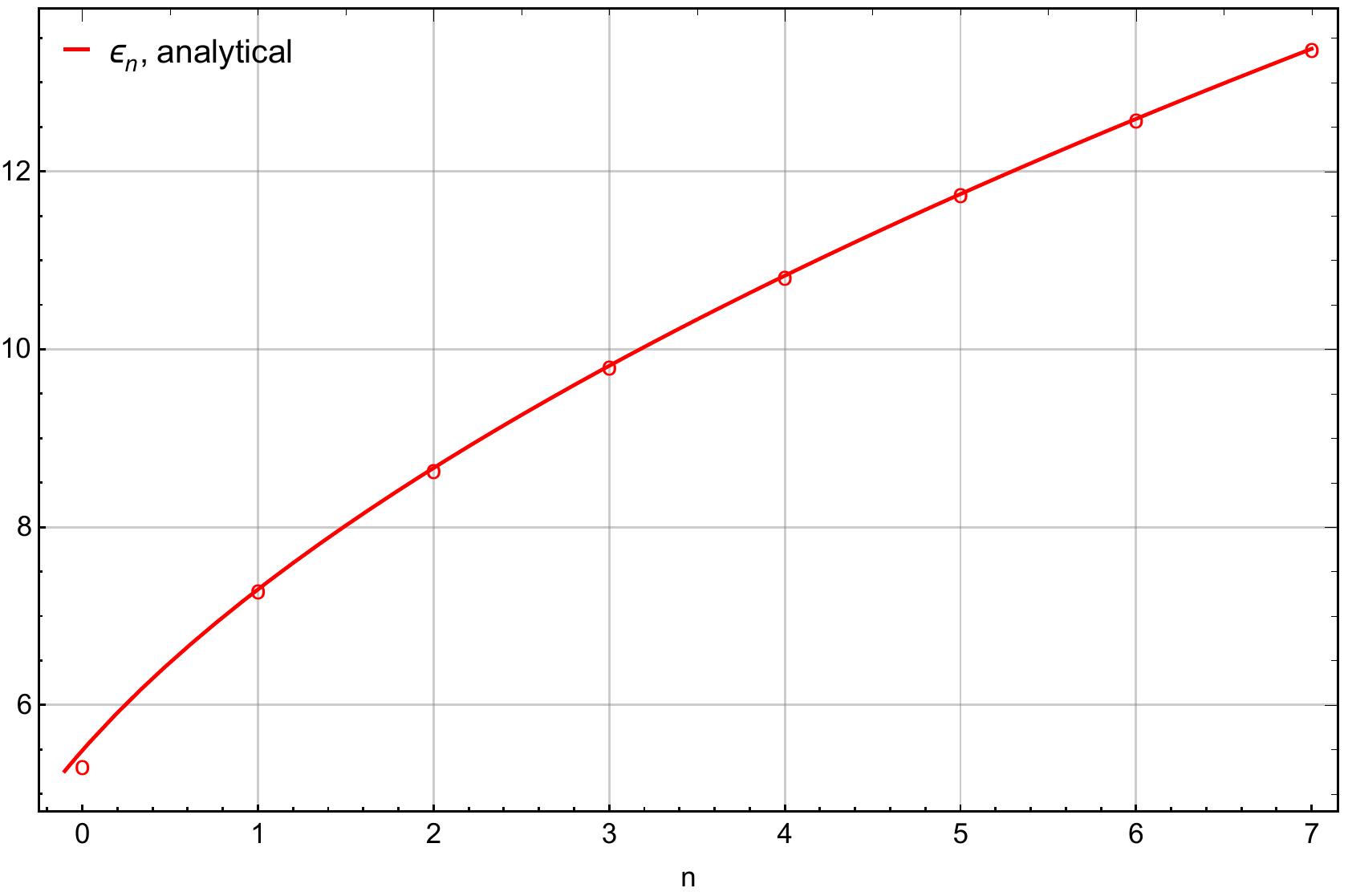}
    \caption{Left: square of the first correction to the meson masses $M_n$ in the heavy-light limit, case $\alpha_1 = 0$, for the first 10 levels, right: first correction to masses for $\alpha_1 = 3$, first 8 levels.  Circles are numerical results; continuous line is the analytic prediction \eqref{epsilon-m=1} and \eqref{epsilon-general-m}. }
    \label{fig:heavy-light}
\end{figure}

This prediction is supported by rough numerical analysis of the heavy-light approximation for `t Hooft equation written in \cite{Burkardt:2000PhysRevD, Kochergin:2024quv}\footnote{We thank Ilia Kochergin for sharing his program to perform the numerics.}. Figure \ref{fig:heavy-light} on the left illustrates this. Again there is a noticeable discrepancy (of order 5 percent) for the first few levels for reasons analogous to ones noted after \eqref{large-alpha-wkb}; for increasingly larger $n$ the agreement becomes better and better. 

For different values of $\beta$, one can note that the leading terms in $\Phi^{(k)}_{\pm}$ coincide with linear in $y$ terms in the coefficients $T^{(k)}(y)$ defined in \eqref{Ts-coefs}. Resumming them thus yields the function $\alpha_1 f\left(\frac{\alpha_2}{2\pi^2 \lambda}, \frac{\alpha_1}{\alpha_2} \right)$ that we know explicitly \eqref{f-funtion-explicit}.

Let us study the heavy-light limit for different values of light quark mass using this observation. Put $\alpha - \beta = \alpha_1$, $\alpha_1$ of order 1. Up to sub-leading orders in $1/\alpha$, sum of phases gives a simple specialization of the function $f$ when the second argument is zero. The quantization condition now looks as follows
\begin{equation}
    \frac{2\alpha}{\pi^2} \left(\frac{2}{A} + \log \frac{A}{2} - 1 \right) - \frac{2 \alpha_1}{\pi^2} \log \left(\sqrt{\alpha}\left(1 - \frac{A}{2}\right) \right)  + \mathcal{O}\left(\frac{1}{\alpha} \right) = n + \frac{5}{8} + \frac{\alpha_1}{\pi^2} \log \left(\frac{8e^{\gamma_E}}{\pi} \right)- \frac{\alpha_1^2 \mathtt{i}_2 (\alpha_1)}{4\pi^2}.
\end{equation}
We denote the r.h.s. as $\mathfrak{n}$. For $n \ll \alpha$ the solution again is expected to take the form 
\begin{equation}
    \lambda_n = \frac{\alpha}{\pi^2}+\sqrt{\frac{\alpha}{\pi^2}}\cdot\delta\lambda_n.
\end{equation}
Expansion in $\delta \lambda$ of the l.h.s. gives an equation that can be solved by iteration method:
\begin{equation} \label{epsilon-general-m}
    \delta \lambda_n^2 - \frac{2\alpha_1}{\pi^2} \log (\pi\, \delta\lambda_n) + \mathcal{O}\left(\frac{1}{\sqrt{\alpha}} \right) = \mathfrak{n}\quad\Rightarrow\quad \delta \lambda_n = \sqrt{\mathfrak{n} + \frac{2\alpha_1}{\pi^2} \log \left(\pi \sqrt{\mathfrak{n}+ \dots} \right)}.
\end{equation}
In \eqref{heavy-light-mass-def} $\epsilon_n$ differs from $\delta \lambda_n$ by the factor $\pi \sqrt{2}$. In \cite{Zhitnitsky:1995qa}, $\epsilon_n=4\sqrt{\pi n}(1+\mathcal{O}(\frac{\log{n}}{n}))$ was announced. The functional dependence on $n$ agrees with our result, but the numerical factor is different. Our analytic prediction, again, is consistent with the numerics (see  Fig. \ref{fig:heavy-light} on the right for the illustration). In this case for finite $\alpha_1$ we do not have a simple formula like \eqref{wkb-nonrel-asympt} that would allow us to calculate more precisely lower levels, in particular, the ground state. Resuming sub-leading corrections to the quantization condition thus becomes even more  interesting problem.
%%%%%%%%%%%%%%%%%%%%%%%%%%%%%%%%%%%%%%%%%%%%%%%%%%%%%%%%%%%%%%%%%%%%%%%%%%%%%%%%%%%%%%%%%%%%%%%%%%%%%%%%%%%%%%%%%%%%%%%%%%%%%%%%%%%%%%%%%%%%%%%%%%%%%%%%%%%%%%%%%%%%%%%%%%%%%%%%%%%%%%%%%%%%%%%%
\section{Numerical results}\label{numerics}
\subsection{Method}
To validate our analytical expressions for the spectral sums and the large$-n$ expansion of the spectrum, we decompose the wave function $\phi(x)$ using the following basis\footnote{Of course, the choice of basis is not the only one. See for example \cite{Chabysheva:2012fe}, where Jacobi polynomials were used.} \cite{Hanson:1976ey} 
\begin{equation}\label{Chebyshev_basis}
    \phi_{12}(x)=\sum_{n=1}^{N}\phi_n h_n(x) ,\quad h_n(x)=\sin{(n\theta)}=2\sqrt{x(1-x)}U_{n-1}(1-2x).
\end{equation}
Here we introduce a new variable $\theta \in [0, \pi]$ with $x = \frac{1 - \cos{\theta}}{2}$, where $U_n(x)$ are the Chebyshev polynomials of the second kind. This transformation allows us to reformulate the 't Hooft integral equation into a matrix problem, which, when truncated to a sufficiently large $N$, becomes suitable for numerical solution
\begin{equation}\label{tHooft-matrixform}
    2\pi^2\lambda\sum_{n=1}^{N} M_{mn}\phi_n =\sum_{n=1}^N H_{mn}\phi_n, \quad M_{mn}=\braket{h_m|h_n},\quad H_{mn}=\bra{h_m}\mathcal{H}\ket{h_n}.
\end{equation}
Here the scalar product takes a conventional form
\begin{equation}\label{scalar-product}
    \braket{h|g}=\int_{0}^{1}\limits dx\; h(x)g(x).
\end{equation}
Let us emphasize that our basis functions do not satisfy the boundary asymptotic behavior \eqref{phi-boundary-conditions}. To address this issue, it is common practice to add two additional functions to the basis \cite{THOOFT1974461}
\begin{equation}\label{extra_functions}
    h_{N+1}(x)=x^{\beta_1}(1-x)^{2-\beta_1},\quad h_{N+2}(x)=x^{2-\beta_2}(1-x)^{\beta_2}.
\end{equation}
However, numerical calculations show (see Table \ref{Table-WKB}) that in the region of intermediate values of masses $m_i$ (i.e.,  $\alpha_i$ not close to $-1$ and not excessively large), it is not necessary to extend the basis to achieve a reasonable accuracy\footnote{The use of functions \eqref{extra_functions} becomes critically important when one of the masses is small (e.g., $m_1\ll g$). This is because, near the boundary point $0$, the wave function $\phi_{12}(0+)\approx \text{const}$ ($\phi_{12}(0) = 0$), while the basis functions $h_n(x)$ vary rapidly.}.

Both matrices $\bm{M}$ and $\bm{H}$ have an integral representation and can be calculated  analytically \cite{Brower:1979PhysRevD}
\begin{equation}
    M_{mn}=\int_0^1\limits dx\;\sin{(m\theta)}\sin{(n\theta)}=-\frac{1+(-1)^{m+n}}{2}\frac{2mn}{m^4+(n^2-1)^2-2m^2(1+n^2)}
\end{equation}
and
\begin{equation}\label{H-matrix-Chebyshev}
    H_{mn}=\underbrace{\int_0^1\limits dx\;\sin{(m\theta)}\sin{(n\theta)}\;\left(\frac{\alpha_1}{x}+\frac{\alpha_2}{1-x}\right)}_{H^{(1)}_{mn}}-\underbrace{\int_0^{1}\limits dx\sin{(m\theta)}\fint_0^1\limits dy\frac{\sin{(n\theta')}}{(x-y)^2}}_{H^{(2)}_{mn},\quad y=\frac{1-\cos{\theta'}}{2}}.
\end{equation}
Note that the matrix $\bm{H}$ is Hermitian. This is due to the fact that we choose the scalar product \eqref{scalar-product} in $x$-space. But it is not the case in $\theta$-space \cite{Brower:1979PhysRevD}.

The first integral $H^{(1)}_{mn}$ in \eqref{H-matrix-Chebyshev} has the form
\begin{equation}
\begin{aligned}
    H^{(1)}_{mn}=&\;\alpha_1\int_0^{\pi}\limits d\theta\;\frac{\sin{(m\theta)}\sin{(n\theta)}\sin{\theta}}{1-\cos{\theta}}+\alpha_2\int_0^{\pi}\limits d\theta\;\frac{\sin{(m\theta)}\sin{(n\theta)}\sin{\theta}}{1+\cos{\theta}}=
    \\=\;&(\alpha_1+(-1)^{n+m}\alpha_2)\sum_{l=\frac{|m-n|}{2}+1}^{\frac{m+n}{2}}\limits\frac{2}{2l-1}=(\alpha_1+(-1)^{n+m}\alpha_2)\left(\psi\left(\frac{m+n}{2}+\frac{1}{2}\right)-\psi\left(\frac{|m-n|}{2}+\frac{1}{2}\right)\right),
    \end{aligned}
\end{equation}
while the second integral $H^{(2)}_{mn}$ is simply 
\begin{equation}
    H^{(2)}_{mn}=\int_0^1\limits dx\sin{(m\theta)}\fint_0^1\limits dy\frac{\sin{(n\theta')}}{(x-y)^2}=-\pi n\int_0^{\pi}\limits d\theta\;\sin{(m\theta)}\sin{(n\theta)}=-\frac{n\pi^2}{2}\delta_{mn}.
\end{equation}
Thus the matrix elements of $\bm{H}$ are  
\begin{equation}
    H_{mn}=\frac{n\pi^2}{2}\delta_{mn}+H^{(1)}_{mn}.
\end{equation}

The described method converges very quickly as $N$ -- the number of basis functions $h_n$ increases. However, the convergence deteriorates for very small nonzero values of $m_j$ and for very large $m_j$. For the numerical analysis of the chiral limit and the heavy-light limit, a different basis of functions must be used (see \cite{Kochergin:2024quv} for details).
%%%%%%%%%%%%%%%%%%%%%%%%%%%%%%%%%%%%%%%%%%%%%%%%%%%%%%%%%%%%%%%%%%%%%%%%%%%%%%%%%%%%%%%%%%%%%%%%%%%%%%%%%%%%%%%%%%%%%%%%%%%%%%%%%%%%%%%%%%%%%%%%%%%%%%%%%%%%%%%%%%%%%%%%%%%%%%%%%%%%%%%%%%%%%%%%
\subsection{Numerical data}
\begin{figure}[h!]
    \centering
    \includegraphics[width=0.6\linewidth]{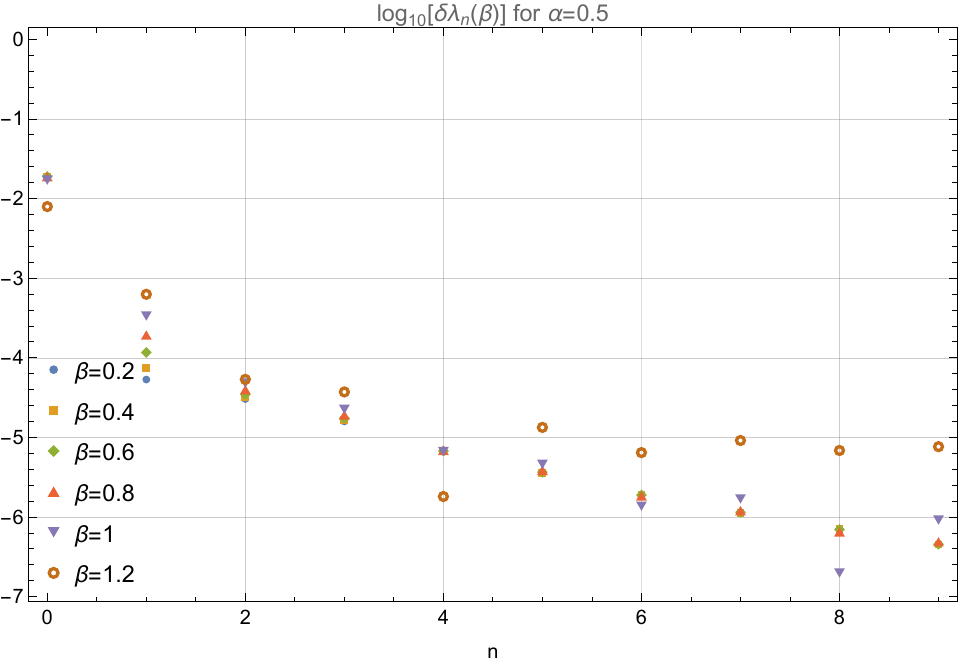}
    \caption{Relative error $\delta\lambda_n(\beta)$ in $\log_{10}$ axes for $\alpha=0.5$. As the number $n$ of the eigenstate increases, the analytical formula \eqref{wkb} becomes more accurate. For the value $\beta=1.2$ we can see that the accuracy decreases significantly (for large $n>4$). This is explained by the fact that the numerical method used ceases to work when one of the masses $m_i \to 0$ ($\alpha_i \to -1$).} 
    \label{fig:d-lambda-error}
\end{figure}

To numerically verify \eqref{wkb}, we employed a basis of $N = 1000$ functions $h_n$ \eqref{Chebyshev_basis}. Table \ref{Table-WKB} displays the numerical and analytical values of the first $10$ eigenvalues for $\alpha=0.5$ and different values of $\beta$, along with their relative errors. Notably, the accuracy is already quite high even at $n = 1$.
\begin{table}[h!]
\begin{center}
    \resizebox{\textwidth}{!}{
    \begin{tabular}{| c | l | c | l|| l | c | l|| l | c | l|| l | c | l|| l | c | l|| l | c | l|}
    \hline \rule{0mm}{3.6mm}
    $\alpha$ & \multicolumn{18}{c|}{$0.5$}\\
    \hline \rule{0mm}{3.6mm}
    $\beta$ & \multicolumn{3}{c||}{$\pm 0.1$} & \multicolumn{3}{c||}{$\pm0.2$} & \multicolumn{3}{c||}{$\pm 0.3$} & \multicolumn{3}{c|}{$\pm0.4$} & \multicolumn{3}{c|}{$\pm0.5$} & \multicolumn{3}{c|}{$\pm0.6$}\\
    \hline
    $n$ & $\lambda_{n}^{(3)}$ & $\delta\lambda$ &  $\lambda_{n}^{(\text{num})}$ & $\lambda_{n}^{(3)}$ & $\delta\lambda$ & $\lambda_{n}^{(\text{num})}$ & $\lambda_{n}^{(3)}$ & $\delta\lambda$ & $\lambda_{n}^{(\text{num})}$ & $\lambda_{n}^{(3)}$ & $\delta\lambda$ & $\lambda_{n}^{(\text{num})}$ & $\lambda_{n}^{(3)}$ & $\delta\lambda$ & $\lambda_{n}^{(\text{num})}$ & $\lambda_{n}^{(3)}$ & $\delta\lambda$ & $\lambda_{n}^{(\text{num})}$\\
    \hline
    $0$ & 0.49373 & $1.8\cdot10^{-2}$ & 0.50280 & 0.49244 & $1.8\cdot10^{-2}$ & 0.50152 & 0.49027 & $1.9\cdot10^{-2}$ & 0.49936 & 0.48720 & $1.9\cdot10^{-2}$ & 0.49630 & 0.48318 & $1.9\cdot10^{-2}$ & 0.49228 & 0.47817 & $1.9\cdot10^{-2}$ & 0.48724 \\
    $1$ & 1.05305 & $4.8\cdot10^{-5}$ & 1.05310 & 1.05156 & $5.3\cdot10^{-5}$ & 1.05162 & 1.04907 & $6.2\cdot10^{-5}$ & 1.04913 & 1.04552 & $7.5\cdot10^{-5}$ & 1.04560 & 1.04086 & $9.2\cdot10^{-5}$ & 1.04096 & 1.03500 & $1.2\cdot10^{-4}$ & 1.03512 \\
    $2$ & 1.57000 & $3.0\cdot10^{-5}$ & 1.56995 & 1.56852 & $3.0\cdot10^{-5}$ & 1.56847 & 1.56603 & $3.1\cdot10^{-5}$ & 1.56598 & 1.56250 & $3.2\cdot10^{-5}$ & 1.56245 & 1.55786 & $3.3\cdot10^{-5}$ & 1.55781 & 1.55204 & $3.5\cdot10^{-5}$ & 1.55198 \\
    $3$ & 2.08652 & $1.6\cdot10^{-5}$ & 2.08656 & 2.08502 & $1.6\cdot10^{-5}$ & 2.08505 & 2.08248 & $1.6\cdot10^{-5}$ & 2.08252 & 2.07888 & $1.6\cdot10^{-5}$ & 2.07892 & 2.07416 & $1.7\cdot10^{-5}$ & 2.07419 & 2.06823 & $1.7\cdot10^{-5}$ & 2.06827\\
    $4$ & 2.59655 & $6.6\cdot10^{-6}$ & 2.59653 & 2.59504 & $6.6\cdot10^{-6}$ & 2.59503 & 2.59251 & $6.7\cdot10^{-6}$ & 2.59249 & 2.58891 & $6.7\cdot10^{-6}$ & 2.58890 & 2.58419 & $6.7\cdot10^{-6}$ & 2.58418 & 2.57827 & $6.8\cdot10^{-6}$ & 2.57826 \\
    $5$ & 3.10652 & $3.5\cdot10^{-6}$ & 3.10653 & 3.10500 & $3.5\cdot10^{-6}$ & 3.10502 & 3.10245 & $3.6\cdot10^{-6}$ & 3.10247 & 3.09883 & $3.6\cdot10^{-6}$ & 3.09884 & 3.09408 & $3.6\cdot10^{-6}$ & 3.09409 & 3.08812 & $3.7\cdot10^{-6}$ & 3.08813 \\
    $6$ & 3.61365 & $1.9\cdot10^{-6}$ & 3.61364 & 3.61213 & $1.9\cdot10^{-6}$ & 3.61213 & 3.60958 & $1.9\cdot10^{-6}$ & 3.60958 & 3.60596 & $1.9\cdot10^{-6}$ & 3.60596 & 3.60121 & $1.9\cdot10^{-6}$ & 3.60121 & 3.59526 & $1.9\cdot10^{-6}$ & 3.59525 \\
    $7$ & 4.12080 & $1.1\cdot10^{-6}$ & 4.12080 & 4.11928 & $1.1\cdot10^{-6}$ & 4.11928 & 4.11672 & $1.1\cdot10^{-6}$ & 4.11672 & 4.11308 & $1.1\cdot10^{-6}$ & 4.11309 & 4.10831 & $1.1\cdot10^{-6}$ & 4.10832 & 4.10233 & $1.1\cdot10^{-6}$ & 4.10234 \\
    $8$ & 4.62634 & $7.2\cdot10^{-7}$ & 4.62633 & 4.62482 & $7.1\cdot10^{-7}$ & 4.62481 & 4.62226 & $7.1\cdot10^{-7}$ & 4.62225 & 4.61862 & $7.1\cdot10^{-7}$ & 4.61862 & 4.61386 & $7.0\cdot10^{-7}$ & 4.61385 & 4.60788 & $7.0\cdot10^{-7}$ & 4.60788 \\
    $9$ & 5.13191 & $4.5\cdot10^{-7}$ & 5.13191 & 5.13038 & $4.5\cdot10^{-7}$ & 5.13038 & 5.12782 & $4.5\cdot10^{-7}$ & 5.12782 & 5.12417 & $4.5\cdot10^{-7}$ & 5.12418 & 5.11940 & $4.5\cdot10^{-7}$ & 5.11940 & 5.11341 & $4.6\cdot10^{-7}$ & 5.11341\\
    \hline \rule{0mm}{3.6mm}
    $\beta$ & \multicolumn{3}{c||}{$\pm 0.7$} & \multicolumn{3}{c||}{$\pm0.8$} & \multicolumn{3}{c||}{$\pm 0.9$} & \multicolumn{3}{c|}{$\pm1$} & \multicolumn{3}{c|}{$\pm1.1$} & \multicolumn{3}{c|}{$\pm1.2$}\\
    \hline
    $n$ & $\lambda_{n}^{(3)}$ & $\delta\lambda$ &  $\lambda_{n}^{(\text{num})}$ & $\lambda_{n}^{(3)}$ & $\delta\lambda$ &  $\lambda_{n}^{(\text{num})}$ & $\lambda_{n}^{(3)}$ & $\delta\lambda$ &  $\lambda_{n}^{(\text{num})}$ & $\lambda_{n}^{(3)}$ & $\delta\lambda$ & $\lambda_{n}^{(\text{num})}$ & $\lambda_{n}^{(3)}$ & $\delta\lambda$ & $\lambda_{n}^{(\text{num})}$ & $\lambda_{n}^{(3)}$ & $\delta\lambda$ & $\lambda_{n}^{(\text{num})}$\\
    \hline
    $0$ & 0.47210 & $1.9\cdot10^{-2}$ & 0.48107 & 0.46488 & $1.9\cdot10^{-2}$ & 0.47364 & 0.45640 & $1.8\cdot10^{-2}$ & 0.46476 & 0.44653 & $1.7\cdot10^{-2}$ & 0.45415 & 0.43518 & $1.4\cdot10^{-2}$ & 0.44137 & 0.42234 & $8.0\cdot10^{-3}$ & 0.42569 \\
    $1$ & 1.02784 & $1.5\cdot10^{-4}$ & 1.02799 & 1.01920 & $1.9\cdot10^{-4}$ & 1.01940 & 1.00885 & $2.5\cdot10^{-4}$ & 1.00911 & 0.99647 & $3.4\cdot10^{-4}$ & 0.99680 & 0.98151 & $4.5\cdot10^{-4}$ & 0.98196 & 0.96310 & $6.2\cdot10^{-4}$ & 0.96371 \\
    $2$ & 1.54492 & $3.7\cdot10^{-5}$ & 1.54486 & 1.53635 & $4.0\cdot10^{-5}$ & 1.53629 & 1.52611 & $2.0\cdot10^{-5}$ & 1.52605 & 1.51388 &  $2.3\cdot10^{-5}$ & 1.51381 & 1.49916 & $3.7\cdot10^{-5}$ & 1.49908 & 1.48111 & $5.4\cdot10^{-6}$ & 1.48103 \\
    $3$ & 2.06099 & $1.8\cdot10^{-5}$ & 2.06103 & 2.05227 & $1.9\cdot10^{-5}$ & 2.05231 & 2.04186 & $2.0\cdot10^{-5}$ & 2.04190 & 2.02943 & $2.3\cdot10^{-5}$ & 2.02948 & 2.01448 & $2.6\cdot10^{-5}$ & 2.01453 & 1.99614 & $3.7\cdot10^{-5}$ & 1.99622 \\
    $4$ & 2.57104 & $6.8\cdot10^{-6}$ & 2.57102 & 2.56234 & $6.9\cdot10^{-6}$ & 2.56232 & 2.55195 & $6.9\cdot10^{-6}$ & 2.55193 & 2.53955 & $6.7\cdot10^{-6}$ & 2.53953 & 2.52464 & $5.2\cdot10^{-6}$ & 2.52463 & 2.50639 & $1.8\cdot10^{-6}$ & 2.50639 \\
    $5$ & 3.08084 & $3.7\cdot10^{-6}$ & 3.08085 & 3.07208 & $3.9\cdot10^{-6}$ & 3.07209 & 3.06162 & $4.1\cdot10^{-6}$ & 3.06164 & 3.04915 & $4.6\cdot10^{-6}$ & 3.04917 & 3.03416 & $6.3\cdot10^{-6}$ & 3.03418 & 3.01581 & $1.3\cdot10^{-5}$ & 3.01585 \\
    $6$ & 3.58798 & $1.9\cdot10^{-6}$ & 3.58797 & 3.57923 & $1.8\cdot10^{-6}$ & 3.57922 & 3.56879 & $1.7\cdot10^{-6}$ & 3.56878 & 3.55633 & $1.4\cdot10^{-6}$ & 3.55632 & 3.54136 & $8.4\cdot10^{-8}$ & 3.54136 & 3.52304 & $6.5\cdot10^{-6}$ & 3.52307 \\
    $7$ & 4.09503 & $1.1\cdot10^{-6}$ & 4.09504 & 4.08625 & $1.2\cdot10^{-6}$ & 4.08626 & 4.07577 & $1.3\cdot10^{-6}$ & 4.07578 & 4.06328 & $1.7\cdot10^{-6}$ & 4.06328 & 4.04827 & $3.1\cdot10^{-6}$ & 4.04828 & 4.02990 & $9.1\cdot10^{-6}$ & 4.02993 \\
    $8$ & 4.60058 & $6.8\cdot10^{-7}$ & 4.60058 & 4.59181 & $6.5\cdot10^{-7}$ & 4.59180 & 4.58133 & $5.5\cdot10^{-7}$ & 4.58133 & 4.56885 & $2.0\cdot10^{-7}$ & 4.56885 & 4.55385 & $1.4\cdot10^{-6}$ & 4.55386 & 4.53550 & $6.9\cdot10^{-6}$ & 4.53554 \\
    $9$ & 5.10609 & $4.6\cdot10^{-7}$ & 5.10609 & 5.09730 & $4.9\cdot10^{-7}$ & 5.09730 & 5.08680 & $5.8\cdot10^{-7}$ & 5.08681 & 5.07429 & $9.2\cdot10^{-7}$ & 5.07430 & 5.05927 & $2.2\cdot10^{-6}$ & 5.05928 & 5.04089 & $7.7\cdot10^{-6}$ & 5.04093 \\
    \hline
    \end{tabular}}
\end{center}
\caption{Values of eigenvalues $\lambda_{n}$ from the large-$\lambda$ expansion for $\alpha=0.5$ and several values of $\beta$. The $\lambda_{n}^{(3)}$ column is obtained from \eqref{wkb} with truncation of the sum behind the term $\propto\lambda^{-3}$. The ratio $\delta\lambda=\frac{|\lambda_{n}^{(3)}-\lambda_{n}^{(\rm{num})}|}{\lambda_{n}^{(3)}}$ shows the relative error.
The column $\lambda_{n}^{(\text{num})}$ shows the eigenvalues calculated by direct numerical solution of \eqref{tHooft-matrixform}.}
\label{Table-WKB}
\end{table}
Fig. \ref{fig:d-lambda-error} shows how the accuracy of large-$n$ expansion \eqref{wkb} grows while $n$ increases.

We also performed a numerical verification of the relation \eqref{dm/dp} (see Table \ref{dm/dp-table}). To do this, we applied the relation \eqref{dp/dm-prod}, using the first $n$ eigenvalues as $\lambda^{(\rm{num})}$. 
\begin{table}[h!]
    \begin{center}
        \begin{tabular}{| c | c | c | c | c |}
        \hline 
        %\rule{0mm}{3.6mm}\\
        \multicolumn{5}{|c|}{$\alpha=0.5$}\\
        \hline
        $\beta $ & $n$ & $(d_-/d_+)^{\rm{num}}$ & $(d_-/d_+)^{\rm{an}}$ & $\delta d$\\
        \hline
        $\pm0.1$ & 312 & 0.550612 & 0.551022 & $7.4\cdot10^{-4}$ \\
        $\pm0.2$ & 312 & 0.549688 & 0.550097 & $7.4\cdot10^{-4}$ \\
        $\pm0.3$ & 311 & 0.548132 & 0.548537 & $7.4\cdot10^{-4}$ \\
        $\pm0.4$ & 310 & 0.545914 & 0.546315 & $7.3 \cdot10^{-4}$ \\
        $\pm0.6$ & 311 & 0.539313 & 0.539699 & $7.2\cdot10^{-4}$ \\
        $\pm0.8$ & 310 & 0.529298 & 0.529664 & $6.9\cdot10^{-4}$\\
        $\pm1.0$ & 313 & 0.514697 & 0.515036 & $6.6\cdot10^{-4}$ \\
        $\pm1.2$ & 312 & 0.492833 & 0.493124 & $5.9\cdot10^{-4}$\\
        \hline 
        \end{tabular}
        \begin{tabular}{| c | c | c | c | c |}
        \hline 
        %\rule{0mm}{3.6mm}\\
        \multicolumn{5}{|c|}{$\alpha=1.5$}\\
        \hline
        $\beta $ & $n$ & $(d_-/d_+)^{\rm{num}}$ & $(d_-/d_+)^{\rm{an}}$ & $\delta d$\\
        \hline
        $\pm0.2$ & 312 & 0.709687 & 0.711192 & $2.1\cdot10^{-3}$ \\
        $\pm0.4$ & 311 & 0.707966 & 0.709466 & $2.1\cdot10^{-3}$ \\
        $\pm0.6$ & 311 & 0.705054 & 0.706543 & $2.1\cdot10^{-3}$ \\
        $\pm0.8$ & 311 & 0.700872 & 0.702344 & $2.1\cdot10^{-3}$ \\
        $\pm1.4$ & 313 & 0.679183 & 0.680560 & $2.0\cdot10^{-3}$ \\
        $\pm1.6$ & 312 & 0.667944 & 0.669279 & $2.0\cdot10^{-3}$ \\
        $\pm1.8$ & 311 & 0.653765 & 0.655049 & $2.0\cdot10^{-3}$ \\
        $\pm2.0$ & 311 & 0.635404 & 0.636620 & $1.9\cdot10^{-3}$ \\
        \hline 
        \end{tabular}
    \end{center}
    \caption{Numerical and analytical \eqref{dm/dp} values for ratio $d_-/d_+$ for several values of $\alpha,\beta$. The ratio $\delta d=\left|\frac{(d_-/d_+)^{\rm{num}}-(d_-/d_+)^{\rm{an}}}{(d_-/d_+)^{\rm{an}}}\right|$ shows the relative error.
    \label{dm/dp-table}}
\end{table}

To numerically verify the spectral sums \eqref{G1pm-analytical}-\eqref{G3pm-analytical}, we utilize the same basis of $N = 1000$ functions $h_n$, which ensures an accuracy limited by a cutoff of at least $\frac{1}{s-1} N^{-s+1} $ (for $s>1$). However, this accuracy is not particularly high, especially for small values of $s$, as the spectral sums exhibit slow convergence. For instance, to achieve an accuracy of $7$ decimal places for $G^{(2)}_{\pm}$, it is necessary to sum the first $10^7$ even or odd spectral values, which requires a basis of $2\cdot10^7$ functions. However, this is not the only challenge; at sufficiently large-$n$, the eigenvalues $\lambda^{(\rm{num})}_n$ obtained through this method start to deviate noticeably from the true eigenvalues. When numerically evaluating \eqref{spectral_sums_def}, starting from a sufficiently large-$n$ (determined through trial and error, we set $n=300$), we use instead their asymptotic form\footnote{We do not employ the highest available precision but rather retain only the leading terms with an accuracy of $\mathcal{O}(\frac{\log n}{n})$. This approximation produces an error of the order of $\mathcal{O}(\frac{\log n}{n^{s+2}})$.} instead of the true values $\lambda^{(\rm{num})}_n$. As a result, we calculated eigenvalues up to the $10^5$th and used them to numerically compute the spectral sums (see Tables \ref{Gpm-table}).
\begin{table}[h!] 
    \begin{center}
        %\resizebox{\textwidth}{!}{
        \resizebox{\textwidth}{4.2cm}{
        \begin{tabular}{| c | c | c | c || c | c | c || c | c | c || c | c | c || c | c | c ||}
        \hline
        $\alpha$ & \multicolumn{12}{c|}{$0.5$}\\
        \hline
        $\beta$ & \multicolumn{6}{c||}{$\pm 0.1$} & \multicolumn{6}{c||}{$\pm0.2$}
        \\
        \hline
        $s$ & $\prescript{\rm{num}}{}G_{+}^{(s)}$ & $\delta G_{+}^{(s)}$ & $\prescript{\rm{an}}{}G_{+}^{(s)}$ & $\prescript{\rm{num}}{}G_{-}^{(s)}$ & $\delta G_{-}^{(s)}$ & $\prescript{\rm{an}}{}G_{-}^{(s)}$ & $\prescript{\rm{num}}{}G_{+}^{(s)}$ & $\delta G_{+}^{(s)}$ & $\prescript{\rm{an}}{}G_{+}^{(s)}$ & $\prescript{\rm{num}}{}G_{-}^{(s)}$ & $\delta G_{-}^{(s)}$ & $\prescript{\rm{an}}{}G_{-}^{(s)}$ \\
        \hline
        $1$ & 1.278027 & $1.3 \cdot10^{-3}$ & 1.279698 & $-$0.130204 & $1.3\cdot10^{-2}$ & $-$0.128546 & 1.284413 & $1.3 \cdot10^{-3}$ & 1.286083 & $-$0.127949 & $1.3\cdot10^{-2}$ & $-$0.126291\\
        $2$ & 4.826191 & $5.3\cdot10^{-6}$ & 4.826217 & 1.507991 & $1.7\cdot10^{-5}$ & 1.508019 & 4.847506 & $5.3\cdot10^{-6}$ & 4.847531 & 1.511076 & $1.7\cdot10^{-5}$ & 1.511101\\
        $3$ & 8.232644 & $6.5\cdot10^{-7}$ & 8.232650 & 1.036883 & $5.3\cdot10^{-6}$ & 1.036888 & 8.293937 & $4.0\cdot10^{-8}$ & 8.293938 & 1.040816 & $3.0\cdot10^{-7}$ & 1.040816\\
        $4$ & 15.84369 & $3.7\cdot10^{-9}$ & 15.84369 & 0.883289 & $4.1\cdot10^{-9}$ & 0.883289 & 16.00479 & $3.4\cdot10^{-9}$ & 16.00479 & 0.888051 & $3.8\cdot10^{-9}$ & 0.888051\\
        \hline
        $\beta$ & \multicolumn{6}{c||}{$\pm0.3$} & \multicolumn{6}{c||}{$\pm0.4$}  
        \\
        \hline
        $s$ & $\prescript{\rm{num}}{}G_{+}^{(s)}$ & $\delta G_{+}^{(s)}$ & $\prescript{\rm{an}}{}G_{+}^{(s)}$ & $\prescript{\rm{num}}{}G_{-}^{(s)}$ & $\delta G_{-}^{(s)}$ & $\prescript{\rm{an}}{}G_{-}^{(s)}$ & $\prescript{\rm{num}}{}G_{+}^{(s)}$ & $\delta G_{+}^{(s)}$ & $\prescript{\rm{an}}{}G_{+}^{(s)}$ & $\prescript{\rm{num}}{}G_{-}^{(s)}$ & $\delta G_{-}^{(s)}$ & $\prescript{\rm{an}}{}G_{-}^{(s)}$ \\
        \hline
        $1$ & 1.295220 & $1.3 \cdot10^{-3}$ & 1.296891 & $-$0.124144 & $1.4\cdot10^{-2}$ & $-$0.122486 & 1.310714 & $1.3\cdot10^{-3}$ & 1.312384 & $-$0.118714 & $1.4\cdot10^{-2}$ & $-$0.117056\\
        $2$ & 4.883734 & $5.2\cdot10^{-6}$ & 4.883760 & 1.516290 & $1.7\cdot10^{-5}$ & 1.516316 & 4.936000 & $5.2\cdot10^{-6}$ & 4.936024 & 1.523759 & $1.7\cdot10^{-5}$ & 1.523785\\
        $3$ & 8.398498 & $9.6\cdot10^{-9}$ & 8.398498 & 1.047484 & $6.2\cdot10^{-8}$ & 1.047484 & 8.550189 & $4.4\cdot10^{-9}$ & 8.550189 & 1.057066 & $2.6\cdot10^{-8}$ & 1.057066\\
        $4$ & 16.28058 & $2.8\cdot10^{-9}$ & 16.28058 & 0.896140 & $3.1\cdot10^{-9}$ & 0.896140 & 16.68283 & $1.7\cdot10^{-9}$ & 16.68283 & 0.907801 & $1.9\cdot10^{-9}$ & 0.907801\\
        \hline
        $\beta$ & \multicolumn{6}{c||}{$\pm 0.6$} & \multicolumn{6}{c||}{$\pm0.8$}
        \\
        \hline
        $s$ & $\prescript{\rm{num}}{}G_{+}^{(s)}$ & $\delta G_{+}^{(s)}$ & $\prescript{\rm{an}}{}G_{+}^{(s)}$ & $\prescript{\rm{num}}{}G_{-}^{(s)}$ & $\delta G_{-}^{(s)}$ & $\prescript{\rm{an}}{}G_{-}^{(s)}$ & $\prescript{\rm{num}}{}G_{+}^{(s)}$ & $\delta G_{+}^{(s)}$ & $\prescript{\rm{an}}{}G_{+}^{(s)}$ & $\prescript{\rm{num}}{}G_{-}^{(s)}$ & $\delta G_{-}^{(s)}$ & $\prescript{\rm{an}}{}G_{-}^{(s)}$ \\
        \hline
        $1$ & 1.357530 & $1.2\cdot10^{-3}$ & 1.359201 & $-$0.102485 & $1.6\cdot10^{-2}$ & $-$0.100827 & 1.430583 & $1.2\cdot10^{-3}$ & 1.432254 & $-$0.077690 & $2.2\cdot10^{-2}$ & $-$0.076031\\
        $2$ & 5.096307 & $5.0\cdot10^{-6}$ & 5.096332 & 1.546280 & $1.7\cdot10^{-5}$ & 1.546306 & 5.353646 & $4.8\cdot10^{-6}$ & 5.353672 & 1.581268 & $1.6\cdot10^{-5}$ & 1.581293\\
        $3$ & 9.021636 & $4.8\cdot10^{-9}$ & 9.021636 & 1.086184 & $1.8\cdot10^{-8}$ & 1.086184 & 9.797605 & $1.2\cdot10^{-7}$ & 9.797606 & 1.132092 & $1.2\cdot10^{-7}$ & 1.132092\\
        $4$ & 17.94894 & $4.0\cdot10^{-9}$ & 17.94894 & 0.943496 & $4.1\cdot10^{-9}$ & 0.943496 & 20.08358 & $1.6\cdot10^{-7}$ & 20.08358 & 1.000555 & $1.5\cdot10^{-7}$ & 1.000555\\
        \hline
        $\beta$ & \multicolumn{6}{c||}{$\pm1.0$} & \multicolumn{6}{c||}{$\pm1.2$}  
        \\
        \hline
        $s$ & $\prescript{\rm{num}}{}G_{+}^{(s)}$ & $\delta G_{+}^{(s)}$ & $\prescript{\rm{an}}{}G_{+}^{(s)}$ & $\prescript{\rm{num}}{}G_{-}^{(s)}$ & $\delta G_{-}^{(s)}$ & $\prescript{\rm{an}}{}G_{-}^{(s)}$ & $\prescript{\rm{num}}{}G_{+}^{(s)}$ & $\delta G_{+}^{(s)}$ & $\prescript{\rm{an}}{}G_{+}^{(s)}$ & $\prescript{\rm{num}}{}G_{-}^{(s)}$ & $\delta G_{-}^{(s)}$ & $\prescript{\rm{an}}{}G_{-}^{(s)}$ \\
        \hline
        $1$ & 1.541743 & $1.1\cdot10^{-3}$ & 1.543419 & $-$0.041157 & $4.2\cdot10^{-2}$ & $-$0.039496 & 1.719609 & $1.0\cdot10^{-3}$ & 1.721355 & 0.014428 & $1.1\cdot10^{-1}$ & 0.016128\\
        $2$ & 5.762229 & $5.9\cdot10^{-6}$ & 5.762263 & 1.634113 & $1.7\cdot10^{-5}$ & 1.634141 & 6.459433 & $3.8\cdot10^{-5}$ & 6.459677 & 1.717562 & $3.9\cdot10^{-5}$ & 1.717628\\
        $3$ & 11.07682 & $2.3\cdot10^{-6}$ & 11.07685 & 1.202961 & $2.0\cdot10^{-6}$ & 1.202964 & 13.38785 & $5.4\cdot10^{-5}$ & 13.38857 & 1.318577 & $4.1\cdot10^{-5}$ & 1.318630\\
        $4$ & 23.73232 & $3.1\cdot10^{-6}$ & 23.73240 & 1.090463 & $2.7\cdot10^{-6}$ & 1.090466 & 30.69774 & $7.3\cdot10^{-5}$ & 30.69996 & 1.241685 & $5.6\cdot10^{-5}$ & 1.241754\\
        \hline
        \end{tabular}}
        \resizebox{\textwidth}{4.2cm}{
        \begin{tabular}{| c | c | c | c || c | c | c || c | c | c || c | c | c || c | c | c ||}
        \hline
        $\alpha$ & \multicolumn{12}{c|}{$1.5$}\\
        \hline
        $\beta$ & \multicolumn{6}{c||}{$\pm 0.2$} & \multicolumn{6}{c||}{$\pm0.4$}
        \\
        \hline
        $s$ & $\prescript{\rm{num}}{}G_{+}^{(s)}$ & $\delta G_{+}^{(s)}$ & $\prescript{\rm{an}}{}G_{+}^{(s)}$ & $\prescript{\rm{num}}{}G_{-}^{(s)}$ & $\delta G_{-}^{(s)}$ & $\prescript{\rm{an}}{}G_{-}^{(s)}$ & $\prescript{\rm{num}}{}G_{+}^{(s)}$ & $\delta G_{+}^{(s)}$ & $\prescript{\rm{an}}{}G_{+}^{(s)}$ & $\prescript{\rm{num}}{}G_{-}^{(s)}$ & $\delta G_{-}^{(s)}$ & $\prescript{\rm{an}}{}G_{-}^{(s)}$ \\
        \hline
        $1$ & 0.296090 & $5.5\cdot10^{-3}$ & 0.297726 & $-$0.594047 & $2.7\cdot10^{-3}$ & $-$0.592423 & 0.303295 & $5.4\cdot10^{-3}$ & 0.304931 & $-$0.590763 & $2.8\cdot10^{-3}$ & $-$0.589140 \\
        $2$ & 2.411201 & $1.1\cdot10^{-5}$ & 2.411226 & 1.020671 & $2.5\cdot10^{-5}$ & 1.020696 & 2.426073 & $1.1\cdot10^{-5}$ & 2.426099 & 1.023934 & $2.5\cdot10^{-5}$ & 1.023959\\
        $3$ & 2.535125 & $2.9\cdot10^{-6}$ & 2.535132 & 0.508592 & $1.4\cdot10^{-5}$ & 0.508600 & 2.562830 & $1.8\cdot10^{-7}$ & 2.562831 & 0.511696 & $8.8\cdot10^{-7}$ & 0.511696\\
        $4$ & 3.160473 & $6.2\cdot10^{-10}$ & 3.160473 & 0.322581 & $8.2\cdot10^{-10}$ & 0.322581 & 3.208432 & $7.4\cdot10^{-10}$ & 3.208432 & 0.325420 & $1.0\cdot10^{-9}$ & 0.325420\\
        \hline
        $\beta$ & \multicolumn{6}{c||}{$\pm0.6$} & \multicolumn{6}{c||}{$\pm0.8$}  
        \\
        \hline
        $s$ & $\prescript{\rm{num}}{}G_{+}^{(s)}$ & $\delta G_{+}^{(s)}$ & $\prescript{\rm{an}}{}G_{+}^{(s)}$ & $\prescript{\rm{num}}{}G_{-}^{(s)}$ & $\delta G_{-}^{(s)}$ & $\prescript{\rm{an}}{}G_{-}^{(s)}$ & $\prescript{\rm{num}}{}G_{+}^{(s)}$ & $\delta G_{+}^{(s)}$ & $\prescript{\rm{an}}{}G_{+}^{(s)}$ & $\prescript{\rm{num}}{}G_{-}^{(s)}$ & $\delta G_{-}^{(s)}$ & $\prescript{\rm{an}}{}G_{-}^{(s)}$ \\
        \hline
        $1$ & 0.315568 & $5.2\cdot10^{-3}$ & 0.317204 & $-$0.585193 & $2.8\cdot10^{-3}$ & $-$0.583569 & 0.333338 & $4.9\cdot10^{-3}$ & 0.334974 & $-$0.577178 & $2.8\cdot10^{-3}$ & $-$0.575554\\
        $2$ & 2.451574 & $1.0\cdot10^{-5}$ & 2.451600 & 1.029492 & $2.5\cdot10^{-5}$ & 1.029518 & 2.488876 & $1.0\cdot10^{-5}$ & 2.488902 & 1.037542 & $2.5\cdot10^{-5}$ & 1.037567 \\
        $3$ & 2.610610 & $3.8\cdot10^{-8}$ & 2.610610 & 0.517002 & $1.9\cdot10^{-7}$ & 0.517002 & 2.681120 & $1.6\cdot10^{-8}$ & 2.681120 & 0.524731 & $7.6\cdot10^{-8}$ & 0.524731\\
        $4$ & 3.291590 & $1.0\cdot10^{-9}$ & 3.291590 & 0.330290 & $1.3\cdot10^{-9}$ & 0.330290 & 3.415336 & $1.5\cdot10^{-9}$ & 3.415336 & 0.337420 & $1.8\cdot10^{-9}$ & 0.337420\\
        \hline
        $\beta$ & \multicolumn{6}{c||}{$\pm 1.4$} & \multicolumn{6}{c||}{$\pm1.6$}
        \\
        \hline
        $s$ & $\prescript{\rm{num}}{}G_{+}^{(s)}$ & $\delta G_{+}^{(s)}$ & $\prescript{\rm{an}}{}G_{+}^{(s)}$ & $\prescript{\rm{num}}{}G_{-}^{(s)}$ & $\delta G_{-}^{(s)}$ & $\prescript{\rm{an}}{}G_{-}^{(s)}$ & $\prescript{\rm{num}}{}G_{+}^{(s)}$ & $\delta G_{+}^{(s)}$ & $\prescript{\rm{an}}{}G_{+}^{(s)}$ & $\prescript{\rm{num}}{}G_{-}^{(s)}$ & $\delta G_{-}^{(s)}$ & $\prescript{\rm{an}}{}G_{-}^{(s)}$ \\
        \hline
        $1$ & 0.428366 & $3.8\cdot10^{-3}$ & 0.430004 & $-$0.535316 & $3.0\cdot10^{-3}$ & $-$0.533691 & 0.479561 & $3.4\cdot10^{-3}$ & 0.481199 & $-$0.513451 & $3.2\cdot10^{-3}$ & $-$0.511826\\
        $2$ & 2.696064 & $9.5\cdot10^{-6}$ & 2.696089 & 1.080590 & $2.4\cdot10^{-5}$ & 1.080616 & 2.813163 & $9.0\cdot10^{-6}$ & 2.813189 & 1.103760 & $2.3\cdot10^{-5}$ & 1.103786\\
        $3$ & 3.085910 & $6.8\cdot10^{-9}$ & 3.085910 & 0.566942 & $3.4\cdot10^{-8}$ & 0.566942 & 3.324329 & $8.0\cdot10^{-9}$ & 3.324329& 0.590270 & $3.3\cdot10^{-8}$ & 0.590270\\
        $4$ & 4.148413 & $8.8\cdot10^{-10}$ & 4.148413 & 0.377141 & $1.0\cdot10^{-9}$ & 0.377141 & 4.597274 & $3.4\cdot10^{-9}$ & 4.597274 & 0.399640 & $3.5\cdot10^{-9}$ & 0.399640\\
        \hline
        $\beta$ & \multicolumn{6}{c||}{$\pm1.8$} & \multicolumn{6}{c||}{$\pm2$}  
        \\
        \hline
        $s$ & $\prescript{\rm{num}}{}G_{+}^{(s)}$ & $\delta G_{+}^{(s)}$ & $\prescript{\rm{an}}{}G_{+}^{(s)}$ & $\prescript{\rm{num}}{}G_{-}^{(s)}$ & $\delta G_{-}^{(s)}$ & $\prescript{\rm{an}}{}G_{-}^{(s)}$ & $\prescript{\rm{num}}{}G_{+}^{(s)}$ & $\delta G_{+}^{(s)}$ & $\prescript{\rm{an}}{}G_{+}^{(s)}$ & $\prescript{\rm{num}}{}G_{-}^{(s)}$ & $\delta G_{-}^{(s)}$ & $\prescript{\rm{an}}{}G_{-}^{(s)}$ \\
        \hline
        $1$ & 0.546211 & $3.0\cdot10^{-3}$ & 0.547850 & $-$0.485691 & $3.4\cdot10^{-3}$ & $-$0.484065 & 0.636197 & $2.6\cdot10^{-3}$ & 0.637840 & $-$0.449451 & $3.6\cdot10^{-3}$ & $-$0.447822\\
        $2$ & 2.971490 & $8.7\cdot10^{-6}$ & 2.971515 & 1.133873 & $2.3\cdot10^{-5}$ & 1.133898 & 3.196019 & $9.4\cdot10^{-6}$ & 3.196049 & 1.174387 & $2.3\cdot10^{-5}$ & 1.174414\\
        $3$ & 3.657424 & $1.2\cdot10^{-7}$ & 3.657425 & 0.621219 & $1.4\cdot10^{-7}$ & 0.621219 & 4.150353 & $2.2\cdot10^{-6}$ & 4.150362 & 0.663976 & $1.9\cdot10^{-6}$ & 0.663977\\
        $4$ & 5.244148 & $1.5\cdot10^{-7}$ & 5.244149 & 0.430069 & $1.4\cdot10^{-7}$ & 0.430069 & 6.240902 & $3.0\cdot10^{-6}$ & 6.240920 & 0.473159 & $2.6\cdot10^{-6}$ & 0.473161 \\
        \hline
        \end{tabular}}
    \end{center}
    \caption{Numerical and analytical values of the spectral sums for $\alpha=0.5,1.5$ and several values of $\beta$ (for the first $n=10^3$ eigenvalues from the Chebyshev polynomial approach). We use the first $300$ eigenvalues and than from $301$ to $10^5$ we use the asymptotic form of large--$n$ \eqref{wkb} (with truncation of the sum behind the term $\propto\lambda^{-1}$).}
    \label{Gpm-table}
\end{table} 
%%%%%%%%%%%%%%%%%%%%%%%%%%%%%%%%%%%%%%%%%%%%%%%%%%%%%%%%%%%%%%%%%%%%%%%%%%%%%%%%%%%%%%%%%%%%%%%%%%%%%%%%%%%%%%%%%%%%%%%%%%%%%%%%%%%%%%%%%%%%%%%%%%%%%%%%%%%%%%%%%%%%%%%%%%%%%%%%%%%%%%%%%%
\section{Discussion}\label{Discussion}
In this work, we have demonstrated that the reformulation of the 't Hooft equation in terms of Baxter's TQ equation remains valid for arbitrary quark/antiquark mass parameters $\alpha_{1,2}$. Thus we developed further the idea originally due to Fateev, Lukyanov and Zamolodchikov \cite{Fateev:2009jf} and used in the case of a single flavour in \cite{Litvinov:2024riz}.

A key outcome of our study is the identification of non-trivial relations \eqref{wronsk-rel}, \eqref{D-nonsymmetric}, \eqref{D+nonsymmetric} that enable us to extract the spectral data directly from the solutions of the TQ equation. As a result, we established non-perturbative in $\alpha_{1,2}$ expressions for the spectral determinants $D_\pm (\lambda)$ and the spectral sums $G^{(s)}_\pm$ that we were able to verify both numerically and (in certain limiting cases) analytically. 

Our results open several directions for future studies.

First, investigating the deeper implications of this integrability structure at the level of form factors, scattering amplitudes and correlation functions could provide valuable insights into the 't Hooft model. As a first step in this direction, it would be useful to determine the asymptotic expansion for the wave functions of mesons $\phi^{(n)}(x|\beta)$ (or their Fourier form $\Psi_n(\nu|\beta)$), and to establish the relation between the linear combinations of coefficients $c^\pm_n$ (it is clear that we need to work with $\bm{Q}_\pm$ functions) in \eqref{Qpm-Kallen–Lehmann} and the current form factors.

Another interesting question within the 2D QCD context is the properties of the spectrum, when we allow for complex masses $m_{1},m_{2}\in\mathbb{C}$. For the $\alpha_1=\alpha_2$ case some results were obtained in \cite{Zamolodchikov:2009pres} and more recently in \cite{Ambrosino:2023dik}. For example, in the special case of tachyonic quark/antiquark masses ($\alpha_{1,2}<-1$), the Hamiltonian $\mathcal{H}$ \eqref{'tHooft-eq} is not hermitian, but commutes with a certain antilinear ``$\mathcal{PT}$-symmetry'' operator. The presence of this symmetry means that the eigenvalues $\lambda_n$ are still real in this regime, except when it is spontaneously broken; then a finite number of eigenvalues develop an imaginary part and split into pairs of complex-conjugate ones. For some complex values of $\alpha_i$ massless degrees of freedom might appear; this manifests itself as a singularity of the spectral sums. One example discussed before in Section \ref{limiting-cases} is the chiral point $\alpha_{1,2}=-1$, in which WZW model is expected to appear as the IR CFT. Note that the spectral sums $G^{(s)}_{\pm}$ are not meromorphic: they exhibit branching points (one of which is the chiral point), so other non-trivial singularities might appear on the non-physical sheets. It would be interesting to understand what kind of CFTs (most likely non-unitary) appear at these critical points (this was first mentioned in \cite{Fateev:2009jf}).  To reliably identify them, one probably needs to be able to compute $1/N_c$ corrections to the critical exponents.

The analytical methods developed in our work are expected to be useful in other theories where similar Fredholm integral equations (with completely integrable kernels \cite{Its:1980,Its:1990MPhysB}) or difference equations akin to TQ equation arise. Examples of such theories include, in particular:
\begin{itemize}
    \item the Ising Field Theory (IFT) in magnetic field --- a different two-dimensional QFT exhibiting confinement. Its particle spectrum in the two-fermion approximation was studied in \cite{Fonseca:2006au}, where the Bethe-Salpeter (BS) equation for this case was derived. Fourier form \eqref{Fourier-def} of this BS equation is very similar to `t Hooft model \cite{Fonseca:2006au}
    \begin{equation}\label{BS-eq-Ising}
        \left(\frac{2\alpha}{\pi}+\nu\tanh{\frac{\pi\nu}{2}}\right)\Psi(\nu) -\frac{1}{16}\frac{\nu}{\cosh{\frac{\pi\nu}{2}}}\int_{-\infty}^{\infty}\limits d\nu'\frac{\nu'}{\cosh{\frac{\pi\nu'}{2}}}\Psi(\nu')=\lambda\int_{-\infty}^{\infty}\limits d\nu'\frac{\pi (\nu-\nu')}{2\sinh{\frac{\pi(\nu-\nu')}{2}}}\Psi(\nu'),
    \end{equation}
    where we have changed the notation compared to \cite{Fonseca:2006au} (FZ)
    \begin{equation}
        \lambda_{FZ}=\frac{f_0}{m^2}=\frac{\pi}{2\alpha},\quad \lambda=\frac{M^2}{4m^2\pi\lambda_{FZ}}.
    \end{equation}
    At this point we have some preliminary results\footnote{Together with Egor Shestopalov.} about the analytical properties of the spectrum in IFT \cite{Litvinov:Ising}. See also \cite{Gao:2025mcg} for a study of another related problem.
    \item topological string theory on toric Calabi-Yau manifolds \cite{Marino:2024tbx, Ambrosino:2023dik}. To a mirror dual of a CY in this class, a certain complex curve in $\mathbb{C}^* \times \mathbb{C}^*$, defined by the equation of the form $P(e^x, e^y) = 0$, is associated. Quantization of this curve then gives a difference operator of TQ type (``quantum spectral curve''). Fredholm determinant associated to this operator is conjectured to compute the full topological string partition function. This statement is referred to as topological string/spectral theory (TS/ST) correspondence.
\end{itemize}

The explicit form of the large-$n$ WKB expansion \eqref{wkb} provides a valuable tool for a wide range of applications, offering analytical insights and facilitating computations in various contexts. For example:
\begin{itemize}
    \item In \cite{Mondejar:2008dt}, using OPE, information about the symmetry of the wave functions \eqref{phi-nontrivial-sym} and knowledge of the first two leading terms in the meson spectrum, the authors managed to calculate a $1/n$ correction to the scalar and pseudoscalar decay constants
    \begin{equation}
        F^{(n)}_S=m_1\sqrt{\frac{N_c}{\pi}}\int_0^1\limits dx\frac{\phi^{(n)}_{12}(x)}{x},\; n=2k+1 \quad \text{and} \quad F^{(n)}_P=m_1\sqrt{\frac{N_c}{\pi}}\int_0^1\limits dx\frac{\phi^{(n)}_{12}(x)}{x},\; n=2k.    
    \end{equation}
    An explicit form of the $1/n$ corrections in WKB will allow us to set the corrections to the decay constants.
    \item Equation \eqref{wkb}, expressing $n$ as a function of $\lambda$, can be thought of as the analogue of Regge trajectory $J(M^2)$, relating spin and hadron masses in higher-dimensional QCD. In particular, our ``Regge trajectory'' is asymptotically linear for energy much larger than quark masses, follows a $n^{\frac{2}{3}}$ behavior in the nonrelativistic (heavy-heavy) limit \eqref{large-alpha-wkb}, and scales as $\sqrt{n}$ in the heavy-light limit \eqref{epsilon-general-m}. The idea of Regge theory is to fit this discrete data by a continuous function $J = \alpha(M^2)$, using analytic continuation. In particular, it is interesting to analytically continue to the negative values of $M^2$. For negative mass squared, $\alpha(M^2)$ can be related to high-energy scattering amplitudes (they were studied for 't Hooft model in \cite{BROWER:1977NPhB,Brower:1977as}). Moreover in asymptotically free theories, such as QCD, there is an expected universal behaviour when $M^2 \to -\infty$: $\lim \limits_{t \to - \infty} \alpha(t) = \text{const}$ \cite{Kirschner:1989pw}. It would be interesting to analyze our formulas from this point of view\footnote{We thank Victor Gorbenko and Jiaxin Qiao for bringing this issue to our attention.}.
\end{itemize}
%%%%%%%%%%%%%%%%%%%%%%%%%%%%%%%%%%%%%%%%%%%%%%%%%%%%%%%%%%%%%%%%%%%%%%%%%%%%%%%%%%%%%%%%%%%%%%%%
%%%%%%%%%%%%%%%%%%%%%%%%%%%%%%%%%%%%%%%%%%%%%%%%%%%%%%%%%%%%%%%%%%%%%%%%%%%%%%%%%%%%%%%%%%%%%%%%
\section*{Acknowledgments}
We acknowledge discussions with Ilia Kochergin, Sergei Lukyanov, Sergei Rutkevich and Alexander Zamolodchikov. A.L. would like to thank all the participants of the conference "Correlation Functions of Integrable Lattice Models and Quantum Field Theories", held at the University of Wuppertal in February 2025, and especially Professor Frank G\"ohmann for creating a wonderful scientific atmosphere and providing with the opportunity to present the results of this paper. P.M. expresses gratitude to Professor Victor Gorbenko and his research group at the \'Ecole Polytechnique F\'ed\'erale de Lausanne for the invitation to give a talk in March 2025 and for their warm hospitality.

The work of A.A. and P.M. performed in Landau Institute has been supported by the Russian Science Foundation under the grant 23-12-00333.
%%%%%%%%%%%%%%%%%%%%%%%%%%%%%%%%%%%%%%%%%%%%%%%%%%%%%%%%%%%%%%%%%%%%%%%%%%%%%%%%%%%%%%%%%%%%%%%%
%%%%%%%%%%%%%%%%%%%%%%%%%%%%%%%%%%%%%%%%%%%%%%%%%%%%%%%%%%%%%%%%%%%%%%%%%%%%%%%%%%%%%%%%%%%%%%%%
\appendix
\section{Analytical structure of $Q$ function}\label{check-analyt-Q} 
Here we provide a proof of the statement about analytic structure of $Q(\nu)$ \eqref{Q-def}
\begin{equation}
    Q(\nu)=\sinh{\frac{\pi\nu}{2}}\left(\frac{2\alpha}{\pi}+\nu\coth{\frac{\pi\nu}{2}}\right)\Psi(\nu).
\end{equation}
The function $Q(\nu)$ has the following zeroes/poles in the strip $\textrm{Im}\;\nu\in[-2,2]$ 
\begin{equation}
    \begin{aligned}
        &\text{necessary zeroes:}\quad &&0,\quad \pm2i,\quad \pm i\nu^*_1(\alpha),\\
        &\text{possible poles:}\quad &&i\nu^*_1(\alpha+\beta),\quad-i\nu^*_1(\alpha-\beta),
    \end{aligned}
\end{equation}
Consider analytical continuation of the equation \eqref{'tHooft-eq-Fourier} (multiplied by $\sinh \frac{\pi \nu}{2}$) in $\nu$ to some point $\nu^*$ in the upper/lower half-strip. As usual, we represent the principal value integral  as a half-sum of integrals with contours just above and just below the real axis
\begin{equation}
    \fint_{-\infty}^{\infty}\limits=\frac{1}{2}\left(\int_{-\infty-i\epsilon}^{\infty-i\epsilon}\limits+\int_{-\infty+i\epsilon}^{\infty+i\epsilon}\limits\right).
\end{equation}
During analytic continuation, the upper/lower contour is deformed and can be represented as the original contour minus/plus the residue at the pole point
\begin{equation}
    \int_{-\infty\pm i\epsilon}^{\infty\pm i\epsilon}  \limits\to \int_{-\infty\pm i\epsilon}^{\infty\pm i\epsilon}\limits \mp 2\pi i\,\underset{\nu'=\nu^*}{\text{Res}}
\end{equation}
This residue only appears from the integral term proportional to $\beta$ and is easy to compute. Then we obtain
\begin{multline} \label{sinh*tHooft-an-cont} \lim_{\nu\to\nu^*}\sinh\frac{\pi\nu}{2}\left(\frac{2(\alpha\pm\beta)}{\pi}+\nu\coth{\frac{\pi\nu}{2}}\right)\Psi(\nu)=\frac{2i\beta}{\pi}\sinh\frac{\pi\nu^*}{2}\int^{\infty}_{-\infty}\limits d\nu'\frac{1}{2\sinh\frac{\pi(\nu^*-\nu')}{2}}\Psi(\nu')+
    \\+\lambda\sinh\frac{\pi\nu^*}{2}\int^{\infty}_{-\infty}\limits d\nu' \frac{\pi(\nu^*-\nu')}{2\sinh{\frac{\pi(\nu^*-\nu')}{2}}}\Psi(\nu').
\end{multline}
Both terms on the r.h.s. cannot be singular at any point in the strip. This means that if $\Psi(\nu^*)$ develops a pole, it can only happen if $\nu^*$ is a zero of the prefactor
\begin{equation} \sinh \frac{\pi \nu^*}{2} \left(
    \frac{2(\alpha\pm\beta)}{\pi}+\nu\coth{\frac{\pi\nu}{2}}\right)=0,
\end{equation}
$\Psi (\nu)$ cannot have a pole at zeroes of $\sinh \frac{\pi \nu}{2}$ ($\nu = 0, \pm 2i$) --- then the l.h.s. of \eqref{sinh*tHooft-an-cont} is finite, but the r.h.s. is zero. $\Psi$ also cannot have a pole when $\nu \coth \frac{\pi \nu}{2} + \frac{2\alpha}{\pi} = 0$ (that is, for $\nu =\pm i\nu^*_1(\alpha)$). Thus, the $Q$-function is necessarily zero at $0,\pm 2i, \pm i\nu^*_1(\alpha)$, as announced. The only other possibility for the pole of $\Psi$ is $\nu^*=i\nu_1^*(\alpha+\beta)$ for the upper half-strip and $\nu^*=-i\nu_1^*(\alpha-\beta)$ for the lower one. Then $Q$-function can only have poles at these points as well; this is what we wanted to prove. 

%%%%%%%%%%%%%%%%%%%%%%%%%%%%%%%%%%%%%%%%%%%%%%%%%%%%%%%%%%%%%%%%%%%%%%%%%%%%%%%%%%%%%%%%%%%%%%

\bibliographystyle{MyStyle}
\bibliography{MyBib}

\providecommand{\href}[2]{#2}\begingroup\raggedright\begin{thebibliography}{10}

\bibitem{THOOFT1974461}
G.~{'t Hooft}, \emph{A two-dimensional model for mesons}, \href{https://doi.org/https://doi.org/10.1016/0550-3213(74)90088-1}{\emph{Nuclear Physics B} {\bfseries 75} (1974) 461}.

\bibitem{tHooft:1973alw}
G.~'t~Hooft, \emph{{A Planar Diagram Theory for Strong Interactions}}, \href{https://doi.org/10.1016/0550-3213(74)90154-0}{\emph{Nucl. Phys. B} {\bfseries 72} (1974) 461}.

\bibitem{Callan:1976PhysRev}
C.~G. Callan, N.~Coote and D.~J. Gross, \emph{{Two-dimensional Yang-Mills theory: A model of quark confinement}}, \href{https://doi.org/10.1103/PhysRevD.13.1649}{\emph{Phys. Rev. D} {\bfseries 13} (1976) 1649}.

\bibitem{Hanson:1976ey}
A.~J. Hanson, R.~D. Peccei and M.~K. Prasad, \emph{{Two-Dimensional SU(N) Gauge Theory, Strings and Wings: Comparative Analysis of Meson Spectra and Covariance}}, \href{https://doi.org/10.1016/0550-3213(77)90167-5}{\emph{Nucl. Phys. B} {\bfseries 121} (1977) 477}.

\bibitem{Brower:1979PhysRevD}
R.~C. Brower, W.~L. Spence and J.~H. Weis, \emph{Bound states and asymptotic limits for quantum chromodynamics in two dimensions}, \href{https://doi.org/10.1103/physrevd.19.3024}{\emph{Physical Review D} {\bfseries 19} (1979) 3024–3049}.

\bibitem{Anand:2021qnd}
N.~Anand, A.~L. Fitzpatrick, E.~Katz and Y.~Xin, \emph{{Chiral limit of 2d QCD revisited with lightcone conformal truncation}}, \href{https://doi.org/10.1007/JHEP01(2024)189}{\emph{JHEP} {\bfseries 01} (2024) 189} [\href{https://arxiv.org/abs/2111.00021}{{\ttfamily 2111.00021}}].

\bibitem{Kochergin:2024quv}
I.~V. Kochergin, \emph{{1/N corrections in QCD$_{2}$: small mass limit and threshold states}}, \href{https://doi.org/10.1007/JHEP02(2025)073}{\emph{JHEP} {\bfseries 02} (2025) 073} [\href{https://arxiv.org/abs/2405.04031}{{\ttfamily 2405.04031}}].

\bibitem{Komargodski:2020mxz}
Z.~Komargodski, K.~Ohmori, K.~Roumpedakis and S.~Seifnashri, \emph{{Symmetries and strings of adjoint QCD$_{2}$}}, \href{https://doi.org/10.1007/JHEP03(2021)103}{\emph{JHEP} {\bfseries 03} (2021) 103} [\href{https://arxiv.org/abs/2008.07567}{{\ttfamily 2008.07567}}].

\bibitem{Popov:2022vud}
F.~K. Popov, \emph{{Supersymmetry in QCD2 coupled to fermions}}, \href{https://doi.org/10.1103/PhysRevD.105.074005}{\emph{Phys. Rev. D} {\bfseries 105} (2022) 074005} [\href{https://arxiv.org/abs/2202.04017}{{\ttfamily 2202.04017}}].

\bibitem{Dempsey:2021xpf}
R.~Dempsey, I.~R. Klebanov and S.~S. Pufu, \emph{{Exact symmetries and threshold states in two-dimensional models for QCD}}, \href{https://doi.org/10.1007/JHEP10(2021)096}{\emph{JHEP} {\bfseries 10} (2021) 096} [\href{https://arxiv.org/abs/2101.05432}{{\ttfamily 2101.05432}}].

\bibitem{Dempsey:2023fvm}
R.~Dempsey, I.~R. Klebanov, S.~S. Pufu and B.~T. S\o{}gaard, \emph{{Lattice Hamiltonian for adjoint QCD$_{2}$}}, \href{https://doi.org/10.1007/JHEP08(2024)009}{\emph{JHEP} {\bfseries 08} (2024) 009} [\href{https://arxiv.org/abs/2311.09334}{{\ttfamily 2311.09334}}].

\bibitem{Damia:2024kyt}
J.~A. Damia, G.~Galati and L.~Tizzano, \emph{{Symmetries, universes and phases of QCD$_{2}$ with an adjoint Dirac fermion}}, \href{https://doi.org/10.1007/JHEP12(2024)230}{\emph{JHEP} {\bfseries 12} (2025) 230} [\href{https://arxiv.org/abs/2409.17989}{{\ttfamily 2409.17989}}].

\bibitem{Asrat:2022aov}
M.~Asrat, \emph{{(1+1)D QCD with heavy adjoint quarks}}, \href{https://doi.org/10.1103/PhysRevD.107.106022}{\emph{Phys. Rev. D} {\bfseries 107} (2023) 106022} [\href{https://arxiv.org/abs/2212.02162}{{\ttfamily 2212.02162}}].

\bibitem{Dubovsky:2018dlk}
S.~Dubovsky, \emph{{A Simple Worldsheet Black Hole}}, \href{https://doi.org/10.1007/JHEP07(2018)011}{\emph{JHEP} {\bfseries 07} (2018) 011} [\href{https://arxiv.org/abs/1803.00577}{{\ttfamily 1803.00577}}].

\bibitem{Donahue:2019adv}
J.~C. Donahue and S.~Dubovsky, \emph{{Confining Strings, Infinite Statistics and Integrability}}, \href{https://doi.org/10.1103/PhysRevD.101.081901}{\emph{Phys. Rev. D} {\bfseries 101} (2020) 081901} [\href{https://arxiv.org/abs/1907.07799}{{\ttfamily 1907.07799}}].

\bibitem{Donahue:2019fgn}
J.~C. Donahue and S.~Dubovsky, \emph{{Classical Integrability of the Zigzag Model}}, \href{https://doi.org/10.1103/PhysRevD.102.026005}{\emph{Phys. Rev. D} {\bfseries 102} (2020) 026005} [\href{https://arxiv.org/abs/1912.08885}{{\ttfamily 1912.08885}}].

\bibitem{Donahue:2022jxu}
J.~C. Donahue and S.~Dubovsky, \emph{{Quantization of the zigzag model}}, \href{https://doi.org/10.1007/JHEP08(2022)047}{\emph{JHEP} {\bfseries 08} (2022) 047} [\href{https://arxiv.org/abs/2202.11746}{{\ttfamily 2202.11746}}].

\bibitem{Fateev:2009jf}
V.~A. Fateev, S.~L. Lukyanov and A.~B. Zamolodchikov, \emph{{On mass spectrum in 't Hooft's 2D model of mesons}}, \href{https://doi.org/10.1088/1751-8113/42/30/304012}{\emph{J. Phys. A} {\bfseries 42} (2009) 304012} [\href{https://arxiv.org/abs/0905.2280}{{\ttfamily 0905.2280}}].

\bibitem{Litvinov:2024riz}
A.~Litvinov and P.~Meshcheriakov, \emph{{Meson mass spectrum in QCD2 't Hooft's model}}, \href{https://doi.org/10.1016/j.nuclphysb.2024.116766}{\emph{Nucl. Phys. B} {\bfseries 1010} (2025) 116766} [\href{https://arxiv.org/abs/2409.11324}{{\ttfamily 2409.11324}}].

\bibitem{Ambrosino:2023dik}
F.~Ambrosino and S.~Komatsu, \emph{{2d QCD and integrability. Part I. \textquoteright{}t Hooft model}}, \href{https://doi.org/10.1007/JHEP02(2025)126}{\emph{JHEP} {\bfseries 02} (2025) 126} [\href{https://arxiv.org/abs/2312.15598}{{\ttfamily 2312.15598}}].

\bibitem{Federbush:1977PRevD}
P.~Federbush and A.~Tromba, \emph{{A Note on 't Hooft's Hamiltonian in Two-Dimensional QCD}}, \href{https://doi.org/10.1103/PhysRevD.15.2913}{\emph{Phys. Rev. D} {\bfseries 15} (1977) 2913}.

\bibitem{Baxter:1972hz}
R.~J. Baxter, \emph{{Partition function of the eight vertex lattice model}}, \href{https://doi.org/10.1016/0003-4916(72)90335-1}{\emph{Annals Phys.} {\bfseries 70} (1972) 193}.

\bibitem{Einhorn:1976uz}
M.~B. Einhorn, \emph{{Form-Factors and Deep Inelastic Scattering in Two-Dimensional Quantum Chromodynamics}}, \href{https://doi.org/10.1103/PhysRevD.14.3451}{\emph{Phys. Rev. D} {\bfseries 14} (1976) 3451}.

\bibitem{Its:1980}
A.~R. Its, A.~G. Izergin and V.~E. Korepin, \emph{{Temperature correlators of the impenetrable Bose gas as an integrable system}}, {\emph{Commun. Math. Phys.} {\bfseries 129} (1990) 205 }.

\bibitem{Its:1990MPhysB}
A.~Its, A.~Izergin, V.~Korepin and N.~Slavnov, \emph{Differential equations for quantum correlation functions}, \href{https://doi.org/10.1142/s0217979290000504}{\emph{International Journal of Modern Physics B} {\bfseries 04} (1990) 1003–1037}.

\bibitem{zbMATH01284258}
P.~Deift, \emph{Integrable operators},  in \emph{Differential operators and spectral theory. M. Sh. Birman's 70th anniversary collection}, pp.~69--84, Providence, RI: American Mathematical Society, (1999).

\bibitem{Mondejar:2009td}
J.~Mondejar and A.~Pineda, \emph{{Deep inelastic scattering and factorization in the 't Hooft Model}}, \href{https://doi.org/10.1103/PhysRevD.79.085011}{\emph{Phys. Rev. D} {\bfseries 79} (2009) 085011} [\href{https://arxiv.org/abs/0901.3113}{{\ttfamily 0901.3113}}].

\bibitem{GEPNER1985481}
D.~Gepner, \emph{{Non-abelian bosonization and multiflavor QED and QCD in two dimensions}}, \href{https://doi.org/https://doi.org/10.1016/0550-3213(85)90458-4}{\emph{Nuclear Physics B} {\bfseries 252} (1985) 481}.

\bibitem{Frishman:2010zz}
Y.~Frishman and J.~Sonnenschein, \emph{{Non-perturbative field theory: From two-dimensional conformal field theory to QCD in four dimensions}}, Cambridge Monographs on Mathematical Physics. Cambridge University Press, 7, 2014, \href{https://doi.org/10.1017/CBO9780511770838}{10.1017/CBO9780511770838}.

\bibitem{Delmastro:2021otj}
D.~Delmastro, J.~Gomis and M.~Yu, \emph{{Infrared phases of 2d QCD}}, \href{https://doi.org/10.1007/JHEP02(2023)157}{\emph{JHEP} {\bfseries 02} (2023) 157} [\href{https://arxiv.org/abs/2108.02202}{{\ttfamily 2108.02202}}].

\bibitem{Zhitnitsky:1985um}
A.~R. Zhitnitsky, \emph{{On Chiral Symmetry Breaking in {QCD} in Two-dimensions ($N_c \to\infty$)}}, \href{https://doi.org/10.1016/0370-2693(85)91255-9}{\emph{Phys. Lett. B} {\bfseries 165} (1985) 405}.

\bibitem{Burkardt:1995eb}
M.~Burkardt, \emph{{Trivial vacua, high orders in perturbation theory and nontrivial condensates}}, \href{https://doi.org/10.1103/PhysRevD.53.933}{\emph{Phys. Rev. D} {\bfseries 53} (1996) 933} [\href{https://arxiv.org/abs/hep-ph/9509226}{{\ttfamily hep-ph/9509226}}].

\bibitem{Gell-Mann:1968hlm}
M.~Gell-Mann, R.~J. Oakes and B.~Renner, \emph{{Behavior of current divergences under $SU(3)\times SU(3)$}}, \href{https://doi.org/10.1103/PhysRev.175.2195}{\emph{Phys. Rev.} {\bfseries 175} (1968) 2195}.

\bibitem{ZIYATDINOV:2010ModPhysA}
I.~Ziyatdinov, \emph{{Asymptotic properties of mass spectrum in 't Hooft's model of mesons}}, \href{https://doi.org/10.1142/S0217751X10050287}{\emph{Int. J. Mod. Phys. A} {\bfseries 25} (2010) 3899} [\href{https://arxiv.org/abs/1003.4304}{{\ttfamily 1003.4304}}].

\bibitem{Burkardt:2000PhysRevD}
M.~Burkardt and N.~Uraltsev, \emph{{Analytical heavy quark expansion in the 't Hooft model}}, \href{https://doi.org/10.1103/PhysRevD.63.014004}{\emph{Phys. Rev. D} {\bfseries 63} (2000) 014004}.

\bibitem{Zhitnitsky:1995qa}
A.~R. Zhitnitsky, \emph{{Lessons from QCD in two-dimensions $N\rightarrow\infty$: Vacuum structure, asymptotic series, instantons and all that...}}, \href{https://doi.org/10.1103/PhysRevD.53.5821}{\emph{Phys. Rev. D} {\bfseries 53} (1996) 5821} [\href{https://arxiv.org/abs/hep-ph/9510366}{{\ttfamily hep-ph/9510366}}].

\bibitem{Bigi:1998kc}
I.~I.~Y. Bigi, M.~A. Shifman, N.~Uraltsev and A.~I. Vainshtein, \emph{{Heavy flavor decays, OPE and duality in two-dimensional 't Hooft model}}, \href{https://doi.org/10.1103/PhysRevD.59.054011}{\emph{Phys. Rev. D} {\bfseries 59} (1999) 054011} [\href{https://arxiv.org/abs/hep-ph/9805241}{{\ttfamily hep-ph/9805241}}].

\bibitem{Chabysheva:2012fe}
S.~S. Chabysheva and J.~R. Hiller, \emph{{Dynamical model for longitudinal wave functions in light-front holographic QCD}}, \href{https://doi.org/10.1016/j.aop.2013.06.016}{\emph{Annals Phys.} {\bfseries 337} (2013) 143} [\href{https://arxiv.org/abs/1207.7128}{{\ttfamily 1207.7128}}].

\bibitem{Zamolodchikov:2009pres}
A.~Zamolodchikov, \emph{{On Confining Interactions in $1+1$. Talk at Conference in the Memory of Aliosha Zamolodchikov, Saclay}}, {\emph{\url{https://indico.in2p3.fr/event/1886/sessions/3945/attachments/17798/21781/Zamolodchikov.pdf}} (2009) }.

\bibitem{Fonseca:2006au}
P.~Fonseca and A.~Zamolodchikov, \emph{{Ising spectroscopy. I. Mesons at T $<$ T$_c$}},  \href{https://arxiv.org/abs/hep-th/0612304}{{\ttfamily hep-th/0612304}}.

\bibitem{Litvinov:Ising}
A.~Litvinov, P.~Meshcheriakov and E.~Shestopalov In preparation, 2025.

\bibitem{Gao:2025mcg}
Y.~Gao, Y.~Jiang and J.~Wu, \emph{{Mesons in a quantum Ising ladder}},  \href{https://arxiv.org/abs/2502.15463}{{\ttfamily 2502.15463}}.

\bibitem{Marino:2024tbx}
M.~Marino, \emph{{Les Houches lectures on non-perturbative topological strings}},  \href{https://arxiv.org/abs/2411.16211}{{\ttfamily 2411.16211}}.

\bibitem{Mondejar:2008dt}
J.~Mondejar and A.~Pineda, \emph{{1/N(c) and 1/n preasymptotic corrections to Current-Current correlators}}, \href{https://doi.org/10.1088/1126-6708/2008/06/039}{\emph{JHEP} {\bfseries 06} (2008) 039} [\href{https://arxiv.org/abs/0803.3625}{{\ttfamily 0803.3625}}].

\bibitem{BROWER:1977NPhB}
R.~Brower, J.~Ellis, M.~Schmidt and J.~Weis, \emph{{Hadron scattering in two-dimensional QCD: (I). Formalism and leading order calculations}}, \href{https://doi.org/https://doi.org/10.1016/0550-3213(77)90303-0}{\emph{Nuclear Physics B} {\bfseries 128} (1977) 131}.

\bibitem{Brower:1977as}
R.~C. Brower, J.~R. Ellis, M.~G. Schmidt and J.~H. Weis, \emph{{Hadron Scattering in Two-Dimensional QCD. 2. Second Order Calculations, Multi-Regge and Inclusive Reactions}}, \href{https://doi.org/10.1016/0550-3213(77)90304-2}{\emph{Nucl. Phys. B} {\bfseries 128} (1977) 175}.

\bibitem{Kirschner:1989pw}
R.~Kirschner and L.~N. Lipatov, \emph{{Bare Reggeons in Asymptotic Free Theories}}, \href{https://doi.org/10.1007/BF01549678}{\emph{Yad. Fiz.} {\bfseries 50} (1989) 1739}.

\end{thebibliography}\endgroup
\end{document}